\documentclass[11pt,a4paper]{article}
\pdfoutput=1

\usepackage{geometry,amsmath,amsfonts}
\usepackage{slashed}
\usepackage{epsfig}
\usepackage{latexsym}
\usepackage{graphicx}
\usepackage{amssymb}
\usepackage{verbatim}
\usepackage{bm}
\usepackage{dsfont}
\usepackage[footnotesize]{caption}
\usepackage{cancel}
\usepackage{cite}
\usepackage{graphicx}
\usepackage[utf8]{inputenc}
\usepackage{epsfig}
\usepackage{soul}
\usepackage[utf8]{inputenc} 
\usepackage{multirow}
\usepackage{tikz}
\usepackage{color}
\usepackage{multirow}
\usepackage{color}
\usepackage{rotating}
\usepackage{ifthen}
\usepackage{subfigure}
\usepackage[utf8]{inputenc}
\usepackage{url}
\usepackage{hyperref}
\hypersetup{
    colorlinks,
    citecolor=black,
    filecolor=black,
    linkcolor=black,
    urlcolor=black
}
\usepackage{bbold}

\def\ee{\end{equation}}
\newcommand{\eeq}{\end{equation}}
\newcommand{\ec}{\end{center}}
\newcommand{\bc}{\begin{center}}

\newcommand{\eea}{\end{eqnarray}}

\newcommand{\eee}[0]{\end{eqnarray}}

\newcommand{\non}[0]{\nonumber \\}

\renewcommand{\tilde}{\widetilde}
\newcommand{\de}{\mbox{d}}

  \usepackage{color}
 \definecolor{cblue}{RGB}{100,5,255}
 \definecolor{cred}{RGB}{180,50,40} 

\begin{document}

\begin{minipage}{0.4\textwidth}
 \begin{flushleft}
CP3-17-23\\
LPT-Orsay-17-35 \\
DESY 17-124
\end{flushleft}
\end{minipage} \hfill
\begin{minipage}{0.5\textwidth}
\begin{flushright} 
\end{flushright}
 \end{minipage}

\vskip 1.1cm

\begin{center}

{{\LARGE\bf  Neutrino masses, leptogenesis  and  dark matter\\ from small lepton number violation?}}

\vskip 1cm

{\large  Asmaa Abada$^a$, Giorgio Arcadi$^b$, Valerie~Domcke$^{c,d}$ and  Michele Lucente$^{e}$ }\\[3mm]

\vspace*{.5cm} 
$^{a}$ {\it
Laboratoire de Physique Th\'eorique, CNRS, \\
Univ. Paris-Sud, Universit\'e Paris-Saclay, 91405 Orsay, France}

${}^b$ {\it Max Planck Institut f\"ur Kernphysik, Saupfercheckweg 1, 69117 Heidelberg, Germany}\\
${}^c$ {\it AstroParticule et Cosmologie (APC)/Paris Centre for Cosmological Physics (PCCP),\\
Universit\'e Paris Diderot, Paris, France}

${}^d$ {\it Deutsches Elektronensynchrotron (DESY), Notkestrasse 85, 22765 Hamburg, Germany}

${}^e$ {\it Centre for Cosmology, Particle Physics and Phenomenology (CP3)\\
Universit\'e catholique de Louvain,
Chemin du Cyclotron 2, 1348 Louvain-la-Neuve, Belgium}\\

\end{center}

\vskip 0.1cm

\begin{abstract}
We consider the possibility of simultaneously addressing the baryon asymmetry of the Universe,  the dark matter problem and the neutrino mass generation in minimal extensions of the Standard Model via sterile fermions with (small) 
total lepton number violation. 
Within the framework of Inverse and Linear Seesaw models, the small lepton number violating parameters set the mass scale of the active neutrinos, the efficiency of leptogenesis through a small mass splitting between pairs of sterile fermions as well as the mass scale of a sterile neutrino dark matter candidate.
We provide an improved parametrisation of these seesaw models taking into account existing experimental constraints and derive a linearised system of Boltzmann equations to describe the leptogenesis process, which allows for an efficient investigation of the parameter space.
This in particular enables us to perform a systematic study of the strong washout regime of leptogenesis.
Our  study reveals that one can have a successful leptogenesis at the temperature of the electroweak scale through oscillations between two sterile states with a natural origin of the (necessary) strong degeneracy in their mass spectrum. The minimal model however requires a non-standard cosmological history to account for the relic dark matter.
Finally, we discuss the prospect for neutrinoless double beta decay and for testing, in future experiments, the values of mass and different active-sterile mixings required for successful leptogenesis.

\end{abstract}

\thispagestyle{empty}

\newpage
\tableofcontents
\newpage

\setcounter{page}{1}
\section{Introduction}\label{introduction}

The generation of the baryon asymmetry of the Universe  (BAU) is one of the major puzzles of modern particle physics and  leptogenesis is among the most popular solutions.  One of the simplest realisations of  leptogenesis is the so called ``thermal leptogenesis"  \cite{Davidson:2008bu}  
, which relies on the out-of-equilibrium decay of heavy right-handed (RH) neutrinos and in which baryogenesis is in general tied to the seesaw mechanism for the light neutrino mass generation~\cite{Minkowski:1977sc,GellMann:1980vs,Yanagida:1979as,Glashow:1979nm,Mohapatra:1979ia}.

At low seesaw scales,  thermal leptogenesis is however very fine-tuned and difficult to implement, and one must consider alternatives to generate any lepton asymmetry.
 An interesting  alternative is the so-called ``ARS" mechanism,  first proposed by Akhmedov, Rubakov and Smirnov~\cite{Akhmedov:1998qx}, in which a lepton asymmetry is produced by the CP-violating oscillations of a pair of heavy sterile neutrinos.
 This mechanism was then implemented in the ~$\nu$-MSM~\cite{Asaka:2005an,Asaka:2005pn,Shaposhnikov:2008pf} with the aim of simultaneously
addressing  the issues of i) neutrino mass generation,  ii) the BAU, and iii) of providing a viable dark matter (DM)
candidate. In this approach, three RH  neutrinos $N_R^{1,2,3}$ were added to the Standard Model  (SM), the lightest of them (with mass at the $\sim$ keV  scale) is almost sterile -  in the sense that its mixings to the active light neutrinos and to the other sterile fermions are negligible - playing thus the role of the DM  candidate. The two other (heavier) RH neutrinos are responsible for the generation of the light neutrino masses and of the lepton asymmetries, both at early times, giving rise to the BAU, and at later times, enabling the production of the correct  relic DM abundance~\cite{Laine:2008pg}. The strong condition  in order to achieve all these tasks simultaneously is that the heavier two RH neutrinos are almost degenerate in mass. 
Variants of this scenario capable of achieving a successful BAU while accommodating neutrino data,  in some cases without in addition providing a DM candidate, have also recently been considered in~\cite{Canetti:2012vf,Canetti:2012kh,Drewes:2012ma,Canetti:2014dka,Hernandez:2015wna,Abada:2015rta,Hernandez:2016kel}.

Remarkably, the crucial condition of degeneracy in the heavy spectrum can find a natural origin in scenarios  in which the smallness of light neutrino masses is due  to a small violation of the total lepton number, $L = L_e + L_\mu + L_\tau$~\cite{Shaposhnikov:2006nn}. This can be achieved when, for instance, the Inverse Seesaw mechanism (ISS)~\cite{Wyler:1982dd, Mohapatra:1986bd} is embedded into the SM.
Here the light neutrino masses $m_\nu$ are proportional a Majorana mass parameter $\mu = \xi \Lambda$, which violates lepton number by two unites ($\Delta L = 2$). The seesaw scale $\Lambda$ sets the mass scale of the additional heavy SM singlets.
In the limit $\xi \rightarrow 0$ lepton number conservation is restored and $m_\nu \rightarrow 0$ (coining the name 'inverse' seesaw). This is achieved by introducing at least two additional sets of SM singlet fermions  (referred to RH neutrinos and sterile fermions) with opposite lepton number assignment. These combine into pseudo-Dirac pairs with masses of ${\cal O}(\Lambda)$ and mass differences of ${\cal O}(\mu)$, and depending on the realisation, may also result in a sterile fermion with a mass scale $\mu$, which can account for the relic DM abundance, see for example,~\cite{Abada:2014vea,Abada:2014zra,DeRomeri:2017oxa}. The available neutrino data is accommodated within the ISS for large values of the Yukawa couplings and a comparatively low seesaw scale $\Lambda$, which renders this mechanism phenomenologically appealing.  
A second low-scale seesaw mechanism based on a small lepton number violation (LNV) is the Linear Seesaw (LSS)~\cite{Barr:2003nn,Malinsky:2005bi}, which also requires the introduction of two types of  fermionic singlets (RH and sterile) with opposite lepton number assignment, and in which the smallness of neutrino masses is also linked to the small $\Delta L=2$ violation of the total lepton number. The difference with respect to the ISS is that the LNV arises from additional small Yukawa couplings of the `sterile' fermions to the left-handed (LH) neutrinos. The resulting light neutrino mass scale is linearly dependent on these Yukawa couplings, coining the name for this mechanism.

In a previous study~\cite{Abada:2015rta} we investigated the generation of the BAU in low energy realisations of the LSS and ISS models. The analysis conducted in~\cite{Abada:2015rta} was mostly focused on a minimal phenomenological model based on adding two singlet fermions with opposite lepton number to the SM, which are almost degenerate in mass and form a pseudo-Dirac pair. Their mass splitting, as well as the masses of the light neutrinos, are determined by two small parameters, a Majorana mass term and a LNV Yukawa coupling, violating the total lepton number by two units; this scenario essentially resembles an ISS realisation extended by a LSS mass term, model we refer to as ``LSS-ISS". 
Ref.~\cite{Abada:2015rta} included a detailed analysis of leptogenesis in the region of parameter space satisfying the ``weak washout'' condition, i.e.\ where the Yukawa couplings are sufficiently weak to strongly suppress any erasing of the generated asymmetry. In the ``strong washout'' regime, the lack of an efficient method to calculate the baryon asymmetry restricted the analysis to a proof of existence of viable solutions. While successfully accounting for the observed baryon asymmetry, the 
  solutions in the weak washout regime predicted active-sterile mixing angles which are too small to be detectable in current and upcoming experiments.
The analysis of~\cite{Abada:2015rta} 
further included the ISS(2,2) model, the most minimal realisation of the ISS, which requires the addition of two RH neutrinos and two steriles to the SM, leading to two pseudo-Dirac pairs in the heavy sector, as shown in~\cite{Abada:2014vea}.  In this case we found that existing neutrino data forces the mass splitting within the two pseudo-Dirac pairs to be too large to achieve successful leptogenesis in the weak washout regime.

The present work aims at refining the analysis and extending the results presented in~\cite{Abada:2015rta};  the distinctive new aspects of the present study are summarised below. 
We have derived a systematic perturbative expansion of the set of coupled  Boltzmann equations (BE) describing the generation of the lepton asymmetry, which is particularly suited to efficiently describe the strong and intermediate washout regime.\footnote{We point out the complementary method presented in Ref.~\cite{Drewes:2016gmt}, which provides an analytical solution deep in the strong washout regime.} 
Moreover, in addition to the production and decay of heavy neutrinos through top quark radiation considered in~\cite{Abada:2015rta}, we include the production via gauge boson radiation through the exchange of a lepton doublet in the t-channel. 
In the limit of vanishing leptonic chemical potentials the rate of these gauge-mediated processes was found to exceed the one associated to the top quark by about a factor of three~\cite{Anisimov:2010gy,Besak:2012qm}. To include these processes in the current study, 
we re-derive the kinetic equations including all  scattering processes considered in~\cite{Anisimov:2010gy,Besak:2012qm}, which we re-evaluate in the presence of small leptonic chemical potentials. We thus complete our previous derivation~\cite{Abada:2015rta} and correct the source term for sterile neutrinos by taking into account thermal effects of  gauge boson interactions in the set of coupled BE, estimating their impact in generating the BAU (see also \cite{Hernandez:2016kel,Drewes:2016gmt} for similar studies); we further take into account the  re-distribution of the asymmetry in the SM sector by so-called spectator processes. Throughout this paper, we have consistently used the Fermi-Dirac statistic rather than the Maxwell-Boltzmann one used in~\cite{Abada:2015rta}.  
Our improved treatment allows, for the scenarios considered in~\cite{Abada:2015rta}, a full coverage of the parameter space corresponding to the ``strong washout'' regime, where we find 
that not only the LSS-ISS model provides successful leptogenesis, but also the ISS(2,2). Moreover, we find that the strong washout regime allows for solutions with sufficiently high active-sterile mixings 
 to be experimentally observable in the near future. 
Furthermore, we extend our study to the ISS(2,3) model, where in addition to two pseudo-Dirac pairs  (already present in the ISS(2,2) model), a lighter mostly sterile state is present in the mass spectrum, which under suitable conditions can play the role of the DM component. We investigate whether viable leptogenesis could be simultaneously compatible with the existence of this stable state and with a solution to the DM problem.

Finally, in these extensions of the SM with extra neutrinos (RH or steriles), the effective mass in neutrinoless double beta decay ($0 \nu \beta \beta$)  is modified and incorporates the additional CP-violating phases and the extra mixing angles. It has been shown in~\cite{LopezPavon:2012zg,Abada:2014nwa,Drewes:2015iva} that if 
sterile neutrinos are present, a signal in $0 \nu \beta \beta$ does not necessarily imply an inverted hierarchy (IH) for the light neutrino spectrum. Part of this project is devoted to study the impact of the additional neutral fermions considered in our minimal scenarios (LSS-ISS , ISS (2,2) and (2,3)) on the effective mass in $0 \nu \beta \beta$  when leptogenesis is at work. 

The remainder of this paper is organised as follows. Section~\ref{Sec:LNVframework} introduces the minimal Inverse and Linear Seesaw models as well as the observational constraints on neutrino mass models. 
The derivation and simplification of the BE responsible for leptogenesis is given in Section~\ref{Sec:mechanism}, additional intermediate results can be found in App.~\ref{app:leptogenesis_details}. Section~\ref{sec:results} is dedicated to the results of the numerical parameter scans, including also our approach to parameterising the above seesaw models in view of existing neutrino data. An analytical approach to understanding these results in the context of the DM problem is given in App.~\ref{app:DM_mixing}. We conclude in Section~\ref{sec:conclusions}.


%
\section{Lepton number violation in minimal low-scale frameworks}\label{Sec:LNVframework}

Adding new neutrinos to the Standard Model leads to a broad range of new phenomenology. Depending on the mass scale of these neutrinos, they may address open questions in cosmology (leptogenesis, dark matter,..) or lead to interesting signals in laboratory experiments (beam-dump experiments, neutrino-less double beta decay,..). In this study, we focus on minimal low-scale seesaw frameworks~\cite{Abada:2014vea} which can account for the observed neutrino masses and mixing with the masses of the neutrinos responsible for generating the BAU not exceeding about 50~GeV. In the following, we will first recall the relevant constraints in this mass range and then turn to the explicit seesaw models. This will lead to the introduction of new fermion fields belonging to two categories: (i) RH neutrinos, which in the interaction basis feature Yukawa interactions with the SM Higgs and lepton doublets and (ii) sterile neutrinos, which have no such couplings. In a slight abuse of notation, we will also apply this categorisation to the LSS, in which case the `sterile' neutrinos in fact have (very suppressed) couplings to the SM. Most of our analysis will be however carried out in the mass basis, where the new states are in general a mixture of the RH and sterile (and active) components. We will thus more generally refer to states dominated by RH and/or sterile components as (SM) singlets.
In all minimal scenarios considered here, the lightest active neutrino will be massless.


\subsection{Constraints}\label{Sec:Constraints}

Beside complying with neutrino oscillation data~\cite{Gonzalez-Garcia:2015qrr,Esteban:2016qun}, the extension of the SM by singlet fermions (RH or sterile ones) is subject to important constraints, which strongly constrain their masses as well as the active-sterile mixing.  We will briefly summarise in this subsection the constraints adopted throughout all our analysis.

Perturbative unitarity~\cite{Chanowitz:1978mv,Durand:1989zs,Korner:1992an,
Bernabeu:1993up,Fajfer:1998px,Ilakovac:1999md} requires  $\frac{\Gamma_{\nu_i}}{m_{\nu_i}}\, < \, \frac{1}{2}\, (i=1,N)$.\footnote{Noticing that the leading contribution to ${\Gamma_{\nu_i}}$
is due to the charged current term, the perturbative
unitarity condition translates into  the following 
bounds:
\begin{equation}\label{eq:sterile:bounds-perturbativity}
m_{\nu_i}^2\,{\bf C}_{ii} \, < 2 \, \frac{M^2_W}{\alpha_w}\, \quad
\quad (i \geq 4)\,,
\end{equation}
where $\alpha_w=g^2_w/4 \pi$, and ${\bf C}_{ii}= \sum_{\alpha=1}^{3} {\cal U}_{\alpha i}^*\,{\cal U}_{\alpha i}\,$, ${\cal U}$ being the lepton mixing matrix. 
} Additional bounds arise from electroweak precision tests~\cite{Akhmedov:2013hec,Basso:2013jka,
Fernandez-Martinez:2015hxa,Fernandez-Martinez:2016lgt,Abada:2013aba} and non-standard interactions~\cite{Antusch:2008tz,Antusch:2014woa,Blennow:2016jkn}.
Sterile fermions can also induce  
potentially large contributions to charged lepton flavour violating (cLFV) observables,
 such as charged lepton flavour violation at low energy like in $\mu-e$ conversion in nuclei, radiative and three-body decays ($\mu\to e\gamma$, $\mu\to eee$)~\cite{Ma:1979px,Gronau:1984ct,Ilakovac:1994kj,Deppisch:2004fa,Deppisch:2005zm,Dinh:2012bp,Alonso:2012ji,Abada:2014kba,Abada:2015oba,Lindner:2016bgg},  as well as cLFV at high energy in Higgs~\cite{Arganda:2014dta,Deppisch:2015qwa,Banerjee:2015gca,BhupalDev:2012zg,Cely:2012bz,Bandyopadhyay:2012px,Das:2012ze} and neutral
 $Z$ boson decays~\cite{Illana:2000ic,Abada:2014cca, Abada:2015zea,DeRomeri:2016gum}; sterile fermions can also impact leptonic and semi-leptonic  meson decays~\cite{Shrock:1980vy,Shrock:1980ct,Atre:2009rg,Abada:2012mc,Abada:2013aba}. They may further contribute to
 lepton flavour conserving but CP-violating observables (due to the additional CP violating phases) such as the charged lepton electric dipole moments~\cite{deGouvea:2005jj,Abada:2015trh,Abada:2016awd}. 
We moreover take into account negative results from 
searches for monochromatic lines in the
spectrum of muons from  $\pi^\pm \to \mu^\pm \nu$
decays~\cite{Kusenko:2009up,Atre:2009rg} as well as those from searches at the LHC~\cite{BhupalDev:2012zg,Cely:2012bz,Bandyopadhyay:2012px}. 
It is worth stressing that particularly severe constraints 
arise from the violation of lepton universality
in leptonic meson decays~\cite{Shrock:1980vy,Shrock:1980ct,Abada:2012mc,Abada:2013aba}.
In the near future,  neutral  fermions with masses in the GeV range can be searched in experiments such as NA62~\cite{Hahn:1404985}, SHiP~\cite{Anelli:2015pba,Alekhin:2015byh}, FCC-ee~\cite{Blondel:2014bra} and LBNF/DUNE~\cite{Adams:2013qkq}.  

The presence of source(s) of LNV in the considered frameworks may also induce consequences on the effective mass in the 
amplitude of the neutrinoless double beta decay rate~\cite{Deppisch:2012nb}, which is defined, in the case where the SM is extended by $N$ sterile fermions as~\cite{Blennow:2010th}: 
\begin{equation} 
\label{eq:0vudbdecay}
 m_{0\nu\beta\beta}\,\simeq \,\sum_{i=1}^{N} {\cal U}_{ei}^2 \,p^2
\frac{m_i}{p^2-m_i^2}\ ,  \end{equation}
where $p^2 \simeq - (125 \mbox{ MeV})^2$ is the virtual momentum (an average estimate over different decaying nuclei) of the propagating neutrino. 
Notice that the additional mixings and possible
new CP-violating Majorana phases might enhance the effective mass, potentially
rendering it within experimental reach, or even leading to the
exclusion of certain regimes due to conflict with the current bounds - the most recent results on neutrinoless double beta decay have been 
 obtained by the EXO-200 experiment~\cite{Albert:2014awa} and by KAMLAND-Zen~\cite{KamLAND-Zen:2016pfg}.

The final constraint is of cosmological origin. The neutrinos involved in the low-energy seesaw cannot be too light, otherwise their lifetimes would be of the same order as the timescale of Big Bang Nucleosynthesis (BBN).  Their decays  into SM states at this time would have severe consequences on the synthesis of the light nuclei. To avoid this possibility, we will assume a conservative lower bound of 100 MeV~\cite{Dolgov:2000jw,Dolgov:2000pj,Hernandez:2013lza,Hernandez:2014fha} on the masses of the new states and impose that their life times do not exceed 1 second.\footnote{Except for the potential DM candidate in the ISS(2,3), whose life time exceeds the age of the Universe.}


\subsection{Minimal particle content: the LSS-ISS model}\label{Sec:LSS-ISS}

The type-I seesaw mechanism provides a simple explanation for the light neutrino masses: Introducing RH neutrinos with a Majorana mass $M_N$ and which share a Dirac mass $m_D$ with the active neutrinos, the diagonalization of the mass matrix yields the light neutrino masses $m_\nu = m_D M_N^{-1} m_D$ which are inversely proportional to the heavy Majorana mass scale. The introduction of additional sterile fermions changes this picture, and depending on the underlying GUT breaking mechanism and the model parameters, different seesaw contributions to $m_\nu$ may be dominant, allowing for small active neutrino masses $m_\nu$ despite lowering the mass scale of the additional SM singlet fermions. In this section we introduce the LSS-ISS model, which is a phenomenological low-scale seesaw model with minimal particle content, based on the introduction of the two small LNV parameters found in the Linear~\cite{Barr:2003nn,Malinsky:2005bi} and Inverse~\cite{Wyler:1982dd, Mohapatra:1986bd} Seesaw models.

The SM spectrum is extended by two RH neutrinos $N_R^{1},N_R^{2}$ at the mass scale $\Lambda$ with opposite lepton number, more specifically $+1$ for $N_R^{1}$ and $-1$ for $N_R^{2}$. In the interaction basis the new states are coupled to the active sector via Yukawa couplings $Y_\alpha$ and $\epsilon Y'_\alpha$, respectively. With $|Y'| \sim |Y|$ and $\epsilon \ll 1$, the latter couplings violates lepton number by a small amount. The second LNV parameter is introduced as a Majorana mass $\mu = \xi \Lambda$ for the sterile neutrinos. The assignment $\epsilon,\xi \ll 1$ is technically natural, since lepton number is restored in the limit of $\epsilon,\xi \rightarrow 0$.
The neutrino global mass term of Lagrangian reads 
\begin{eqnarray}
- \mathcal{L}_{m_\nu} = n_L^T\ C\ { \mathcal{M}}\ n_L + \mathrm{h.c.},
\label{eq:nulagrangian}
\end{eqnarray}
where
\begin{eqnarray}
\begin{array}{ccc}
n_L \equiv  \left(  \nu_L^e,\nu_L^\mu, \nu_L^\tau, N_R^{1\,c},N_R^{2\,c}
	    \right)^T, & \rm{and} &  C = i \gamma^2\gamma^0\ , 
\end{array}
\end{eqnarray}
and with the full mass matrix of the neutrino sector
\begin{equation}
 \mathcal{M}^{(\nu)} =\ \Lambda \, \begin{pmatrix}
      0 & 0 & 0 & \frac{1}{\sqrt{2}}Y_1  v/ \Lambda & \frac{1}{\sqrt{2}} \epsilon Y'_1  v/\Lambda \\
      0 & 0 & 0 & \frac{1}{\sqrt{2}} Y_2  v/\Lambda & \frac{1}{\sqrt{2}} \epsilon Y'_2  v/\Lambda \\
      0 & 0 & 0 & \frac{1}{\sqrt{2}} Y_3  v/\Lambda& \frac{1}{\sqrt{2}} \epsilon Y'_3  v/\Lambda\\
      \frac{1}{\sqrt{2}} Y_1  v/\Lambda &  \frac{1}{\sqrt{2}} Y_2  v/\Lambda &  \frac{1}{\sqrt{2}} Y_3  v/\Lambda & 0 & 1 \\
    \frac{1}{\sqrt{2}}  \epsilon Y'_1  v/\Lambda &  \frac{1}{\sqrt{2}} \epsilon Y'_2  v/\Lambda  &   \frac{1}{\sqrt{2}} \epsilon Y'_3  v/\Lambda  & 1  & \xi
     \end{pmatrix}  \,, \label{eq_Mpertexp}
\end{equation} 
where $v$ is the Higgs boson vacuum expectation value.

To illustrate the structure of Eq.~\eqref{eq_Mpertexp}, let us consider a toy model  with a single generation for the active neutrinos. In this case the mass matrix has the following form,
 \begin{equation}
 {\mathcal{M}}^{(\nu)} = \begin{pmatrix}
0 &  Y v /\sqrt{2} & \epsilon Y' v /\sqrt{2} \\
Y v /\sqrt{2} & 0 & \Lambda \\
\epsilon Y' v /\sqrt{2} & \Lambda & \xi \Lambda
\end{pmatrix}\, ,\label{eq_Mpertexp:1gen}
\end{equation} 
which can be decomposed as
\begin{equation}
{\mathcal{M}}^{(\nu)}={\mathcal{M}}_0 + \Delta {\mathcal{M}}_{ISS}+ \Delta {\mathcal{M}}_{LSS}\ .\end{equation}
The lepton number conserving mass matrix ${\mathcal{M}}_0$ is given by 
 \begin{equation}{\mathcal{M}}_0= \left(\begin{array}{ccc} 
0 & \frac{1}{\sqrt{2}}Y v & 0\\
\frac{1}{\sqrt{2}} Y v & 0 & \Lambda \\
0 & \Lambda & 0
\end{array}\right) \ ,\label{M0} \end{equation}
and its diagonalization gives rise to the mass spectrum 
\begin{equation}
m_\nu =0 \,,\quad
M_{1,2} = \sqrt{|\Lambda|^2 +\frac{1}{2}|Y v|^2}\ .
\label{eq_M}
\end{equation}
The perturbation of the latter matrix by two sources of lepton number violation encoded in $\Delta {\mathcal{M}}_{ISS}$ (proportional to the LNV parameter $\xi$) and  $\Delta {\mathcal{M}}_{LSS}$ (proportional to the LNV parameter $\epsilon)$ 
leads to the matrix of Eq.~({\ref{eq_Mpertexp:1gen}).
Considering only the perturbation $\Delta {\mathcal{M}}_{ISS}$ leads to the Inverse Seesaw pattern~\cite{Wyler:1982dd,Mohapatra:1986bd} while considering the second perturbation, $\Delta {\mathcal{M}}_{LSS}$,  leads to a Linear  Seesaw pattern~\cite{Malinsky:2005bi}.\footnote{
A non-vanishing value of the $(2,2)$ entry of the matrix Eq.~(\ref{eq_Mpertexp:1gen}), ${\mathcal{M}}^{(\nu)}_{22} = \xi' \Lambda$, would correspond to an additional LNV violation by two units, which does not generate neutrino masses at tree level but does it  only at loop level\cite{Dev:2012sg,LopezPavon:2012zg}. These loop corrections will be relevant only  if $\xi' \gtrsim 1$, meaning for  regimes of a large lepton number violation. However,   as in our approach to leptogenesis we focus  on models with an approximate lepton number conservation,  we will not pursue this option any further here.}
Allowing both $\xi \neq 0$ and $\epsilon \neq 0$ leads to a mixed Linear and Inverse Seesaw mechanism, a model we refer to as ``LSS-ISS". Compared to Eq.~\eqref{eq_M}, the mass scale $m_\nu$ of the lightest neutrino and the mass splitting $\Delta m^2_\text{PD}$ between the two heavy neutrinos are now non-zero,
\begin{equation}
m_\nu \simeq 2 \epsilon \frac{m_D^2}{\Lambda} \,, \qquad \Delta m^2_\text{PD} = M_2^2 - M_1^2 = 2 \xi \Lambda^2 \,,
\label{eq:LISStoymasses}
\end{equation}
where $m_D = Y v /\sqrt{2} \simeq Y' v/\sqrt{2}$. This structure immediately generalises to the full mass matrix of Eq.~\eqref{eq_Mpertexp}: the mass scale of the active neutrinos is set by the LNV parameter $\epsilon$, whereas the second LNV parameter $\xi$ controls the small mass splitting within the heavy pseudo-Dirac pair - a crucial parameter for ARS leptogenesis. Further details of this model, including the perturbative diagonalization of the full matrix~\eqref{eq_Mpertexp}, can be found in Ref.~\cite{Abada:2015rta}. This neutrino mass model can account for the observed neutrino oscillation data, a suitable parametrisation of the remaining free parameters will be introduced in Sec.~\ref{sec:parISSLSS}.

\subsection{Minimal realisations of the Inverse Seesaw: ISS(2,2) and ISS(2,3)}\label{Sec:MISS}

Next we turn to the pure Inverse Seesaw mass generation mechanism~\cite{Wyler:1982dd, Mohapatra:1986bd}. Compared to the previous subsection, this implies $\epsilon = 0$ and we will moreover allow for an independent number of RH and sterile neutrinos ($\#\nu_R\ne 0$ and $\#s\ne 0$).
In the case of $\# s=0$, one recovers the usual type I  seesaw realisation, which could account for neutrino masses and mixings provided that the number of right-handed neutrinos is at least $\#\nu_R=2$. 

The neutrino mass term has the same structure as in Eq.~\eqref{eq:nulagrangian}, with 
\begin{equation}
n_L \equiv \left(\nu_L^e, \, \nu_L^\mu, \, \nu_L^\tau,\, \nu_{R,i}\, ^c,s_j\right)^T \,,
\end{equation}
where $\nu_{R,i}^c$ ($i=1,  .. \#\nu_R$) and $s_j$ ($j=1, .. \# s$) are RH  
neutrino fields and additional fermionic gauge singlets, respectively.  The  neutrino mass matrix in Eq.~\eqref{eq:nulagrangian} has the form,
\begin{equation}\label{general-iss}
{\mathcal{M}} \equiv
\left(
\begin{array}{ccc}
0 & d & 0 \\
d^T & 0 & n \\
0 & n^T & \xi \Lambda
\end{array}
\right)\ ,
\end{equation}
where $d, n, \xi\Lambda$ are complex mass matrices.
The Dirac mass matrix $d$ arises from the Yukawa couplings to the SM Higgs boson, $\tilde{H} =i  \sigma^2 H$,
\begin{equation}
d_{\alpha i} = \frac{v}{\sqrt{2}}Y_{\alpha i}^*,\hspace{1cm} \mathcal{L} \ni Y_{\alpha i} \overline{\ell_L^\alpha}\tilde{H }\nu_R^i+\text{h.c., }\,\,\,\,\,\,\ell_L^\alpha=\left(
\begin{array}{c}
\nu_{L}^\alpha \\
e^\alpha_L
\end{array}
\right)\ ,
\end{equation}
while the matrix $\xi \Lambda$, instead, contains the Majorana mass terms for the sterile fermions $s_j$. This is the only LNV parameter in the ISS models. The sub-matrix $n$ is $\# \nu_R \times \# s$ matrix with entries of order $\Lambda$.
By assigning a opposite leptonic charge to $\nu_R^c$ and $s$, one ensures that the off-diagonal terms are lepton number conserving, while $s^T Cs$ violates the lepton number by two units. The  feature of the ISS is that the entries of the matrix  $\xi$ can be made small in order to accommodate for ${\mathcal{O}}(\text{eV}$) masses of (mostly) active neutrinos, while having large Yukawa couplings.  This is not in conflict with naturalness since the lepton number is restored in the limit of $\xi \rightarrow 0$ and all along  this work we impose the above matrices to fulfil a naturalness criterion, $|\xi \Lambda|\ll |d|<|n|$~\cite{Abada:2014vea}.

Concerning the singlet fermions, $\nu_R$ and $s$, since there is no direct evidence for their existence (and because they do not contribute to anomalies), their number is unknown. 
 In ~\cite{Abada:2014vea}  it was shown that it is possible to construct several distinct realisations of the ISS, reproducing the correct neutrino mass spectrum while complying with 
the constraints listed above, thereby strongly preferring a normal ordered active neutrino spectrum.
More specifically, it was shown that, depending on the number of additional fields, the neutrino mass spectrum obtained for each ISS realisation is characterised by either 2 or 3 mass scales, corresponding to the light neutrino mass scale $m_\nu$, the mass scale of the heavy pseudo-Dirac pair(s) $m_\text{PD}$ and, only if $\# s > \#\nu_R$, an intermediate scale $m_\text{DM}$:
\begin{equation}
m_\nu \simeq \xi \Lambda \frac{d^2}{n^2} = \frac{\xi (Y v)^2}{2 m_\text{PD}} \,, \quad m_\text{PD} \simeq n \simeq \Lambda \,, \quad m_\text{DM} \simeq \xi \Lambda \,.
\label{eq:ISStoym1}
\end{equation}
The mass splitting within the pseudo-Dirac pair(s) is given by
\begin{equation}
\Delta m^2_\text{PD} \simeq 2 \, \xi\,  m_\text{PD}^2 \,. 
\label{eq:ISStoym2}
\end{equation}
This allows to identify two truly minimal ISS realisations~\cite{Abada:2014vea}, 
the ISS (2,2) model, which corresponds to the SM extended by two RH neutrinos and two additional sterile fermions,  and the ISS (2,3) model, where the SM is extended by two RH neutrinos and three sterile states. In agreement with the discussion above, the physical mass spectrum of both models presents two pairs of pseudo-Dirac neutrinos. The ISS(2,3) features, in addition, an intermediate mass scale mostly sterile neutrino. As extensively discussed in~\cite{Abada:2014vea,Abada:2014zra}, this additional state can have mass both in the eV range, possibly accommodating a 3+1-mixing scheme at low energies which can be used to interpret the short baseline (reactor/accelerator) anomalies~\cite{Gariazzo:2015rra}, or in the keV range. In this case the mostly sterile neutrino could be a  DM candidate providing interesting phenomenology related to structure formation~\cite{Klypin:1999uc,Moore:1999nt,Strigari:2010un,BoylanKolchin:2011de,Baur:2015jsy,Merle:2015oja,Konig:2016dzg,Murgia:2017lwo} and to indirect detection~\cite{Pal:1981rm,Boyarsky:2005us,Boyarsky:2012rt}. Related to this last point, it is worth mentioning the hint (not confirmed) of the detection of an X-ray line at approximately the energy of 3.5~keV\cite{Bulbul:2014sua,Boyarsky:2014jta}. In view of more recent analyses both the hints related to reactor anomalies and to the X-ray line appear increasingly disfavoured. The reference to them is only intended to highlight the rich phenomenology of the scenario under study. A suitable parameterisation of the ISS models, taking into account observational constraints, will be introduced in Sec.~\ref{sec:iss_parametrization}. 
We finally remark that in the ISS(2,3) the mass of the intermediate sterile state, relevant in our work mostly for DM phenomenology, is tightly related to the mass splitting of the pseudo-Dirac pairs. As will be discussed more extensively in the following this is a key parameter for the achievement of a viable BAU.


\section{Leptogenesis}\label{Sec:mechanism}

The neutrino mass models of the previous section feature one (or several) pair(s) of pseudo-Dirac neutrinos, whose mass splitting(s) is (are) governed by a small LNV parameter, and whose overall mass scale can be set to the GeV - TeV range. These are the crucial ingredients to implement leptogenesis through neutrino oscillations~\cite{Akhmedov:1998qx}: Starting from a negligible abundance of these heavy neutrinos in the early Universe, a pseudo-Dirac pair is thermally produced  just before the electroweak 
(EW) phase transition. Due to the small mass splitting, rapid oscillations of the neutrinos within the pseudo-Dirac pair occur, entailing a CP-violating background for the active neutrinos. The active neutrinos hence experience an effective CP-violating potential, similar to the MSW effect of matter~\cite{Wolfenstein:1977ue}. This induces a lepton asymmetry in the active species, which in turn back-reacts to the pseudo-Dirac pairs, further enhancing the CP asymmetry of this sector. This way an asymmetry is generated both in the sector of the SM states and in the sector of the new SM singlet fermions. For simplicity we will label, here and in the next sections, these two sectors as active and singlet sectors, respectively.

Note that the leptogenesis process occurs in the highly relativistic regime for the new neutrinos, $m_\text{PD}/T \ll 1$. Extending the usual definition of lepton number to the different helicity states of the new neutrinos, we can define a generalised lepton number which is conserved to leading order in  $m^2_\text{PD}/T^2$~\cite{Shaposhnikov:2008pf}. In this work, we hence neglect generalised lepton number violating (GLNV) processes.  It has recently been pointed out that there exists a region of parameter space where GLNV processes are in fact dominant in the generation of a lepton asymmetry~\cite{Hambye:2016sby}: this happens when washout processes for the generalised lepton number conserving (GLNC) rates are already effective  above the sphaleron freeze-out temperature, while at the same temperature the GLNV rates are far from thermal equilibrium. To assess the relative importance of the GLNV processes with respect to the GLNC ones it is necessary to overlay this parameter region with the one where  successful baryogenesis via leptogenesis can be achieved: the resulting intersection, for the frameworks considered in the present work, shrinks by increasing the ratio $\Delta m_\text{PD}/m_\text{PD}$, and appears to cover only a small portion of the total parameter space for the value $\Delta m_\text{PD}/m_\text{PD} = 10^{-6}$~\cite{Hambye:2017elz}. In this work we consider values of relative mass splitting much larger (cf. Fig.~\ref{fig_LISS_asymmetry} and Fig.~\ref{fig_ISS22_asymmetry}). We thus do not expect any relevant change in our conclusions\footnote{Notice that the two asymmetries (GLNC and GLNV) can in general contribute constructively, provided the correct combination of CP phases is realised: thus the inclusion of GLNV processes will in general enlarge the viable region of solutions. Such an analysis is however beyond the scope of the current paper.} (see also~\cite{Eijima:2017anv,Ghiglieri:2017gjz}). 

In the absence of GLNV processes, the total asymmetry summed over both the active and singlet sectors must vanish. The asymmetries produced in the active and singlet sector are hence of equal value, but have opposite sign. However, sphaleron processes act only on the asymmetry in the SM sector, (partially) converting it into the baryon asymmetry we observe today. In this way, leptogenesis occurs even if the total generalised lepton number is (approximately) conserved. In this section, we first summarise the key equations describing these processes following a series of earlier studies~\cite{Asaka:2005an,Asaka:2005pn,Shaposhnikov:2008pf,Asaka:2011wq,Canetti:2012kh, Drewes:2012ma,Abada:2015rta, Hernandez:2015wna, Hernandez:2016kel, Drewes:2016gmt} which have lead to an improved understanding of many aspects 
in recent years. Some technical details are relegated to Appendix~\ref{app:leptogenesis_details}. In Section~\ref{sec:weakwashout} we summarise the results of our earlier work~\cite{Abada:2015rta} on the weak washout regime, before developing a new method of solving the differential equations in the full parameter space of interest, including the strong washout regime, in Section~\ref{sec:strongwashout}.

The processes sketched above can be described in the density matrix formalism by two differential matrix equations, one for the singlet neutrinos~$N$ and one for the active species~$L$~\cite{Sigl:1992fn},
\begin{align}
\label{eq:starta}
\frac{d\rho_N}{dt}&=-i \left [H_N,\rho_N\right]-\frac{1}{2}\left \{ \Gamma_N^d,\rho_N \right \}+\frac{1}{2}\left \{ \Gamma_N^p,I-\rho_N\right \}  \,, \\
\frac{d\rho_L}{dt}&=-i \left [H_L,\rho_L\right]-\frac{1}{2}\left \{ \Gamma_L^d,\rho_L\right \}+\frac{1}{2}\left \{ \Gamma_L^p,I-\rho_L\right \} \,. \label{eq:startb}
\end{align}
Here $\rho_{N,L}$ denote the density matrices of the singlet and active species, $\Gamma_{N,L}^{p,d}$ are the respective production and decay rates and $H_{N,L}$ are the corresponding Hamiltonians, containing a vacuum part $H_{N,L}^0$ describing the propagation and oscillations as well as an effective potential $V_{N,L}$. All quantities are functions of the wave numbers $k_{N,L}$, the temperature $T$ (or equivalently the cosmic time $t$) and the chemical potential $\mu_L$ of the active flavours (arising in a CP-violating background). The equations for the corresponding anti-particles are obtained by substituting: $L \leftrightarrow \bar L$, $N \leftrightarrow \bar N$, $F \leftrightarrow F^*$ and $\mu_L \leftrightarrow -\mu_L$. Here $F_{\alpha I}$, contained in the production and decay rates, denotes the Yukawa coupling of the singlet neutrinos in their mass eigenbasis,
\begin{equation}
F_{\alpha I} = Y_{\alpha i} \;{\cal U}_{i I}\,,
\label{eq:effYukawa}
\end{equation}
with ${\cal U}$ denoting the unitary matrix which diagonalises $({\mathcal{M}}^{(\nu)})^\dagger {\mathcal{M}}^{(\nu)}$. Here $\alpha$ and $i$ run over the active and SM singlet neutrino flavours, respectively, while $I$ runs  over the heavy mass eigenstates. Due to the 
unitarity of the matrix ${\cal U}$, $ \sum_{\alpha, I} |F_{\alpha I}|^2 \simeq \sum_{\alpha ,i} |Y_{\alpha,i}|^2$. This Yukawa coupling is moreover the key ingredient of the effective potential contained in $H_N$,
\begin{equation}
 V_N = \frac{N_D T^2}{16 \, k_N} F^\dagger F \,,
\end{equation}
where $N_D = 2$ is an $SU(2)$ factor.

These equations can be significantly simplified by taking the active species to be in thermal equilibrium with a chemical potential $\mu_L$, $\rho_L(k_L,T,\mu_L) = N_D \,  f_F(k_L/T, \mu_L)\,  I$, with $f_F(k_L/T,\mu_L)={\left[\exp\left(\frac{k_L}{T}-\mu_L\right)\right]}^{-1}$ denoting the Fermi-Dirac distribution function with momentum $k_L$ and chemical potential\footnote{Below we will perform a perturbative expansion with respect to the chemical potential; for this reason it is appropriate to define it as a dimensionless quantity by reabsorbing the temperature factor.} $\mu_L$, computed at the thermal bath temperature $T$.

We further note that the system studied here contains two small parameters which may be exploited for a perturbative analysis: the entries of the Yukawa matrix $F_{\alpha I}$ and the chemical potentials ${\mu_L}_\alpha$. The latter are directly related to the generated baryon asymmetry, which is why they are expected to be small for all viable parameter points. 

Performing these expansions and after some additional manipulations detailed in Appendix~\ref{app:leptogenesis_details}, the production and destruction rates of the singlet neutrinos can be rewritten, to first order in $\mu_L$, as:
\begin{align}
& \Gamma_N^p=f_F^0(y_N) \gamma_N^0 F^\dagger F + \delta \gamma_N^p F^\dagger \mu_L F \,, \nonumber\\
& \Gamma_N^d=(1-f_F^0(y_N)) \gamma_N^0 F^\dagger F + \delta \gamma_N^d F^\dagger \mu_L F\,,
\label{eq:decayrates}
\end{align}
where $y_N = k_N/T$ and where we have defined $f_F^0(y_N)\equiv f_F(y_N,0)$.  Moreover,
\begin{align}
\gamma^0_N &=\frac{T^3}{64 \pi^3 k_N^2} \Sigma(y_N)\,, 
\label{eq:gammaN0}\\
 \delta \gamma^p_N &=- \frac{T^3}{64 \pi^3 k_N^2} \left(f_F^{'}(y_N) \Sigma(y_N) +f_F^0(y_N)\Psi(y_N)\right) \,, \\
 \delta \gamma^d_N &=\frac{T^3}{64 \pi^3 k_N^2}  \left( f_F^{'} (y_N) \Sigma(y_N)+ (1-f_F^0(y_N)) \Psi(y_N)\right) \,,
 \label{eq:dgammaNd}
\end{align}
where $\Sigma(y_N)$ and $\Psi(y_N)$ are integrals containing the matrix element of the interaction process, (their formal expression is provided in Appendix~\ref{app:A1}), while $f_F^{'}=\frac{df_F^0 (y)}{dy}$.

The corresponding decay and production rates of the active species (describing the exact same processes from the point of view of these particles), can be related to Eqs.~\eqref{eq:gammaN0} - \eqref{eq:dgammaNd} by exchanging the order of integration in the integrated decay rates: 

\begin{align}
\int \frac{d^3 k_L}{(2 \pi)^3} & \Gamma_L^d(k_L) f_F(k_L/T,\mu_L) \nonumber \\
&= \frac{1}{N_D}\int \frac{d^3 k_N}{(2 \pi)^3} f^0_F(k_N/T) \gamma^0_N F (I-\rho_N(k_N)) F^{\dagger}+  \delta \gamma^p_N  \mu_L F (I-\rho_N(k_N)) F^{\dagger} \,,  \nonumber\\
\int \frac{d^3 k_L}{(2 \pi)^3} & \Gamma_L^p(k_L) (1-f_F(k_L/T,\mu_L)) \nonumber \\
&= \frac{1}{N_D}\int\frac{d^3 k_N}{(2 \pi)^3} (1-f_F^0(k_N/T)) \gamma^0_N F \rho_N(k_N) F^{\dagger}+ \delta \gamma^d_N \mu_L F \rho_N(k_N) F^{\dagger} \,.
\end{align}

The expressions above show that, in general, one has to solve a system of coupled integro-differential equations. It can be however reduced to a system of ordinary differential equations by assuming that the heavy neutrinos fulfil the weaker condition of kinetic equilibrium, $\rho_N(k_N,T)_{IJ} = R_N(T)_{IJ} f_F(k_N/T, \mu_L = 0)$.  With this we can factor out the momentum-independent variable $R_N (T)$ in the integrals above, and replace the integrated rates with thermally averaged destruction and production rates:
\begin{equation}
\langle \gamma (T) \rangle =\frac{\int d^3 p \, \gamma(p,T) f_F^0(p/T)}{\int d^3 p \,f_F^0(p/T)} \,.
\end{equation}
Additionally substituting the lepton number densities by an equation directly for the chemical potential, 
\begin{align}
\int \frac{d^3 k_L}{(2 \pi)^3}\left[ f_F(k_L/T,\mu_L)-f_F(k_L/T,-\mu_L)\right] &\approx -2\mu_L \int \frac{d^3 k_L}{(2 \pi)^3} f^{'}(k_L/T) = \frac{T^3}{6 } \mu_L \\
 \rightarrow {\mu_L}_{\alpha}&=\frac{6}{T^3}\int\frac{d^3 k_L}{(2 \pi)^3} \left(\rho_L - \rho_{\bar L} \right)_{\alpha \alpha} \frac{1}{N_D} \,,
\end{align}
we obtain to first order in $\mu_L$
\begin{align}
\label{eq:Rnstart}
\frac{d R_N}{dt}= & -i \left[\langle H \rangle ,R_N\right]-\frac{1}{2}\langle \gamma^{(0)} \rangle \left \{ F^{\dagger}F, R_N-I\right \}- \frac{1}{2} \langle \gamma^{(1b)} \rangle \left \{ F^{\dagger} \mu_L F, R_N \right \}+\langle \gamma^{(1a)} \rangle F^{\dagger} \mu_L F \,, \\
\label{eq:mustart}
 \frac{d {\mu_L}_{\alpha}}{dt} = & \,  \frac{9 \,  \zeta (3)}{2 N_D\, \pi^2} \left \{ \langle \gamma^{(0)} \rangle \left(F R_N F^{\dagger}-F^{*} R_{\bar N} F^T\right) -2  \langle \gamma^{(1a)}  \rangle \mu_L F F^{\dagger} + \right.  \nonumber \\   & +  \left.\langle \gamma^{(1b)} \rangle \mu_L  \left(F R_N F^{\dagger}+F^{*} R_{\bar N}F^T\right)\right \}_{\alpha \alpha} \,,
\end{align}
with
\begin{equation}
\mu_L = \text{diag}({\mu_L}_\alpha) \,, \quad \gamma^{(0)} \equiv \gamma_N^0 \,, \qquad f_F^0(k_N/T) \, \gamma^{(1a)} \equiv \delta \gamma^p_N \,, \qquad \gamma^{(1b)} \equiv \delta \gamma_N^p + \delta \gamma_N^d \,. 
\end{equation}
Note that the off-diagonal elements of $\rho_L, \rho_{\bar L}$ do not enter Eq.~\eqref{eq:Rnstart}, and hence it is sufficient to solve Eq.~\eqref{eq:mustart} for the diagonal components only, implying that the commutator in Eq.~\eqref{eq:startb} can be dropped. The leading order decay rate $\gamma^{(0)}$ was recently re-evaluated in Ref.~\cite{Besak:2012qm}, taking into account, not only scattering processes involving the top quark, but also processes involving soft gauge bosons of the thermal plasma. This work was extended to account for a finite chemical potential in Ref.~\cite{Hernandez:2016kel}, thus determining 
 $ \gamma^{(1a),(1b)} $  (there labeled $\gamma^{(1),(2)}$, respectively.). The resulting thermally averaged rates are found to be
\begin{equation}
\langle \gamma^{(i)} \rangle = A_i \left[ c^{(i)}_\text{LPM} + y_t^2 c_Q^{(i)} + (3 g^2 + g'^2) \left( c_V^{(i)} - \ln(3 g^2 + g'^2) \right) \right]\ ,
\end{equation}
where $g, g'$ denote the (temperature-dependent) SM $SU(2)$ and $U(1)$ gauge couplings, $y_t$ is the top Yukawa coupling, and
\begin{equation}
A_0 = 2 A_{1a} = -4 A_{1b} = \frac{\pi T}{2304 \, \zeta(3)} \,.
\end{equation}
The numerical values of $c^{(i)}_{LPM,Q,V}$ are reported in Tab.~1 of Ref.~\cite{Hernandez:2016kel}. Both $c^{(i)}_Q$ and $c^{(i)}_V$ are found to be $T$-independent, the temperature dependence of $c^{(i)}_{LPM}$ is so mild that we will neglect it in the following, using $c^{(i)}_{LPM}(T = 10^4~\text{GeV})$ as a reference value, leading to
\begin{align}
 c^{(0)}_{LPM} &= 4.22\,,    &c^{(0)}_{Q} = 2.57\,, &&   c^{(0)}_{V} &= 3.17\,,  \nonumber \\
 c^{(1a)}_{LPM} &= 3.56\,,  &c^{(1a)}_{Q}  = 3.10\,, &&c^{(1a)}_{V} &= 3.83\,, \nonumber \\
 c^{(1b)}_{LPM} &= 4.77\,,  &c^{(1b)}_{Q} = 2.27\,, && c^{(1b)}_{V} & = 2.89\,. \label{tab:ci}
\end{align}

So far, we have focused on interactions between the various active and singlet neutrino species which are mediated by the Yukawa coupling $F$ and which modify $\mu_\alpha$, i.e.\ the total lepton number of the active sector. However, as pointed out in Refs.~\cite{Barbieri:1999ma,Buchmuller:2001sr}, a further important role is played by the so-called spectator processes. Before the EW phase transition, sphaleron 
and SM Yukawa mediated processes 
distribute the asymmetry among the various species of the thermal bath, thereby conserving $B-L$ but violating $B+L$. 
A simple way to incorporate these processes is to work directly with the differential equation for $B-L$. The production and decay terms for the lepton doublet on the right-hand side of Eq.~\eqref{eq:mustart} (now producing $L =  - (B-L)$) are in fact the only terms which change $B-L$. Labelling the chemical potential associated with $B-L$ as $\mu_\Delta$, this yields
\begin{align}
 \frac{d {\mu_\Delta}_\alpha}{dt} =& -  \frac{9 \zeta (3)}{2 N_D\, \pi^2} \left \{ \langle \gamma^{(0)} \rangle \left(F R_N F^{\dagger}-F^{*} R_{\bar N} F^T\right) -2  \langle \gamma^{(1a)}  \rangle \mu_L F F^{\dagger} \right.  \nonumber \\   & +  \left.\langle \gamma^{(1b)} \rangle \mu_L  \left(F R_N F^{\dagger}+F^{*} R_{\bar N}F^T\right)\right \}_{\alpha \alpha} \,,
\end{align}
with $\mu_L$ and $\mu_{\Delta}$ related~\cite{Harvey:1990qw,Drewes:2016gmt} 
through
\begin{align}
{\mu_L}_\alpha = A_{\alpha \beta} {\mu_\Delta}_\beta \,, \quad \quad A = \frac{1}{711} \begin{pmatrix}
-221  & 16 &  16 \\
16 & -221 & 16 \\
16 & 16 & -221 
\end{pmatrix} \,, \label{eq:muDelta}
\end{align}
for $T \lesssim 10^5$~GeV. The conversion of the $B-L$ asymmetry to the observed baryon asymmetry finally introduces the usual sphaleron conversion factor $28/79$.

The above simplifications preserve a crucial consistency feature of the framework: the total asymmetry generated in both the active and singlet sectors vanishes, i.e.\
\begin{align}
&0 = \left( \frac{d n_N}{dt} - \frac{d n_{\bar N}}{dt}\right) - \mbox{Tr}\left[\frac{d {n_\Delta}_{\alpha}}{dt}\right]\nonumber\\
&=\int \frac{d^3 k_N}{(2 \pi)^3} f_F^0(k_N) \mbox{Tr} \left[{\left(\frac{d R_N}{dt}-\frac{d R_{\bar N}}{dt}\right)}_{\alpha \alpha}\right]+ 2 N_D \int \frac{d^3 k_L}{(2 \pi)^3} f^{'}(k_L) \mbox{Tr}\left[\frac{d {\mu_\Delta}_{\alpha}}{dt}\right] \,. \label{eq:consistency}
\end{align}

We can now distinguish two phenomenologically different regimes. The weak washout regime, obtained for $|F_{\alpha I}| \lesssim 10^{-7}$, is characterised by $R_N \ll 1$ and $\mu_L \lll 1$, which allows for a perturbative analytical solution  of Eqs.~\eqref{eq:Rnstart} and \eqref{eq:mustart}~\cite{Abada:2015rta}. For larger values of $|F_{\alpha I}|$, $R_N$ grows from initially small values to $|R_N| \sim 1$, inducing sizeable washout effects on the final asymmetry. Consequently, the asymmetry $\mu_L$ reaches a peak value at intermediate time-scales before washout-processes reduce the value to the one observed today. We find that $\mu_L$ is still small enough to serve as an expansion parameter, however the larger values we find here compared to the weak washout regime require a more careful and rigorous treatment of the expansion.

\subsection{Weak washout regime \label{sec:weakwashout}}
The weak washout regime was studied in detail in Refs.~\cite{Asaka:2005pn,Shaposhnikov:2008pf,Abada:2015rta}. Starting from Eqs.~\eqref{eq:Rnstart} and \eqref{eq:mustart}, an iterative process allows a fast determination of the final baryon asymmetry: in a first step, Eq.~\eqref{eq:Rnstart} is solved in the limit $\mu_L \rightarrow 0, R_N \ll 1$; this is inserted into Eq.~\eqref{eq:mustart} (neglecting again the terms proportional to $\mu_L$ on the right-hand side); the resulting expression for $\mu_{L_\alpha}$ is finally re-inserted into Eq.~\eqref{eq:Rnstart}, now evaluated to first order in $\mu_L$. As long as $\mu_L$ is sufficiently small, this decoupling of the equations is justified and the resulting asymmetry matches the asymmetry obtained in the full system to good accuracy. Moreover, this procedure allows for an analytical estimate of the final asymmetry (see Appendix~\ref{app:weakwo}):
\begin{equation}
\label{eq:baryo_analytical}
Y_{\Delta B}=\frac{n_{\Delta B}}{s}= \frac{2835}{5056} \frac{1}{\pi^{17/6}  \,  \Gamma(5/6)} \frac{1}{g_s}\sin^3 \phi \, \frac{M_0}{T_{\rm W}} \frac{M_0^{4/3}}{ \left(\Delta m_{PD}^2\right)^{2/3}} \, Tr\left[ F^\dagger A_{\alpha \beta} \delta_\beta F\right] \ ,
\end{equation}
where $\Delta m_{PD}^2$ is the difference between the squared masses of  the nearly-degenerate heavy neutrinos, $T_{\rm W} = 140$~GeV is the temperature of the EW phase transition, $g_s$ counts the degrees of freedom in the thermal bath at $T = T_\text{W}$, $M_0 \approx 7 \times 10^{17}\,\mbox{GeV}$, $\sin\phi \sim 0.004$ and $\delta = \text{diag}(\delta_\alpha)$ is the CP asymmetry in the oscillations defined as:
\begin{equation}\label{eq:deltaCP}
\delta_{\alpha}=\sum_{i >j} \text{Im}\left[F_{\alpha i} \left(F^{\dagger} F\right)_{ij} F^{\dagger}_{j\alpha}\right]\ .
\end{equation}
Equation~\eqref{eq:baryo_analytical} has an analogous functional form as found in~\cite{Abada:2015rta}.
To facilitate the comparison of \cite{Abada:2015rta} with the results presented here, we highlight the three most important refinements of the present work: Firstly, we are here working with the full Fermi-Dirac distribution function  whereas Ref.~\cite{Abada:2015rta} uses the Maxwell-Boltzmann distribution, which leads to a different overall factor in Eq.~\eqref{eq:baryo_analytical}. Secondly, we are now taking into account soft scatterings of gauge bosons of the thermal plasma on the production and decay rates, whereas Ref.~\cite{Abada:2015rta} estimated these rates based on top-quark scattering only. These changes are encoded in the definition of $\sin \phi$, which hence takes a different numerical value here compared to \cite{Abada:2015rta}.
Thirdly, we take into account the re-distribution of the asymmetry in the active sector through spectator processes. Taking the Maxwell-Boltzmann limit of Eqs.~\eqref{eq:Rnstart} and \eqref{eq:mustart}, keeping only the top-quark contribution to the scattering rates and taking $A \rightarrow I$ in Eq.~\eqref{eq:muDelta}, one recovers precisely the system of equations used in Ref.~\cite{Abada:2015rta}.

\subsection{Beyond the weak washout regime \label{sec:strongwashout}}

To solve Eqs.~\eqref{eq:Rnstart} and \eqref{eq:mustart} outside the weak washout regime, we will linearise these equations in the small parameters ${\mu_L}_{\alpha}$ and $(\Delta R_N)_{ij} = (R_N - R_{\bar N})_{ij}$, which parametrise the asymmetry in the system.

\vspace{0.5cm}

\noindent\textbf{Zeroth order}

\noindent Let us first consider the equation for the singlet states, Eq.~\eqref{eq:Rnstart}. We will first solve it at zeroth order in $\mu_\alpha$, which will provide some useful insight on how to treat the linearised system: 
\begin{equation}
\frac{d R_N^{(0)}}{dt}=  -i \left[\langle H \rangle ,R_N^{(0)}\right]-\frac{1}{2}\langle \gamma^{(0)} \rangle \left \{ F^{\dagger}F, R_N^{(0)}-I\right \} \,.
\label{eq:Rnstart0}
\end{equation}
We can now conduct a series of simplifications. First, we perform a change of variables, namely $t \rightarrow x \equiv T_\text{EW}/T$ with $dt = (M_0/T_\text{EW}^2)\, x \, dx$. Second, we note that 
\begin{equation}
\langle V_N \rangle=\frac{N_D T}{16} F^{\dagger}F \frac{\int d y_N \, y_N f(y_N)}{\int d y_N \, y_N^2 f(y_N)}=\frac{N_D T}{16}\frac{4 \pi^2}{3 \, \zeta(3)}\frac{1}{24}F^{\dagger}F = \frac{\langle \gamma^{(0)} \rangle}{2 \phi^{(0)}}  F^\dagger F \,,
\end{equation}
which introduces
\begin{align}
\phi^{(0)} & = \frac{144 \, \zeta(3)}{N_D \pi^2 T} \langle \gamma^{(0)} \rangle \\
& =\frac{1}{16 \pi} \left[c_Q^{(0)} h_t^2+c_{LPM}^{(0)}+(3 g^2+g^{\,2})\left(c_V^{(0)}+\log\left(\frac{1}{3 g^2+g^{'\,2}}\right)\right)\right] \,.
\end{align}
Third, we perform a change of basis to absorb the oscillations induced by the vacuum Hamiltonian and in order to simultaneously diagonalise all remaining operators on the right-hand side of Eq.~\eqref{eq:Rnstart0}:
 \begin{equation}\label{eq:basis}
R_N^{(0)} \mapsto S^{(0)} = V_\alpha^\dagger R_N^{(0)} V_\alpha \,.
\end{equation}
The derivation and explicit form of the $x$-independent unitary matrix $V_\alpha$ is given in Appendix~\ref{app:A3}. It is of the form 
\begin{equation}
V_\alpha = \begin{pmatrix} e^{i \alpha} f_{11} & e^{i \alpha} f_{12} \\ f_{21} & f_{22} \end{pmatrix}\ ,
\label{eq:Ualpha}
\end{equation}
where $f_{ij}$ are time-independent combinations of the absolute values of the matrix elements of $F^\dagger F$, and $\alpha$ denotes the phase of the matrix element $(F^\dagger F)_{12}$.

With this, Eq.~\eqref{eq:Rnstart0} can be expressed as
\begin{eqnarray}\label{eq:finalR0}
\frac{\de S^0(x)}{\de x} = S^0(x) \left( (i- \phi^{(0)}) Y + x^2 D\right) - \left((i+\phi^{(0)}) Y + x^2 D\right) S^0(x)
+  2 \phi^{(0)} Y \,,
\end{eqnarray}
where $Y = M_0/(T \, T_\text{EW}) \, V_\alpha^\dagger  \langle V_N \rangle  V_\alpha$ encodes the eigenvalues of $V_N$ and $D$ is defined in Appendix~\ref{app:A3}. Both $D$ and the diagonal matrix $Y$ are $x$-independent.  Equation~\eqref{eq:finalR0} is moreover invariant under $F \leftrightarrow F^*$, i.e.\ in this basis, particles and anti-particles are described by the same quantity $S^0$ (to 0th order in $\mu_\alpha$). This makes this basis highly suitable for linearising our system of differential equations. Note that the expressions for $R_N^{(0)}$ and $R_{\bar N}^{(0)}$ in the original basis however differ, as encoded in the transformation matrix $\bar V_{\alpha}(\alpha) = V_\alpha(-\alpha)$, see Eq.~\eqref{eq:Ualpha}. 

\vspace{0.5cm}

\noindent\textbf{First order}

\noindent We now turn to linearising the full equation for $R_{N, \bar N}$ in this basis, see Eq.~(\ref{eq:basis}). To expand around the 0th order solution, we moreover change variables to 
\begin{equation}
S_+ = S_N + S_{\bar N} = 2 S^0 + \Delta S_+ \,, \quad S_- = S_N - S_{\bar N} = \Delta S_- \ ,
\end{equation}
where $S_{N, \bar N} = V^\dagger_\alpha(\pm \alpha)R_{N, \bar N}V_\alpha(\pm \alpha)$ and $\Delta S_\pm$ denote the contributions arising due to the presence of the terms proportional to $\mu_L$ in Eq.~\eqref{eq:Rnstart}. With this,
\begin{align}
\frac{d \Delta S_-}{dx} &= - x^2 [D, \Delta S_-] - i[Y, \Delta S_-] - \phi^{(0)} \{Y, \Delta S_-\} + \phi^{(1a)} O_\mu + \frac{\phi^{(1b)}}{2} \{O_\mu, S_0\} + {\cal O}(\mu_L \Delta S_- ) \label{eq:S-} \,,
\end{align}
with
\begin{align}
\phi^{(1a)} & \equiv  \frac{144 \, \zeta(3)}{\pi^2 T} \gamma^{(1a)} \nonumber \\
& = \frac{1}{32 \pi} \left[c_Q^{(1)} h_t^2+c_{LPM}^{(1)}+(3 g^2+g^{\,2})\left(c_V^{(1)}+\log\left(\frac{1}{3 g^2+g^{'\,2}}\right)\right)\right] \,, \\
\phi^{(1b)} & \equiv - \frac{144 \, \zeta(3)}{\pi^2 T} \gamma^{(1b)} \nonumber \\
&  =\frac{1}{64 \pi} \left[c_Q^{(2)} h_t^2+c_{LPM}^{(2)}+(3 g^2+g^{\,2})\left(c_V^{(2)}+\log\left(\frac{1}{3 g^2+g^{'\,2}}\right)\right)\right] \,,
\end{align}
and
\begin{equation}
O_\mu = \frac{\pi^2}{144 \, \zeta(3)} \frac{M_0}{T_{EW}} V_\alpha^\dagger [ F^\dagger \mu_L F + \Phi^* F^T \mu_L F^* \Phi] V_\alpha \,,
\end{equation}
where $\Phi = \text{diag}(\exp(- 2 i \alpha), 1)$. 

Similarly, the equation for the asymmetry in the active sector can be cast as
\begin{align}
  16 N_D \frac{T_{EW}}{M_0} \frac{d \mu_\alpha}{dx} =  \left[- \frac{N_D}{2} \phi^{(0)} (F U_c S^\text{aux} U_c^\dagger F^\dagger) + \phi^{(1a)} \mu_L F^\dagger F +  \phi^{(1b)} \mu_L (F  U_c \text{Re}[S_0] U_c^\dagger F^\dagger) \right]_{\alpha \alpha},
\label{eq:mulin}
\end{align}
with 
\begin{align}
 S^\text{aux} & =  2 \, i \, \text{Im}[S_0] +  \text{Re}[\Delta S_-] \,.
\end{align}
Note that the equation for  $\Delta S_+$ decouples from the equations for $\Delta S_-$ and $\mu$ at linear order. It is thus sufficient to solve Eqs.~\eqref{eq:finalR0},  \eqref{eq:S-}  and \eqref{eq:mulin}. In total this enables a strong simplification of the system of differential equations, empowering a fast numerical solution and thus allowing to use this framework for a numerical scan of the parameter space.

The final asymmetries $Y_x = (n_x - n_{\bar x})/s$ are then obtained by evaluating the solutions of Eqs.~\eqref{eq:finalR0}, \eqref{eq:S-}  and \eqref{eq:mulin} at $T = T_\text{EW}$:
\begin{align}
Y_N &= \frac{1}{s} \int \frac{d^3 k_N}{(2 \pi)^3} f_F^0(k_N) \text{Tr}[\Delta R] = \frac{3}{8}\frac{45 \zeta(3)}{\pi^4 g_s} \, \text{Tr}[V_\alpha \Delta S_- V_\alpha^\dagger] \,, \\
Y_{B-L} &= \frac{N_D}{s} \int \frac{d^3 k_L}{(2 \pi)^3} (f_L - f_{\bar L}) = - \frac{2 N_D}{s} \text{Tr}[\mu_\Delta] \int \frac{d^3 k_L}{(2 \pi)^3} f_F^{'}(k_L/T) = 
\frac{45 N_D}{12 \pi^2 g_s} \text{Tr} \mu_\Delta \,, \\
Y_B &= \frac{28}{79} Y_{B-L} \,, 
\end{align}
where $s =  \frac{2 \pi^2 g_s}{45} T^3$ is the entropy density. With Eq.~\eqref{eq:consistency}, one immediately sees that $Y_N = Y_{B-L}$.

The time evolution of the above system is depicted in Fig.~\ref{fig_leptogenesis} for two benchmark points distinguished by the value of the Yukawa coupling $F$. The first benchmark, characterised by $|F|=1.5\times 10^{-7}$, is essentially a weak washout scenario. Once the neutrino oscillations become effective, the asymmetries of both singlet and active sector grow monotonically until $T=T_{\rm EW}$ is reached. The second benchmark solution, given a higher value of the Yukawa coupling, $|F|=1.4 \times 10^{-6}$, shows the characteristic behaviour of strong washout. After reaching a peak asymmetry of $\mathcal{O}(10^{-8})$ around $x\simeq 0.4$, the asymmetry is subsequently reduced by washout processes by about an order of magnitude. The final asymmetry is nevertheless sizeable enough to comply with the observed value.

\begin{figure}
  \centering
\subfigure{%
   \includegraphics[height=4.8cm]{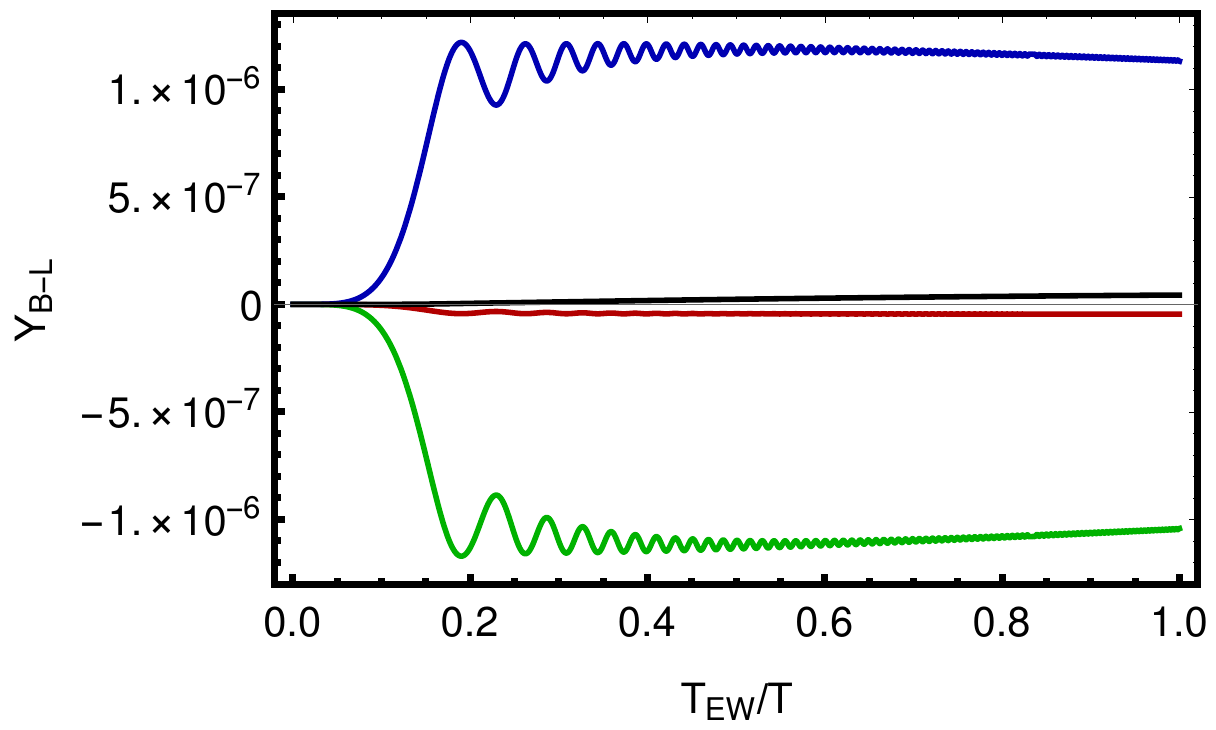}
}\hfill
\subfigure{%
   \includegraphics[height=4.8cm]{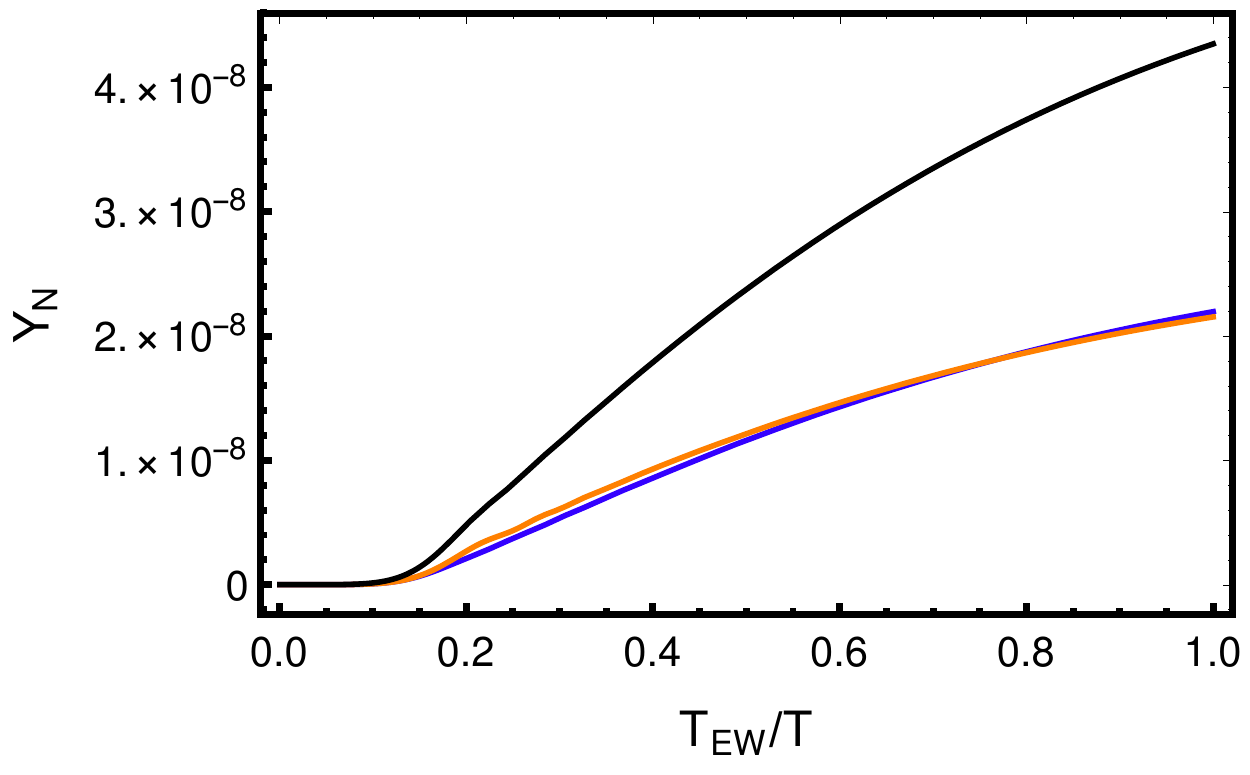} 
} 
\subfigure{%
   \includegraphics[height=4.8cm]{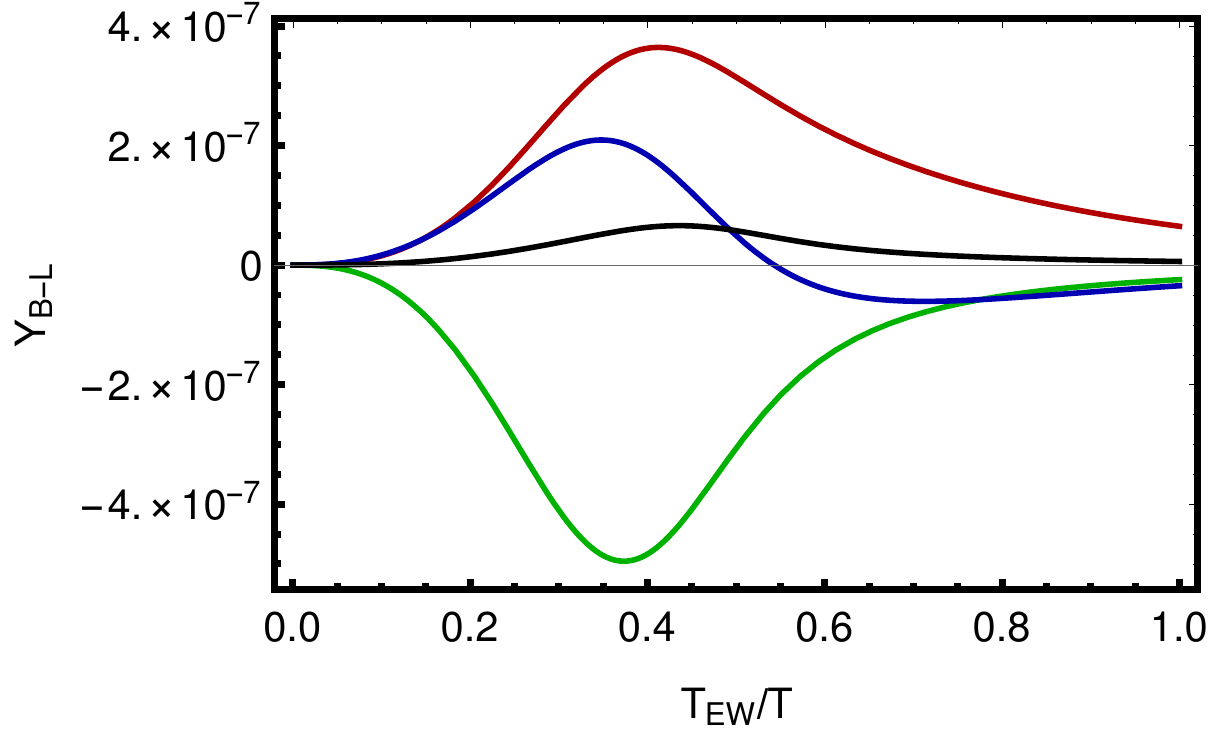}
}\hfill
\subfigure{%
   \includegraphics[height=4.8cm]{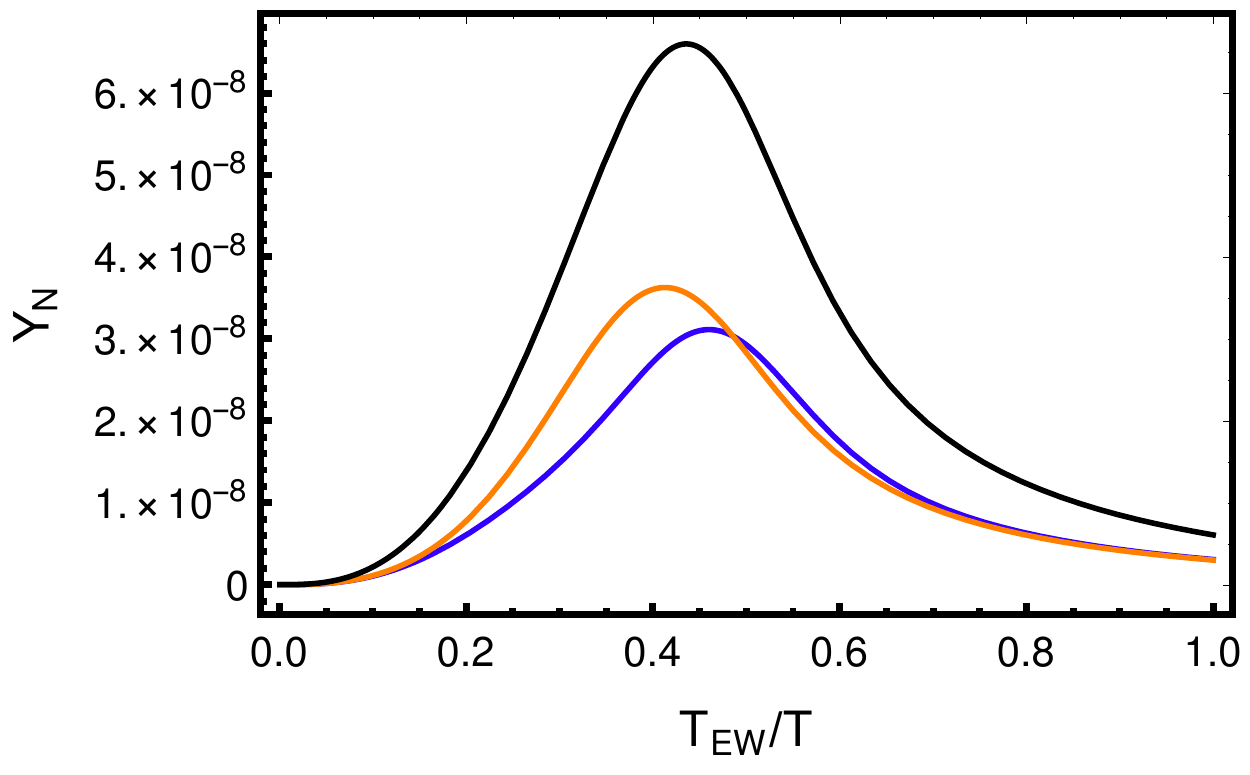} 
}
\caption{Asymmetries in the active and singlet sectors for different values of the Yukawa coupling $|F|$. \textbf{Left panel}: $B$$-$$L$ asymmetries in the three active flavours (coloured) and total asymmetry (black). \textbf{Right panel}: asymmetries of the two singlet flavours (coloured) and the total asymmetry (black). Since lepton number is conserved, the black curve in the left column identical to the black curve in the corresponding right panel. The values for the norm of the Yukawa coupling are (top to bottom) are $|F| = 1.5 \times 10^{-7}$ and $1.4 \times 10^{-6}$. Both examples fulfil all the low-energy neutrino constraints within the LSS-ISS setup.}
\label{fig_leptogenesis}
\end{figure}



\section{Numerical analysis and results \label{sec:results}}

In this section we perform a numerical analysis of the parameter space of the models presented in the Section~\ref{Sec:LNVframework}, taking into account available constraints (discussed in Section~\ref{Sec:Constraints}) and requiring successful leptogenesis (Section~\ref{Sec:mechanism}). In the case of the LSS-ISS scenario, our results represent a direct extension of the analysis conducted in~\cite{Abada:2015rta}, which was limited to the weak washout regime. 
In contrast, our numerical study in this section is focused on the strong washout scenario.  As already anticipated in~\cite{Abada:2015rta}, in this case the correct baryon asymmetry is obtained for relatively large mixing angles between the heavy and the active neutrinos, testable with future experiments such as NA62~\cite{Hahn:1404985}, SHiP~\cite{Alekhin:2015byh,Anelli:2015pba}, FCC-ee~\cite{Blondel:2014bra} and LBNF/DUNE~\cite{Acciarri:2015uup}. We will then extend our analysis to the more refined models ISS(2,2) and ISS(2,3). Also in these cases our study will be focused on  the strong washout regime since it was already found in~\cite{Abada:2015rta} that viable leptogenesis cannot be achieved in the weak washout regime for these models. Although the two ISS setups feature a similar outcome in terms of allowed masses and mixing angles of the neutrinos responsible for the leptogenesis process, the ISS(2,3) receives additional constraints due to the presence of a potential DM candidate.

To ease the notation, we will refer to the mass scale of the pseudo-Dirac pair involved in leptogenesis as $m_\text{PD}^2$, $\Delta m^2$ referring to the splitting between the two squared masses of this pair and $\Delta m \equiv \sqrt{\Delta m^2}$. If a second (heavier) pseudo-Dirac pair is present (as in the case of the ISS), we denote the corresponding mass scale as $M_\text{PD}$.

\subsection{The LSS-ISS model}\label{sec:Results.LSS + ISS}
To analyse our parameter space, we proceed in two steps. In the first one, we generate parameter points within the LSS-ISS neutrino mass model which reproduce the low-energy neutrino observables, i.e.\ the mass splittings and mixing angles observed in neutrino oscillations. We also impose the bounds from direct and indirect searches for singlet neutrinos discussed in Section~\ref{Sec:Constraints}. In a second step, we calculate the resulting baryon asymmetry, based on the differential equations given in Section~\ref{Sec:mechanism}. Here we briefly outline both procedures.

\subsubsection{Parameter space}\label{sec:parISSLSS}

For the LSS-ISS case, we adopt the parametrisation of Ref.~\cite{Gavela:2009cd}. For a normal-ordered hierarchy among the active neutrinos, the six Yukawa couplings in Eq.~\eqref{eq_Mpertexp} are obtained as
\begin{align}
Y_\alpha &= \frac{y}{\sqrt{2}} \left[ U^{(\nu)\, *}_{\alpha 3} \sqrt{1 + \rho} + U^{(\nu)\, *}_{\alpha 2}\right] \,, \\
Y'_\alpha &= \frac{y'}{\sqrt{2}} \left[ U^{(\nu)\, *}_{\alpha 3} \sqrt{1 + \rho} - U^{(\nu)\, *}_{\alpha 2}\right] + \frac{k}{2} Y_\alpha\ ,
\end{align} 
with $U^{(\nu)}_{\alpha i}$ denoting the entries of the $3 \times 3$ PMNS matrix\footnote{Similar to the unitary matrix ${\cal U}$, the PMNS matrix $U^{(\nu)}$ is obtained by diagonalising the neutrino mass matrix, however in this case after integrating out the SM singlet states.} and
\begin{align}
\rho \equiv \frac{\sqrt{1 + r} - \sqrt{r}}{\sqrt{1 + r} + \sqrt{r}} \,, \quad  r \equiv \frac{|\Delta m^2_\text{solar}|}{|\Delta m^2_\text{atm}|} \,, \quad  k \equiv \frac{\xi}{\epsilon} \,.
\end{align}
$y$ and $y'$ are two positive real parameters characterising the size of the Yukawa couplings, which in the spirit of this model we will assume to be of similar size.
In the case of an inverted hierarchy among the active neutrinos, one needs to replace
\begin{equation}
\rho \mapsto \frac{\sqrt{1 + r} - 1}{\sqrt{1 + r} + 1}\,, \quad  U^{(\nu)}_{\alpha 3} \mapsto U^{(\nu)}_{\alpha 2} \,, \quad U^{(\nu)}_{\alpha 2} \mapsto U^{(\nu)}_{\alpha 1} \,.
\end{equation}
This parametrisation conveniently encodes the observed mixing angles in the PMNS matrix. Since one of the active neutrinos remains massless, we can directly associate the masses of the active neutrinos,
\begin{equation}
m_1 = 0 \,, \quad m_2 = \left|\frac{\epsilon y y' (1 - \rho) v^2}{2 \Lambda} \right| \,, \quad m_3 = \left|\frac{\epsilon y y' (1 + \rho) v^2}{2 \Lambda} \right|\ ,
\end{equation}
with the measured mass splittings, eliminating a further parameter. The masses of the two heavy neutrinos are given by $m_{4,5} \simeq m_\text{PD} (1 \mp \xi)$ with $m_\text{PD} = |\Lambda|$. In this parametrisation, the Dirac phase $\delta_{CP}$ and the Majorana phase $\alpha^{(\nu)}$ appear in $Y$ and $Y'$, whereas the third `high-energy' phase is assigned to $\Lambda$.

With this we perform a systematic scan covering the parameter ranges
\begin{align}
& 100~\text{MeV} \leq m_\text{PD} \leq 50~\text{GeV} \,,  \label{eq:rangeISS_Lambda}\\
& 10^{-7} \leq y, y' \leq 10^{-4} \,, \\
& 10^{-7} \leq k \leq 1 \,. \label{eq:rangeISS_k}
\end{align}
Here the range of $m_\text{PD}$ is bounded from below by the requirement that the singlet neutrinos should decay before BBN and from above by the assumption that the singlet neutrinos are ultra-relativistic, implying that lepton number is approximately conserved. The range of $y$ and $y'$ selects the strong washout regime ($|F| \gtrsim 10^{-7}$) where we omit too large Yukawa couplings since in this case the strong washout processes will erase all the previously produced asymmetry.\footnote{Note that within the framework of two heavy neutrinos and small LNV parameters presented here, we will typically obtain at most a moderate hierarchy between Yukawa couplings to different active flavours, $Y_{e} \sim Y_{\mu} \sim Y_{\tau}$. This provides a contrast to Ref.~\cite{Hernandez:2015wna}, where also significant hierarchies were considered. This constraint does not apply to the hierarchy between the Yukawa couplings associated with the two different singlet states, which is governed by the parameter $\epsilon$.} The range of $k$ reflects that on the one hand, we expect $\epsilon$ and $\xi$ to be of similar size (both violate lepton number by two units) while on the other hand, a mild hierarchy $\xi < \epsilon $ is preferred to simultaneously reproduce the light neutrino mass scale and obtain a sufficiently small mass splitting between the singlet states. Note that the case $k \gg 1$ (and $k \ll \xi^2$) corresponds to the limit of the pure Inverse Seesaw, which will be discussed below.\footnote{Swapping the labels of the fourth and fifth column in the mass matrix corresponds to $\epsilon \mapsto 1/\epsilon$, i.e. $k \rightarrow \xi^2/k$. In this sense, $k \ll \xi^2$ also corresponds to the pure ISS limit.} 
For each parameter point the three CP phases are chosen randomly. All parameter points are furthermore checked for consistency with the bounds from direct and indirect searches for singlet neutrinos discussed in Section~\ref{Sec:Constraints}.

Considering the parameters relevant for leptogenesis, the choice of small LNV parameters $\epsilon$ and $\xi$ has interesting consequences for the matrix  $V_\alpha$ introduced in Eq.~\eqref{eq:Ualpha}. Recall that $V_\alpha$ is the unitary matrix diagonalising the operators in the 0th order differential equation for the singlet neutrinos, i.e.\ diagonalising $F^\dagger F$. Let us investigate the properties of $V_\alpha$ in a toy model with a single active neutrino, where the neutrino  mass matrix is given by Eq.~(\ref{eq_Mpertexp:1gen}). 
In the limit where $\epsilon, \xi \rightarrow 0$, the matrix ${\cal U}$ diagonalising the symmetric matrix ${\mathcal{M}}^{(\nu)}$, ${\cal U}^T {\mathcal{M}}^{(\nu)} {\cal U} = m_\text{diag}$, is given by\footnote{In the limit $\epsilon = \xi = 0$, ${\mathcal{M}}^{(\nu)}$ has two degenerate eigenvalues and ${\cal U}$ is not unique. The solution presented here is distinguished since it continuously maps to the solution for small but finite $\epsilon$ and $\xi$.}
\begin{equation}
{\cal U} = \begin{pmatrix}
1 & 0 & 0 \\
0 & i e^{i \beta/2} /\sqrt{2} & e^{- i  \beta/2} /\sqrt{2} \\
0 & - i /\sqrt{2} & 1/\sqrt{2}  
\end{pmatrix}\ . \label{eq:Utoy}
\end{equation}
Here $\beta$ is the phase of $Y'$, whereas $Y$ can be taken to be real and positive without loss of generality.  The columns of ${\cal U}$ are the eigenvectors of 
$({\mathcal{M}}^{(\nu)})^\dagger {\mathcal{M}}^{(\nu)}$, and the requirement of $m_\text{diag} > 0$ determines the phase of these vectors (up to an ambiguous unphysical sign).
With this, we can determine the Yukawa couplings in the mass eigenbasis as
\begin{equation}
F^{\alpha I} = Y_{\alpha j} {\cal U}_{j I} = e^{- i \beta/2} Y /\sqrt{2} \, (\pm i \,, \; \pm 1) \,.
\end{equation}
We can now determine the unitary matrix $V_\alpha$ which diagonalises $F^\dagger F$. Parameterising $V_\alpha $ as 
\begin{equation}
V_\alpha = \begin{pmatrix} 
- e^{i \alpha} \cos \theta & e^{i \alpha} \sin \theta \\
\sin \theta & \cos \theta
\end{pmatrix}\ ,
\end{equation}
we can immediately identify $\alpha = \pi/2$ and $\theta = \pi/4$. Thus, in the limit of vanishing LNV parameters, the eigenvectors of the effective potential $V_N$ are maximally mixed combinations of the degenerate mass eigenstates, and the associated CP-phase indicates maximal CP-violation; the two Majorana states pair form a massive Dirac particle. The same conclusion can be shown to hold in the full model with 3 active neutrino species. 

Switching on $\epsilon$ and $\xi$ in the full model with 3 active neutrino species enables a deviation from the above results $\alpha = \pi/2$ and $\theta =\pi/4$. However, we stress that for small LNV parameters, values of $\alpha \simeq \pi/2$ and $\theta \simeq \pi/4$ are the generic expectation. To avoid entering into too fine-tuned regions of the parameter space, we will thus impose the additional restriction that $\alpha$ and $\theta$ must lie within 10$\%$ of the values derived above.

\begin{figure}
  \centering
\subfigure{%
   \includegraphics[height=5.4cm]{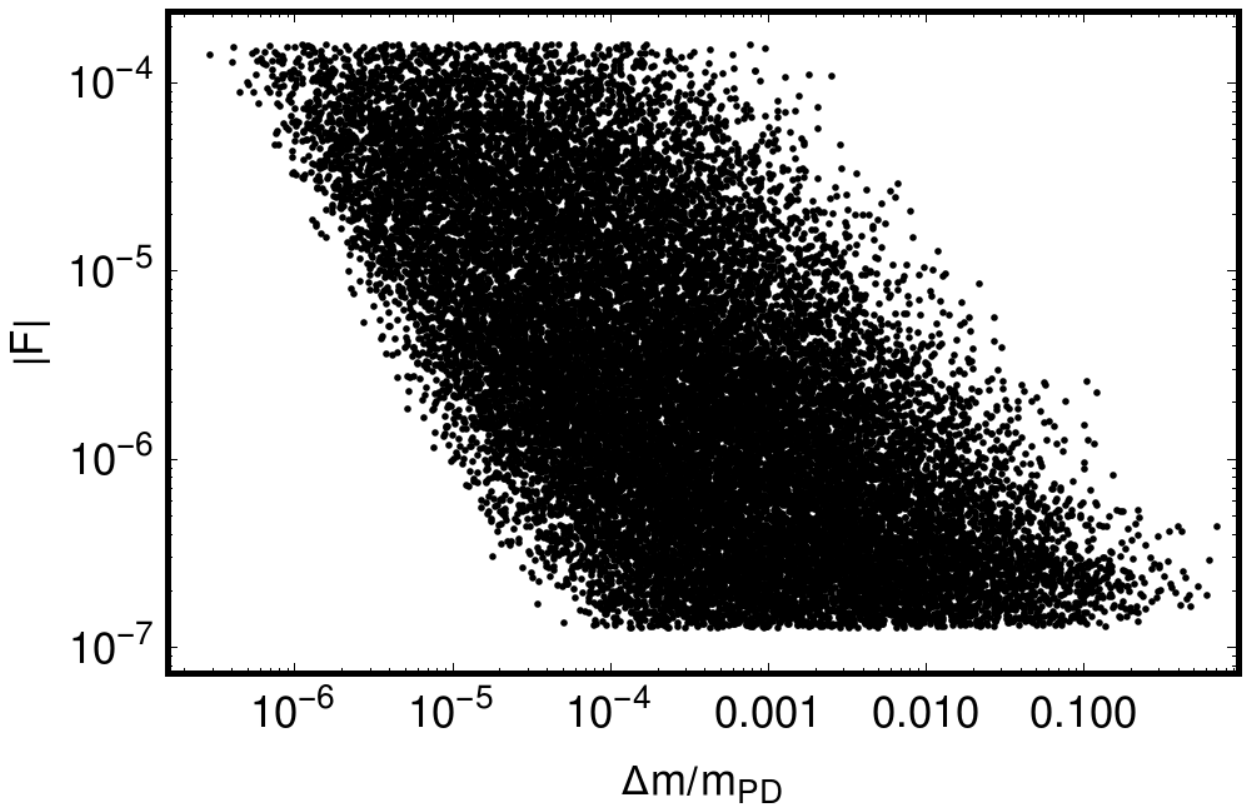}
}\hfill
\subfigure{%
   \includegraphics[height=5.6cm]{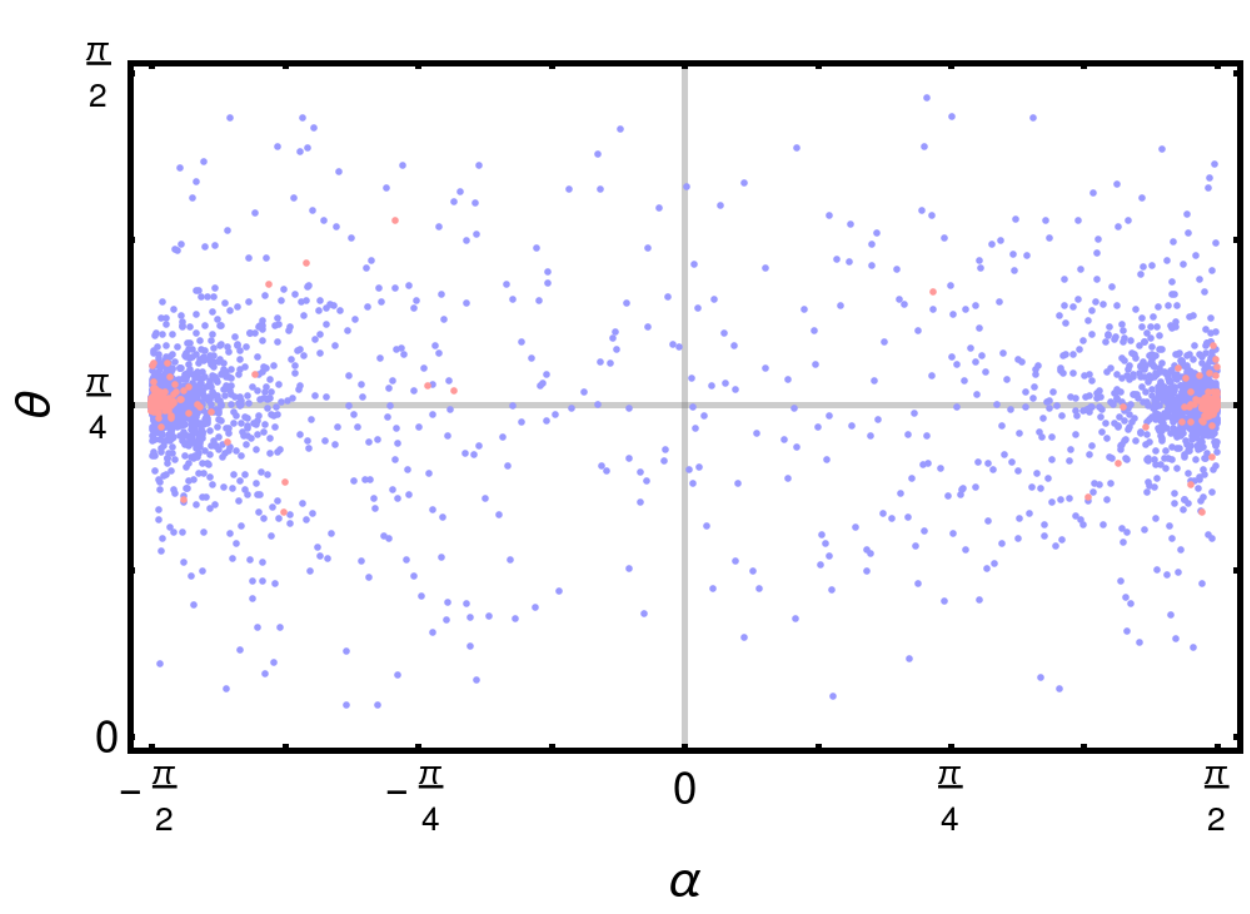} 
}
\caption{Characteristic properties of the relevant parameter space in the LSS-ISS model. 
 \textbf{Left panel:} Norm of the $2\times3$ Yukawa matrix compared to the relative mass splitting. \textbf{Right panel:} The mixing angle $\theta$ and the CP-violating phase $\alpha$ describe the transition from the mass eigenbasis to the eigenbasis of the effective potential for the singlet neutrinos. For small LNV parameters, these are pushed to $\alpha \simeq \pm \pi/2$ and $\theta \simeq \pi/4$. In the right panel, we distinguish between normal hierarchy (blue) and inverted hierarchy (red) in active neutrino mass spectrum. Quantities insensitive to this distinction (as in the left panel) are shown in black.}
\label{fig_LISS_points}
\end{figure}

In Fig.~\ref{fig_LISS_points} we demonstrate some of the key properties of this parameter space, focusing on the parameters which will be relevant for leptogenesis. In the left panel, we show the dependence of the overall scale\footnote{{Here we define $|F|$ (norm of $F$) as the largest singular value of the matrix $F$.}} of the Yukawa coupling on the relative mass splitting between the two heavy neutrinos. Small relative mass splittings, which render leptogenesis through neutrino oscillations particularly efficient, are obtained for large Yukawa couplings - this emphasises why the strong washout regime is of particular interest for this scenario. The depicted dependence can be easily understood from Eq.~\eqref{eq:LISStoymasses}: The ratio $|F|^2/m_\text{PD} \simeq |Y|^2/\Lambda$ is fixed by the light neutrino mass up to a factor of $\epsilon$ (for $k < 1$). The relative mass splitting above the EW phase transition is determined by $\xi$, leading to 
$|F| \simeq \sqrt{2 \, k \, m_\nu \, m_\text{PD}}/(v \Delta m/m_\text{PD})$. Taking into account the ranges of $k$ and $m_\text{PD}$ in Eqs.~\eqref{eq:rangeISS_Lambda} and \eqref{eq:rangeISS_k}, this explains the depicted relation between $|F|$ and $\Delta m/m_\text{PD}$.
 Finally, the right panel of Fig.~\ref{fig_LISS_points} illustrates how for small LNV parameters, the parameters of the mixing matrix in the heavy neutrino sector are  pushed to $\alpha \simeq \pm \pi/2$ and $\theta \simeq \pi/4$. This effect is especially pronounced in the case of the inverted hierarchy.

Additional singlet neutrinos may lead to observable effects not only in direct and indirect searches (described in Section~\ref{Sec:Constraints}) and in leptogenesis (described below), but also in neutrinoless double beta decay. 
Notice however that the contribution to the neutrinoless effective mass $m_{0 \nu \beta\beta}$ from a pseudo-Dirac pair is characterised by two terms which are similar in modulus but opposite in sign (see Eqs.~\eqref{eq:0vudbdecay} and \eqref{eq:Utoy}), with an exact cancellation realised in the limit of vanishing LNV parameters (when the pseudo-Dirac pair reduces to a lepton number conserving Dirac state). For that reason, in the LSS-ISS setup discussed in this paper,
we find no significant enhancement of the SM contribution to neutrinoless double beta decay in any part of the parameter space; see also the discussions in Refs.~\cite{Bezrukov:2005mx,Asaka:2013jfa,Abada:2014vea,Hernandez:2016kel,Drewes:2016lqo}.\footnote{Refs.~\cite{Drewes:2016lqo,Asaka:2016zib,Hernandez:2016kel} recently pointed out that an enhancement of the $0\nu\beta\beta$ decay rate due to two additional heavy neutrinos can be achieved in a specific {part} of the parameter space, characterised by relatively small $m_\text{PD}$, large $\Delta m$ and very different mixings of the two
 heavy neutrinos  to $\nu_e$. 
Note that under the addition of three singlet neutrinos, successful leptogenesis via neutrino oscillation and a sizeable enhancement of the neutrinoless double beta decay rate are  simultaneously possible. This can be traced back to the observation that in the case of three singlet neutrinos, leptogenesis is possible without a high degree of mass degeneracy~\cite{Drewes:2012ma}. While this is surely a very attractive scenario, in this case the connection to (small) LNV is lost.  }


\subsubsection{Leptogenesis in the LSS-ISS \label{sec:BAUinLISS}}

We now turn to the baryon asymmetry in this model, implementing the procedure described in Section~\ref{Sec:mechanism}. In particular, we numerically simultaneously solve the differential equations~\eqref{eq:finalR0}, \eqref{eq:S-} and \eqref{eq:mulin}, starting from vanishing abundances of the singlet neutrinos at $x = 0.5 \times 10^{-3}$ and then evolving the system until the EW phase transition at $x = 1$. We stress that the simplifications discussed in Section~\ref{Sec:mechanism} are crucial to speed up the numerical computation, which can now easily be performed on an ordinary desktop computer.

\begin{figure}
  \centering
\subfigure{%
   \includegraphics[width=8.1cm]{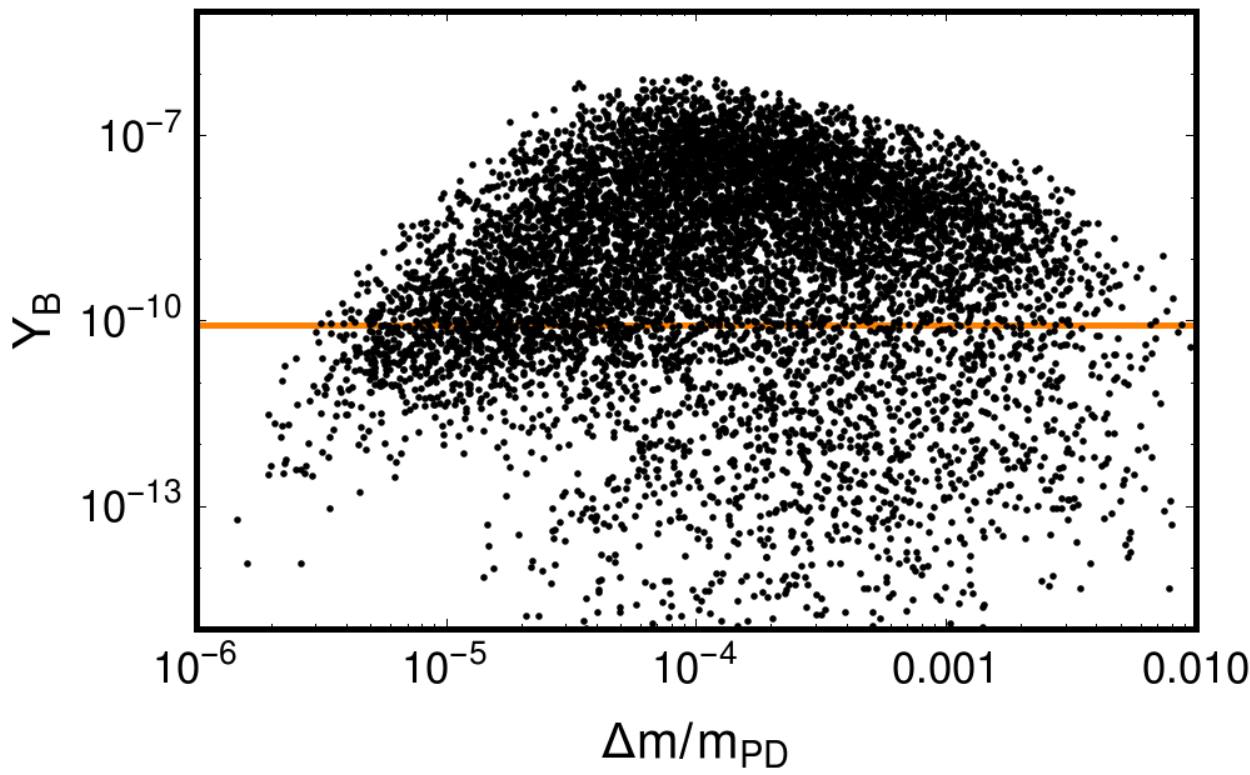}
}\hfill
\subfigure{%
   \includegraphics[width=7.8cm]{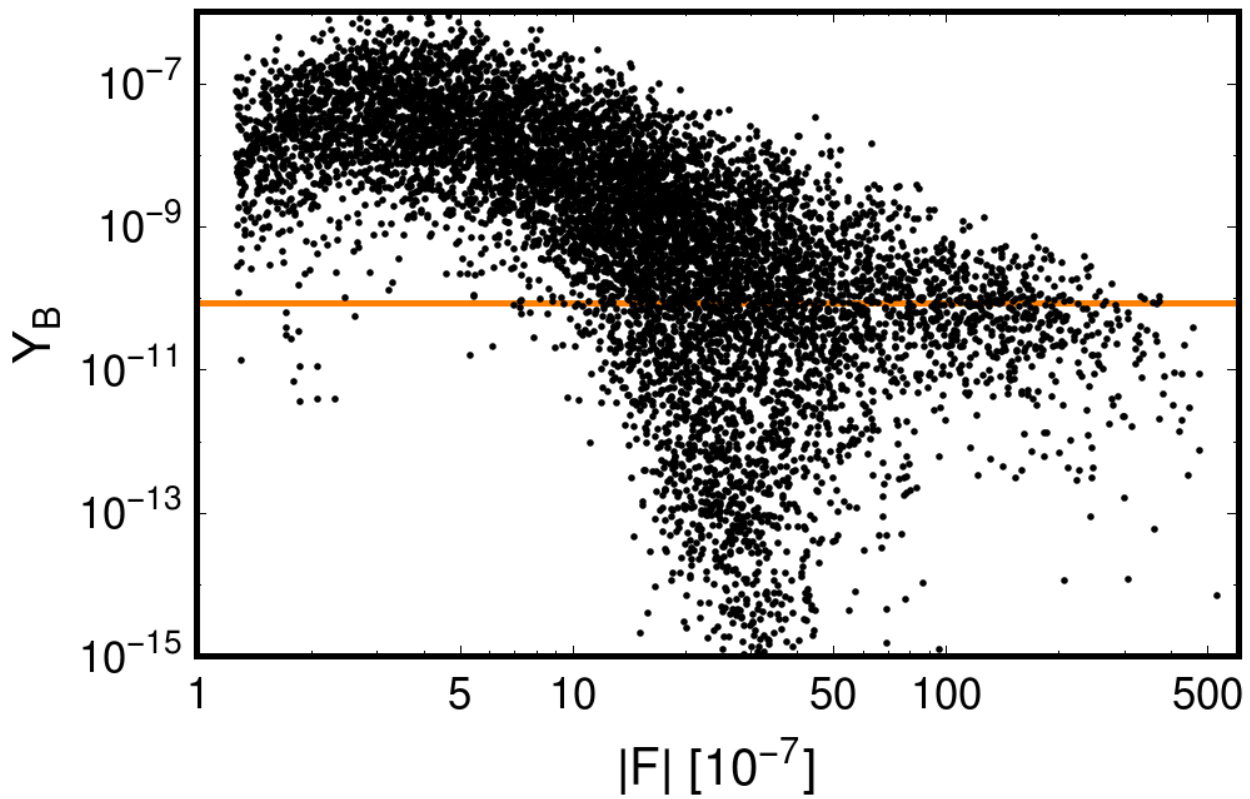} 
}
\caption{Dependence of the generated baryon asymmetry on the mass splitting and on the norm of the Yukawa couplings in the LSS-ISS model. The  horizontal orange line denotes the observed baryon asymmetry, $Y_B = 8.6 \times 10^{-11}$~\cite{Ade:2015xua}.}
\label{fig_LISS_asymmetry}
\end{figure}

In Fig.~\ref{fig_LISS_asymmetry} we show the resulting asymmetry as a function of the relative mass splitting and of the norm of the $2\times 3$ Yukawa matrix. We find a preference for a relative mass splitting around $\Delta m/m_\text{PD} \sim 10^{-4} - 10^{-3}$ and for $|F| \lesssim 10^{-5}$. Larger relative mass splittings render leptogenesis through neutrino oscillations inefficient. Smaller relative mass splittings come with larger Yukawa couplings (see Fig.~\ref{fig_LISS_points}), resulting in a too efficient washout of the generated asymmetry before the EW phase transition. These results confirm that the parameter ranges specified above indeed cover all the parameter space relevant for leptogenesis in the strong washout regime.

\begin{figure}
  \centering
   \includegraphics[width=8.5cm]{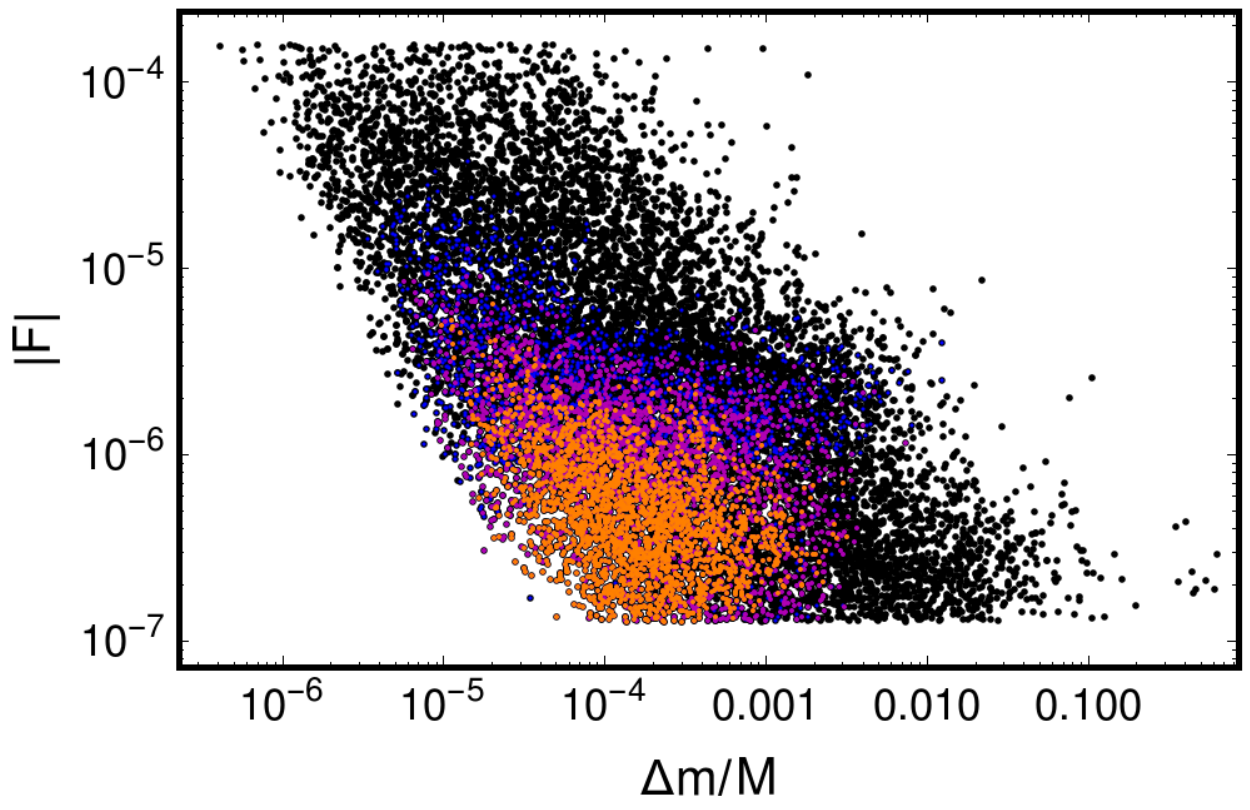}
\caption{Generated baryon asymmetry in the LSS-ISS model in terms of the relative mass splitting in the heavy neutrino pair and the norm of the corresponding Yukawa couplings. The orange (purple, blue) points mark values of the asymmetry larger than $10^{-8}$ ($10^{-9}, 10^{-10}$).} 
\label{fig_LISS_asymmetry_contours}
\end{figure}

These results are further emphasised in Fig.~\ref{fig_LISS_asymmetry_contours}, where the different colours indicate the level of asymmetry achieved in different parts of the parameter space. 
The depicted region is bounded to the bottom left by the lower bound on $k$ and $m_\text{PD}$ in Eqs.~\eqref{eq:rangeISS_Lambda} and \eqref{eq:rangeISS_k}, see Fig.~\ref{fig_LISS_points}. To the right, the relative mass splitting becomes too large to yield effective leptogenesis, whereas from above, too large Yukawa couplings impose a too strong washout of the generated asymmetry.

\begin{figure}
  \centering
\subfigure{%
   \includegraphics[width=8cm]{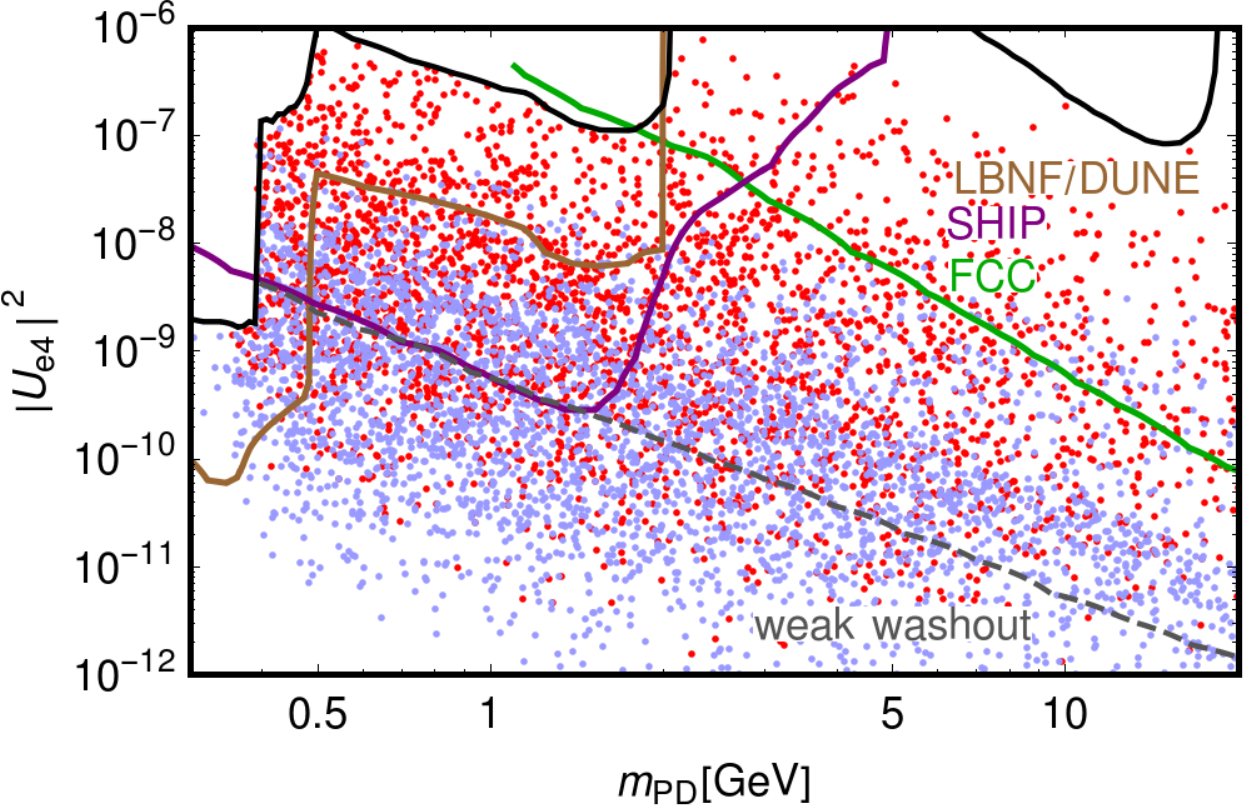}
}\hfill
\subfigure{%
   \includegraphics[width=8cm]{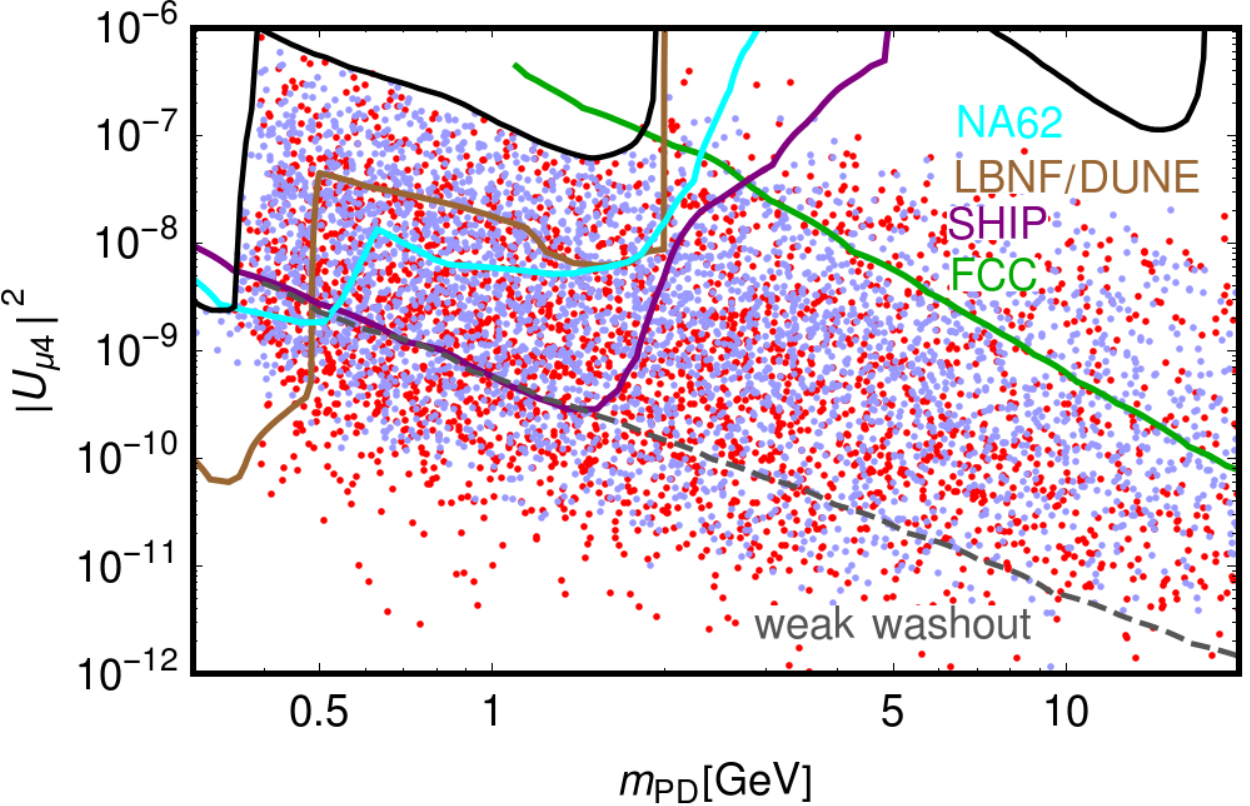} 
}
\caption{Mixing between the active and singlet neutrino sector in the LSS-ISS model for viable leptogenesis solutions. The black line denotes the existing bounds discussed in Section~\ref{Sec:Constraints}, the coloured lines refer to the sensitivity curves of the planned experiments NA62, LBNF/DUNE, FCC-ee and SHiP. For comparison, the dashed grey line indicates the largest mixing found in the weak washout regime in Ref.~\cite{Abada:2015rta}.
As in Fig.~\ref{fig_LISS_points}, the blue (red) points correspond to normal (inverted) hierarchy. The difference between inverted and normal hierarchy is most evident in the mixing with the electron neutrino, here the inverted hierarchy leads to a significantly larger mixing. }
\label{fig_LISS_mixing}
\end{figure}

In Fig.~\ref{fig_LISS_mixing} we depict the mixing between the active and the singlet sector as a function of the heavy neutrino mass scale - for the parameter points which yield successful leptogenesis. Here we consider any parameter point leading to $|Y_B| > Y_B^\text{obs}$ as a viable parameter point for leptogenesis, since for any given parameter point, we can always modify the phases $\delta_\text{CP}$, $\alpha^{(\nu)}$ and $\text{arg}(\Lambda)$ to reduce the asymmetry ($Y_B \rightarrow 0$ if the CP-violating phases vanish) or to flip its sign.
 This mixing is parametrised by the corresponding element ${\cal U}_{\alpha I}$ in the unitary matrix ${\cal U}$ which diagonalises the total neutrino mass matrix ${\mathcal{M}}^{(\nu)}$. Here we show the mixing between the lightest of the singlet neutrinos ($I = 4$) and the electron ($\alpha = e$) / muon ($\alpha = \mu$) neutrino. For comparison we show the reach of future experiments such as NA62~\cite{Ceccucci:NA62}, SHiP~\cite{Anelli:2015pba,Alekhin:2015byh}, FCC-ee~\cite{Blondel:2014bra} and LBNF/DUNE~\cite{Adams:2013qkq}. We note that a sizeable part of the parameter space can be probed by these experiments, in agreement with earlier studies, see e.g.~\cite{Drewes:2016gmt,Drewes:2016jae}. This result is in contrast with what was found in the case of the weak washout regime in~\cite{Abada:2015rta}, where the mixing angle were found to be too small to be probed experimentally.
 Upcoming experiments hence have the potential to discriminate between the weak and strong washout regimes in the context of the LSS-ISS model.


\subsection{The Inverse Seesaw}
\subsubsection{Parameter space \label{sec:iss_parametrization}}
To perform the numerical exploration of the parameter space of the minimal Inverse Seesaw models discussed in Sec.~\ref{Sec:MISS} we adopt a parametrisation inspired by the Casas-Ibarra one~\cite{Casas:2001sr}, but adapted for the ISS(2,2) and ISS (2,3) models. 
In the framework of a generic ISS mechanism, the low-energy effective neutrino mass matrix $m_\nu$ is given by the relation
\begin{equation} \label{eq:ISS_mnu}
d \left(n^{-1}\right)^T \xi \, \Lambda \left(n^{-1}\right) d^T = m_\nu = U^* \hat{m}_\nu U^\dagger,
\end{equation}
where $\hat{m}_\nu$ is a diagonal matrix containing the physical neutrino masses and $U$ is a unitary matrix, which approximately coincides with the PMNS mixing matrix $U^{(\nu)}$ measured in experiments.\footnote{The two matrices are related by
\begin{equation}
U^{(\nu)} = \left( 1 -\frac{1}{2} \Theta \Theta^\dagger\right)U+\mathcal{O}(\Theta^3),
\end{equation}
where the matrix $\Theta$ parametrises the deviation from unitarity of the PMNS matrix. Given the strong experimental constraints on it, $\Theta$ can be neglected in the present discussion.} By working in a basis in which the sub-matrix $\xi \, \Lambda$ in Eq.~\eqref{general-iss} is real and diagonal, it is possible to rewrite Eq.~(\ref{eq:ISS_mnu}) as
\begin{equation}\label{eq:ISS22_low}
\underbrace{U^T d \left(n^{-1}\right)^T \sqrt{\xi \, \Lambda}}_{K} \underbrace{\sqrt{\xi \, \Lambda} \left(n^{-1}\right) d^T U}_{K^T} = \hat{m}_\nu,
\end{equation}
where we have defined a complex ($3\times 2$)-dimensional matrix $K$. The relation $\sum_{i=1,2} K_{\alpha i} K_{\beta i} = \delta_{\alpha \beta} m_\alpha$, with the additional constraint $m_{\alpha} = 0$ for $\alpha =1$ ($\alpha =3$) for normal (inverted) hierarchy  (we recall that in the ISS (2,2) and (2,3) models, the lightest neutrino is massless) provides 10 independent conditions for the entries in $K$, leaving only 2 free parameters. Consequently the matrix $K$ can be parametrised as
\begin{equation}\label{eq:K}
K_{N,I} = \sqrt{\hat{m}_\nu}\ R_{N,I},
\end{equation}
where the ``orthogonal'' matrix $R$ reads 
\begin{equation}\label{eq:Rorth}
\begin{array}{lr}
R_N = \left(
\begin{array}{cc}
 0 & 0 \\
 \cos \gamma  & \sin \gamma \\
 -\sin \gamma & \cos \gamma \\
\end{array}
\right),& R_I = \left(
\begin{array}{cc}
 \cos \gamma & \sin \gamma \\
 -\sin \gamma & \cos \gamma \\
 0 & 0 \\
\end{array}
\right),
\end{array}
\end{equation}
for normal and inverted hierarchy, respectively, and where $\gamma$ is a complex angle.
By inverting the definition for $K$ in Eq.~(\ref{eq:ISS22_low}), it is possible to parametrise the Dirac (and hence the Yukawa) matrix $d$ as
\begin{equation} \label{eq:pard22} 
d_{(2,2)} = U^* \sqrt{\hat{m}_\nu}\ R_{N,I} \sqrt{(\xi \, \Lambda)^{-1}} n^T.
\end{equation}
Equations~(\ref{eq:K}-\ref{eq:Rorth}) ensure the relation in Eq.~(\ref{eq:ISS22_low}) to hold for arbitrary values of $\gamma$. However, the imaginary part of $\gamma$ cannot be too large, since in the present parametrisation the Yukawa couplings are linearly proportional to the functions $\cos\gamma$ and $\sin\gamma$, and large Yukawa entries can violate the perturbativity of couplings, or the seesaw condition $||d|| << ||n||$ (interpreted as a condition on the magnitude of the entries in the $d$ and $n$ matrices, see Eq.~\eqref{general-iss}), rendering in either case the relation in Eq.~(\ref{eq:ISS_mnu}) not suitable to account for low energy phenomenology in neutrino experiments. We thus conduct our scan in the range
\begin{equation}
0\leq \rho\le  2\pi,\hspace{1cm} 0\le \phi \le 2\pi,\hspace{1cm} \textrm{ with } \gamma = \rho e^{i \phi},
\end{equation}
and we perform a consistency check on each realisation of the model, explicitly diagonalising the full $(7\times 7)$ mass matrices constructed with the present parametrisation, and verifying their agreement with neutrino data.

For the (2,3) ISS model an analogous parametrisation can be derived: in this case, however, since $n$ is not squared the matrix $n^{-1}$ is not well defined, and a more general version of Eq.~(\ref{eq:ISS_mnu}) holds:
\begin{equation}
d\ a\ d^T = m_\nu = U^* \hat{m}_\nu U^\dagger,
\end{equation}
where $a$ is the $(2\times 2)$-dimensional submatrix defined as
\begin{equation}
M^{-1} = \left(\begin{array}{cc} a_{2\times 2} & \cdots\\
\vdots &\ddots \\
\end{array}
\right),\hspace{2cm} \textrm{ with } M = \left(
\begin{array}{cc}
 0 & n \\
 n^T & \xi \, \Lambda  \\
\end{array}
\right).
\end{equation}
By diagonalising $a$ with the help of a unitary matrix $W$, $a = W^* \hat{a} W^\dagger$, we obtain
\begin{equation}\label{eq:ISS23_low}
\underbrace{U^T d\ W^* \sqrt{\hat{a}}}_{K_{(2,3)}} \underbrace{\sqrt{\hat{a}}\ W^\dagger d^T U}_{K^T_{(2,3)}} = \hat{m}_\nu,
\end{equation}
from which, analogously to the derivation of eq.~(\ref{eq:pard22}), we can write
\begin{equation} \label{eq:pard23} 
d_{(2,3)} = U^* \sqrt{\hat{m}_\nu}\ R_{N,I} \sqrt{\hat{a}^{-1}}\ W^T.
\end{equation}

To efficiently explore the full parameter space of interest we perform a grid-based numerical scan: for each phenomenologically relevant parameter in the model we chose physically motivated upper and lower bounds, and divide the resulting interval in a number of steps, equally distributed on a logarithmic scale.

The mass scales of the model can be easily linked to the order of magnitude of the sub-matrices in the full ISS mass matrix: the first (second) row of the submatrix $n$ determines the mass scale for the lightest (heavier) pseudo-Dirac pair, while the submatrix $\xi \, \Lambda$ determines the mass splittings within the pseudo-Dirac states, as well as the mass scale for the lightest sterile state in the ISS(2,3). For each point in the sampling of the parameter space of the model, we fix a value for each of these three parameters and generate, in the corresponding sub-matrices, random entries; these entries are of the same order of magnitude than the reference parameter in the scan of the ISS(2,2). As will be discussed extensively in the following, the ISS(2,3) requires a certain amount of hierarchy in the entries of the submatrix $n$ in order to accommodate viable active neutrino-DM mixing angles; we will thus consider, in its scan, random entries that span up to 3 orders of magnitude around the reference parameter. 
Once the sub-matrices $n$ and $\xi \, \Lambda$ are generated in this way, the submatrix $d$ is determined following Eqs.~(\ref{eq:pard22}) or~(\ref{eq:pard23}).
We scan over the following range of masses:
\begin{eqnarray}
m_\text{PD} &\in & \left[ 0.1 - 40 \right] \textrm{ GeV},\non 
M_\text{PD} &\in & \left[ 125 -  10^6 \, \right] \textrm{ GeV},\non 
m_\text{DM} &\in & \left[ 0.1 - 50 \right] \textrm{ keV},
\label{eq:ISSranges}
\end{eqnarray}
where $m_{PD}$ ($M_{PD}$) represent the mass of the lightest (heavier) pseudo-Dirac pair and $m_{DM} \simeq \Delta m$ corresponds to mass splitting in the pairs, or equivalently the mass of the DM candidate in the ISS(2,3). Here the range of $m_\text{PD}$ is determined as in Eq.~\eqref{eq:rangeISS_Lambda}, while  $M_\text{PD}$ is bounded from above by the perturbative unitarity condition, see Eq.~\eqref{eq:sterile:bounds-perturbativity}, and from below 
by requiring that the generated lepton asymmetry is not washed out by the heavier pseudo-Dirac pair (see below). 
Finally for the intermediate scale $m_\text{DM}$, we concentrate on the viable mass range for sterile neutrino DM found in Ref.~\cite{Abada:2014zra}.

\subsubsection{Leptogenesis in the ISS(2,2)}

Ref.~\cite{Abada:2015rta} demonstrated that the minimal ISS models are not capable of reproducing the observed baryon abundance in the weak washout regime. This can be understood by considering a toy model with one active flavour and one heavy pseudo-Dirac pair with mass scale $m_\text{PD}$ and mass splitting $\Delta m$. In this case, the mass scale of the active neutrino and the mass splitting within the pseudo-Dirac pair are given by Eqs.~\eqref{eq:ISStoym1} and \eqref{eq:ISStoym2}, 
implying a relative mass splitting of
\begin{equation}
\frac{\Delta m}{m_\text{PD}} \simeq 0.6 \left( \frac{10^{-7}}{|F|} \right) \left( \frac{m_\text{PD}}{\text{GeV}} \right)^{1/2} \left( \frac{m_\nu}{0.05 \text{ eV}} \right)^{1/2} \,.
\label{eq:nogoISS}
\end{equation}
In the weak washout regime, $|F| \lesssim 10^{-7}$, this is much larger than mass splitting $\Delta m/m_\text{PD}\sim 10^{-6} - 10^{-2}$ required for successful leptogenesis, see Fig.~\ref{fig_LISS_asymmetry}. Ref.~\cite{Abada:2015rta} generalised this argument to realistic models of more active and singlet neutrino flavours, confirming the above naive reasoning also in these cases. However, Eq.~\eqref{eq:nogoISS} also illustrates that these difficulties may be overcome in the strong washout regime with $|F| \gg 10^{-7}$. In this section we demonstrate how indeed low-scale leptogenesis can be successfully implemented within the minimal realistic ISS framework.

\begin{figure}[t]
  \centering
   \includegraphics[width=8cm]{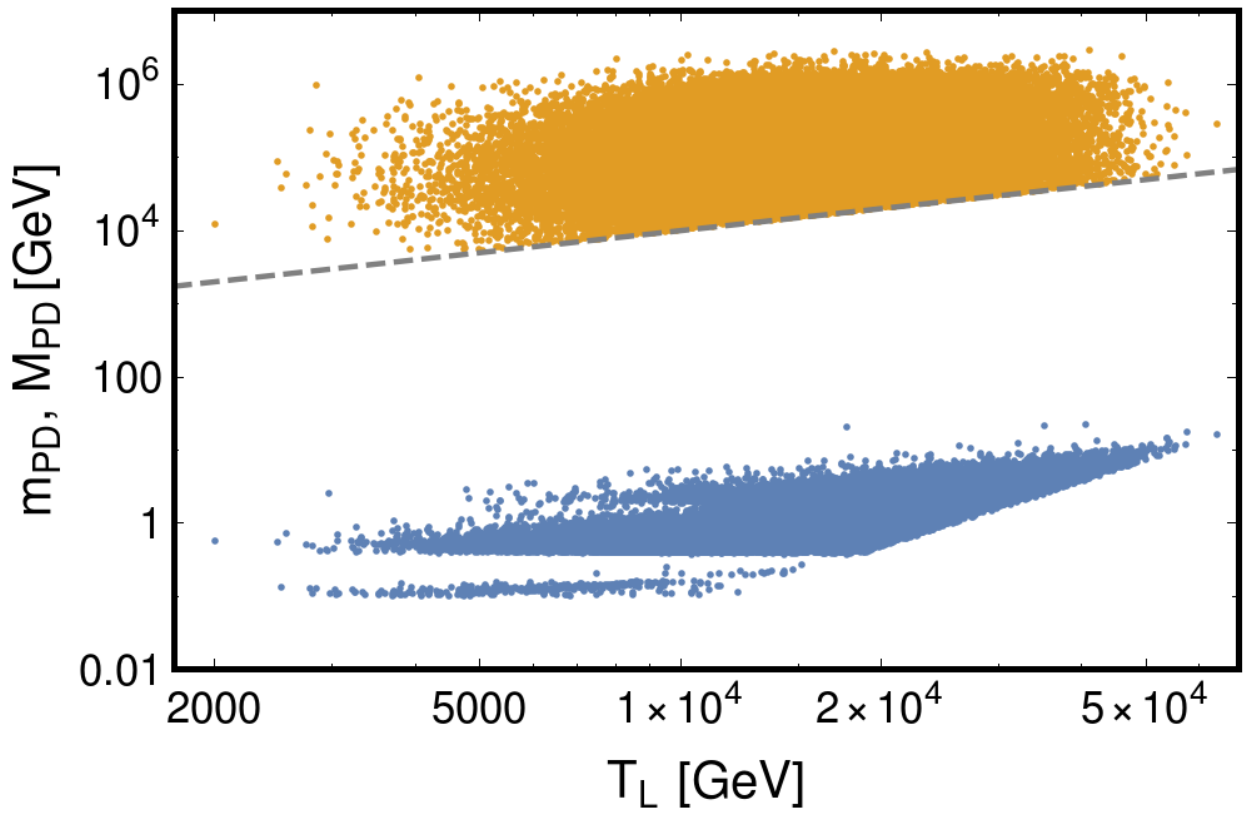}
\caption{Mass scales for the lighter (blue) and heavier (orange) pseudo-Dirac pair in the ISS(2,2) as a function of the leptogenesis temperature $T_L$. Orange points below the dashed line $T_L = M_{PD}$ would lead to a too strong washout of the generated asymmetry.}
\label{fig:ISS22_TL}
\end{figure}

The minimal ISS mechanism which can reproduce the observed neutrino masses and mixings is the ISS(2,2), containing two additional pairs of pseudo-Dirac neutrinos~\cite{Abada:2014vea}. In this section we focus on the possibility that the lighter pseudo-Dirac pair generates the lepton asymmetry as described in Section~\ref{Sec:mechanism}, whereas the mass scale of the second pseudo-Dirac pair is taken to be much heavier, so that it effectively decouples during leptogenesis. This will set the stage for the following section, where in the context of the ISS(2,3), we consider the possibility of simultaneously accounting for (a fraction of) dark matter in the form of sterile neutrinos. We point out that one could also consider the case in which the generation of the baryon asymmetry is accounted by only the heavier pseudo-Dirac pair or by both pairs. We postpone the discussion of these cases to a future study.

Focusing on leptogenesis through the lighter pseudo-Dirac pair requires nevertheless control over the washout rates induced by the heavier pair. Typically, the heavier pair will come with larger Yukawa couplings, thus thermalising earlier, and its interactions with the SM thermal bath can wash out any asymmetry generated by the lighter pair. If however the heavier pair is non-relativistic, its abundance and accordingly the washout processes are exponentially Boltzmann suppressed. Specifically, we will require that at the characteristic leptogenesis temperature $T_L$ (representing in good approximation the temperature at which most of the asymmetry is produced, even if eventually depleted by washout at later times (see also Appendix~\ref{app:leptogenesis_details})),
\begin{equation}
T_L = \left( \frac{\pi^2}{54 \, \zeta(3)} M_0 \; m_{PD} \; \Delta m \right)^{1/3}\,,
\end{equation}
the number density of the lighter pair is larger than that of the heavier one,
\begin{equation}
1 \geq \max\{R_N^{11},R_N^{22}\} > \exp(- M_{PD}/T) \quad  \text{at } T = T_L\,.
\label{cond:washout}
\end{equation}
In Fig.~\ref{fig:ISS22_TL} we show the masses of the heavier pair (in orange) and of the lighter pair (in blue) in terms of the corresponding leptogenesis temperature $T_L$. The dashed line denotes $T_L = M_{PD}$, orange points below this line will not obey Eq.~\eqref{cond:washout}. This sets the lower bound for the range of $M_{PD}$ in Eq.~\eqref{eq:ISSranges}.

\begin{figure}[t]
  \centering
\subfigure{%
   \includegraphics[width=8cm]{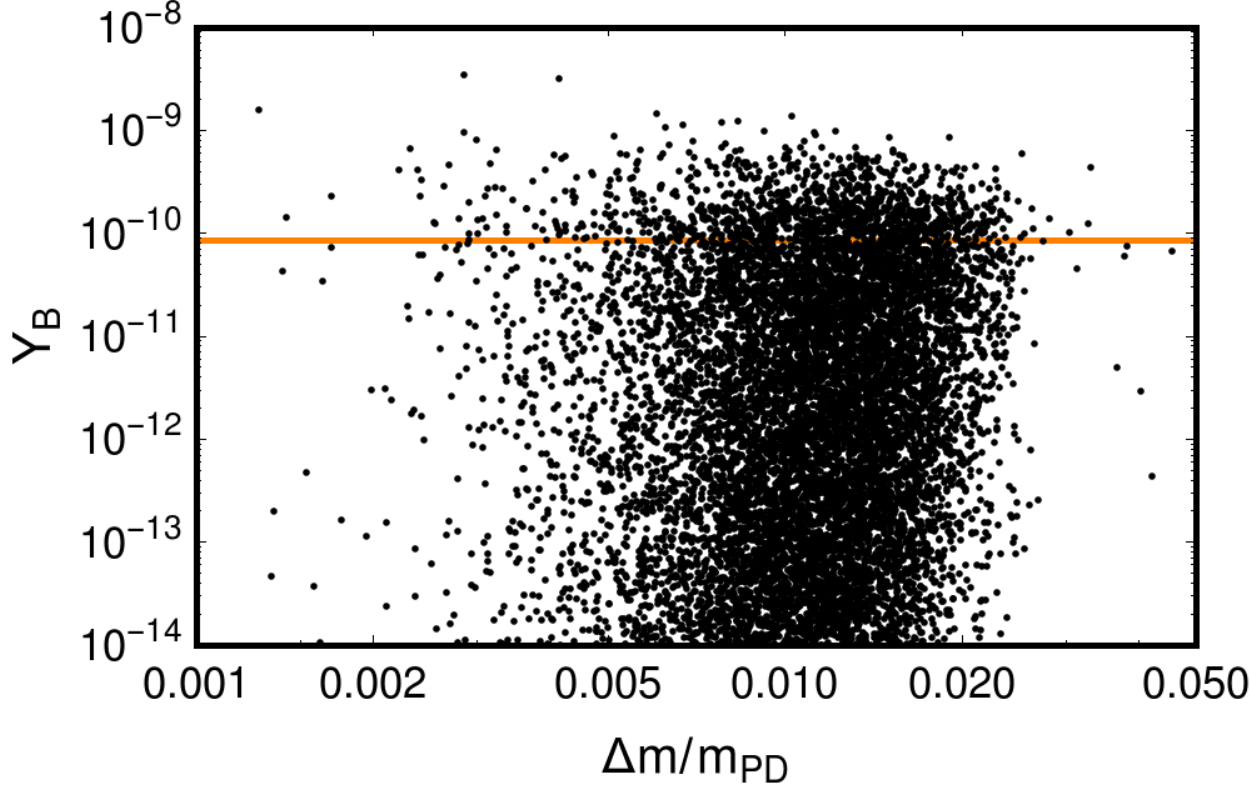}
}\hfill
\subfigure{%
   \includegraphics[width=8cm]{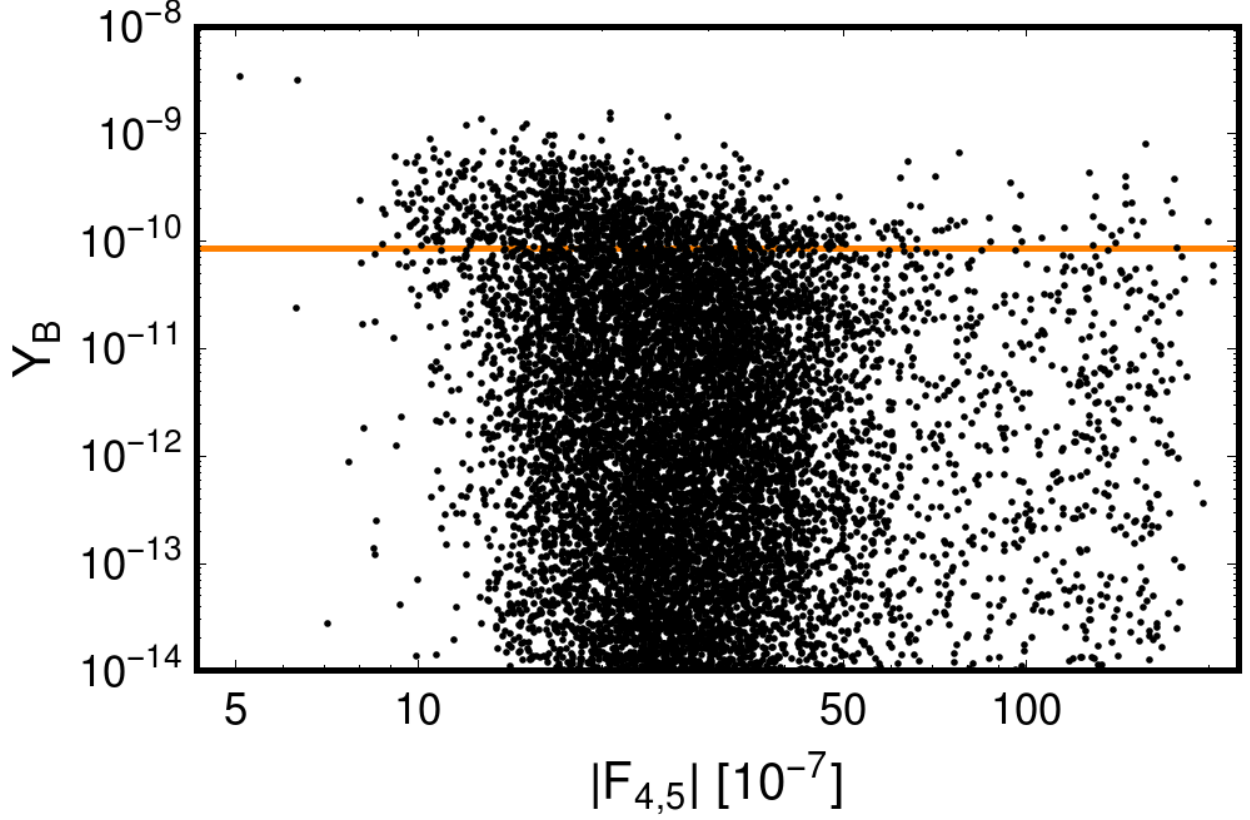} 
}
\caption{Dependence of the produced baryon asymmetry in the ISS(2,2) on the mass splitting and on the magnitude of the Yukawa couplings of the lighter pseudo-Dirac pair. The horizontal orange line indicates the observed asymmetry.}
\label{fig_ISS22_asymmetry}
\end{figure}

Restricting ourselves to points which do obey the condition~\eqref{cond:washout} and for which the washout due to the heavier pair is thus negligible, we proceed as in Section~\ref{sec:BAUinLISS} to calculate the resulting baryon asymmetry, applying the formalism of Section~\ref{Sec:mechanism} to the lighter pseudo-Dirac pair. In Fig.~\ref{fig_ISS22_asymmetry} we show the resulting asymmetry as a function of the mass splitting and the Yukawa coupling. Compared to the LSS-ISS model of Fig.~\ref{fig_LISS_asymmetry}, we note that the mass splitting and the Yukawa couplings are pushed to larger values, reducing the generated asymmetry. While we still find points which produce a sufficient amount of baryon asymmetry, this is more difficult than in the LSS-ISS case. This is the result of the restriction schematically given by Eq.~\eqref{eq:nogoISS} together with the observation that too large Yukawa couplings lead to a too strong washout. Figure~\ref{fig_ISS22_asymmetry_contours} summarises these results in the Yukawa coupling versus mass-splitting plane.

As a result of this tension (Eq.~\eqref{eq:nogoISS} prefers $|F| \gg 10^{-6}$, the preferred range for leptogenesis is $10^{-7} < |F| < 5 \cdot 10^{-6}$), we find a preference for parameter points which feature a (mildly) hierarchical Yukawa spectrum with respect to the active flavour index $\alpha$. A typical example of this type is depicted in Fig.~\ref{fig_ISS22_leptogenesis}. While the Yukawa coupling to the $\tau$-flavour is relatively large, well in the strong washout regime, the coupling to the $\mu$-flavour is much smaller, experiencing only marginal washout (green curve in the left panel). Since the total asymmetry summed over both sectors always vanishes, the asymmetry stored in the active $\mu$-flavour induces asymmetries in the singlet flavours as well as in the other active flavours. This is similar to the situation in flavoured leptogenesis~\cite{Drewes:2012ma}. For the parameter point depicted in Fig.~\ref{fig_ISS22_leptogenesis}, we find a mass splitting of $\Delta m/m_{PD} \simeq 0.01$ and an asymmetry of $|Y_B| \simeq 5.7 \cdot 10^{-10}$.

\begin{figure}
  \centering
   \includegraphics[width=9cm]{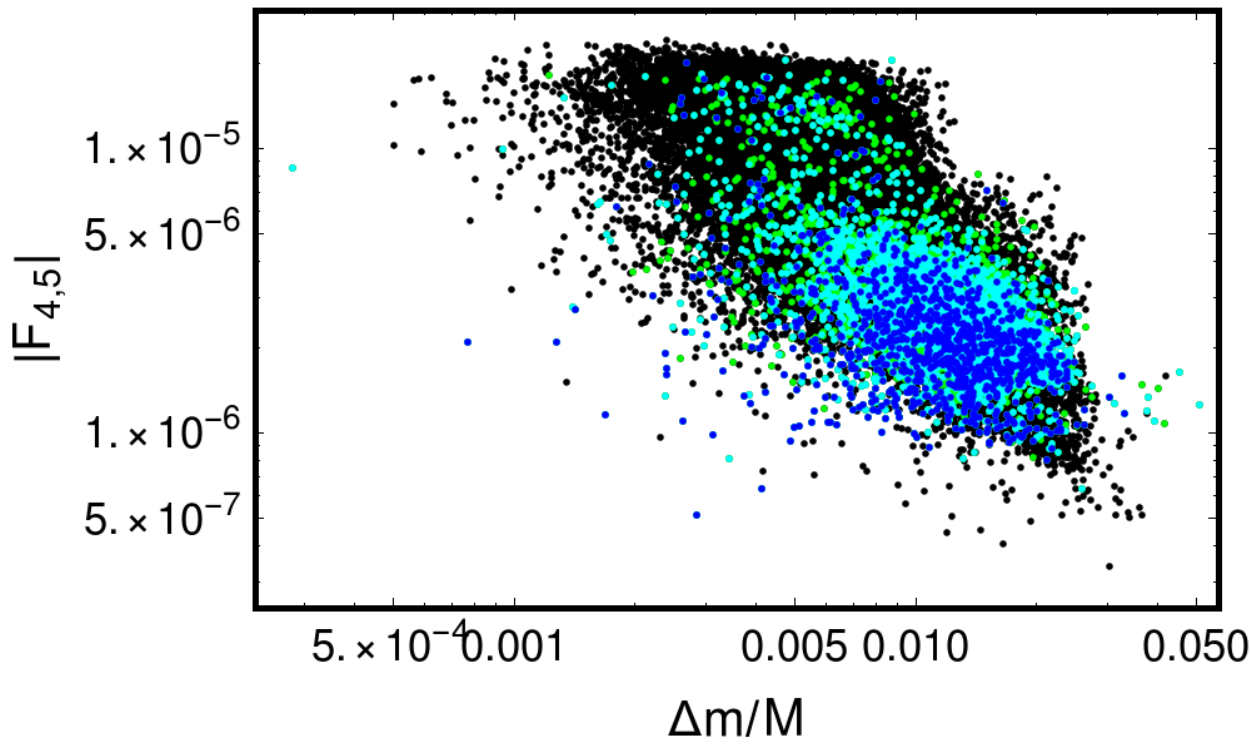}
\caption{Generated baryon asymmetry in the IS(2,2) in terms of the relative mass splitting in the lighter pseudo-Dirac neutrino pair and the absolute value of the corresponding Yukawa coupling. The blue (cyan, green) points mark values of the asymmetry larger than $10^{-10}$ ($10^{-11}, 10^{-12}$).} 
\label{fig_ISS22_asymmetry_contours}
\end{figure}

\begin{figure}
  \centering
\subfigure{%
   \includegraphics[width=8cm]{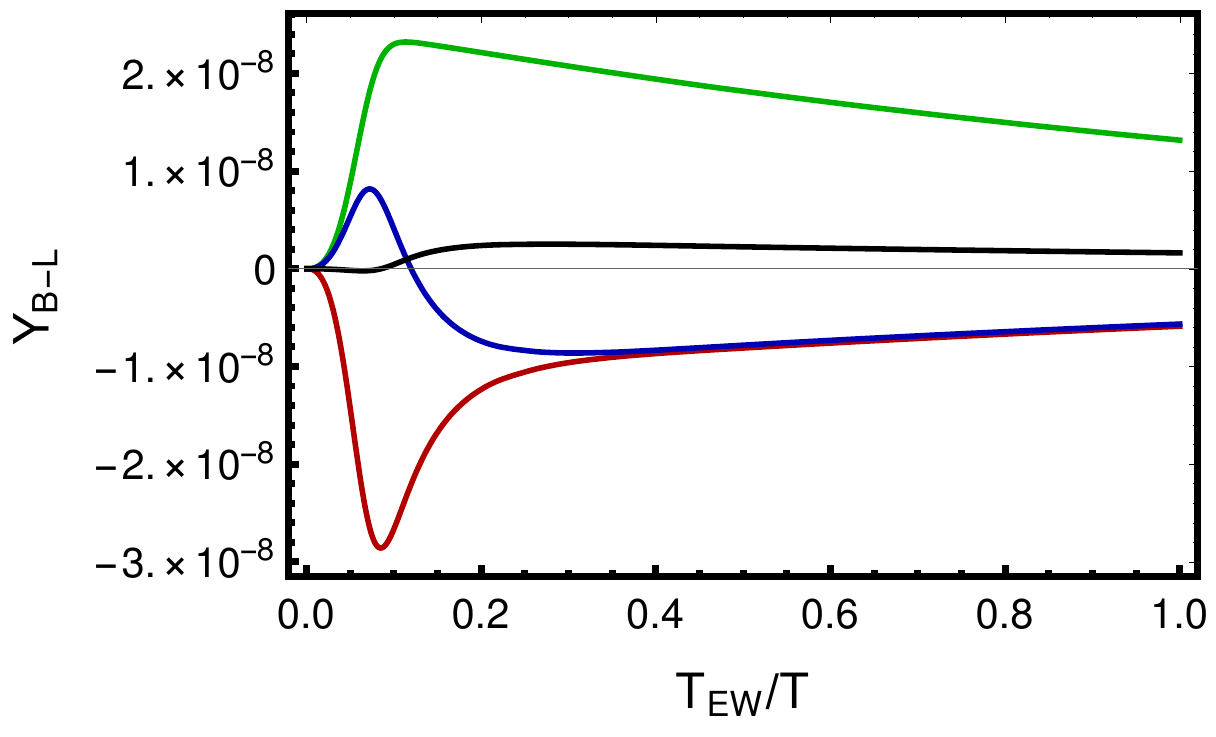}
}\hfill
\subfigure{%
   \includegraphics[width=8cm]{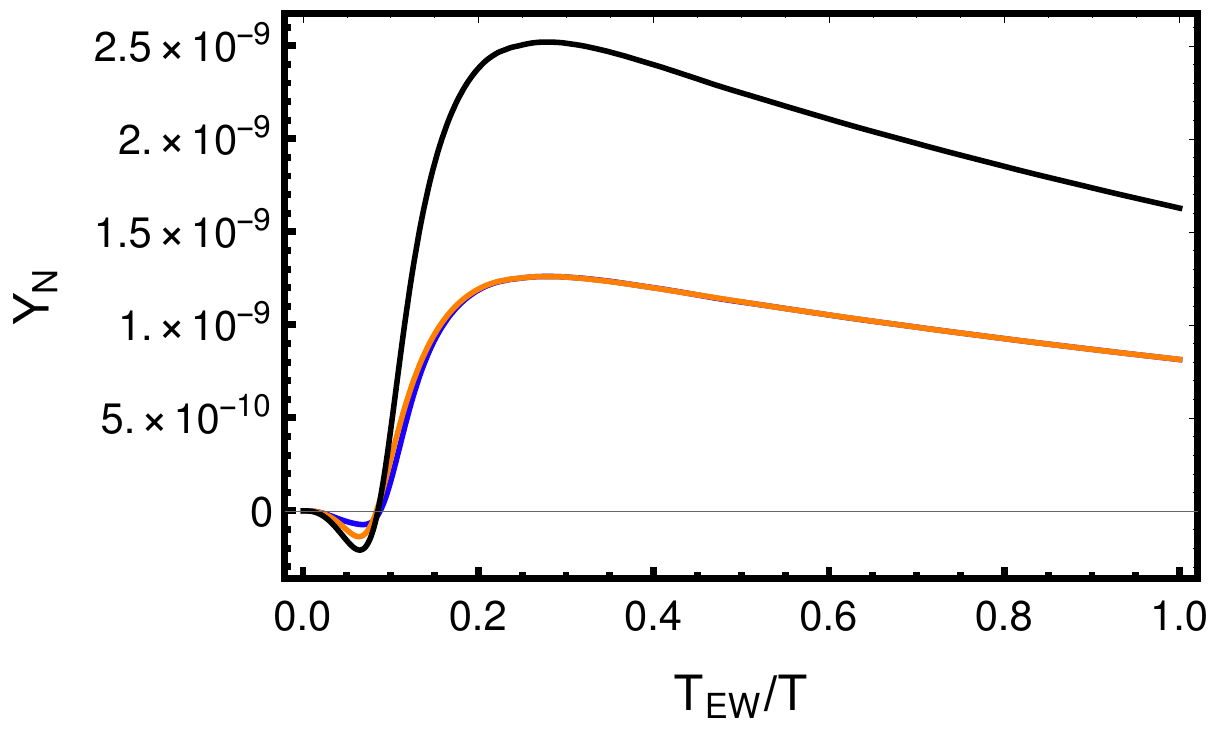} 
}
\caption{Asymmetries in the active and singlet sector in the ISS(2,2), here for an example with a sizeable hierarchy in the Yukawa couplings $|F|$, $|F_{\tau,i}| \simeq 1.2 \cdot 10^{-6} > |F_{e,i}| \gg |F_{\mu,i}| \simeq 1.2 \cdot 10^{-7}$. \textbf{Left panel:} $B$$-$$L$ asymmetries in the active flavours $e, \, \mu, \, \tau$ (red, green, blue) and total asymmetry (black). \textbf{Right panel:} asymmetries in the two sterile flavours of the lighter pseudo-Dirac pair (coloured) and the total asymmetry (black). For this parameter point, the total baryon asymmetry is found to be $|Y_B| = 5.7 \cdot 10^{-10}$.}
\label{fig_ISS22_leptogenesis}
\end{figure}

In analogy with Fig.~\ref{fig_LISS_mixing}, Fig.~\ref{fig_ISS22_mixing} illustrates the mixing between the lighter pseudo-Dirac pair and the active sector, compared to the corresponding expected  sensitivities of NA62, LBNF/DUNE, FCC-ee and SHiP (the heavier pseudo-Dirac pair is not visible in these experiments). Notice that, since the region of viable leptogenesis in the ISS covers a smaller range of masses and mixings with respect to the LSS-ISS case, future experiments can probe almost the all of this space. The lower abundance of points associated with the inverted hierarchy is due to the observation that the ISS setup for neutrino mass generations generally disfavours the inverted hierarchy~\cite{Abada:2014vea}.

\begin{figure}
  \centering
\subfigure{%
   \includegraphics[width=8cm]{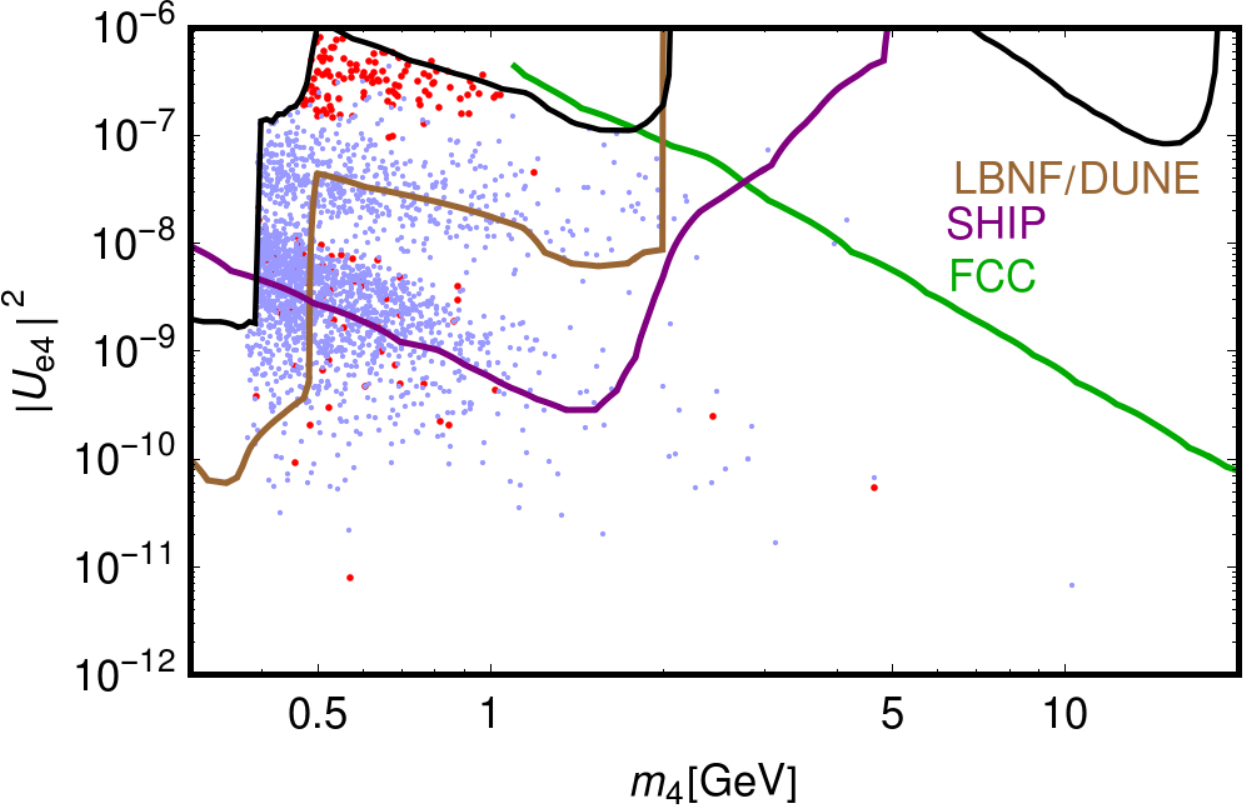}
}\hfill
\subfigure{%
   \includegraphics[width=8cm]{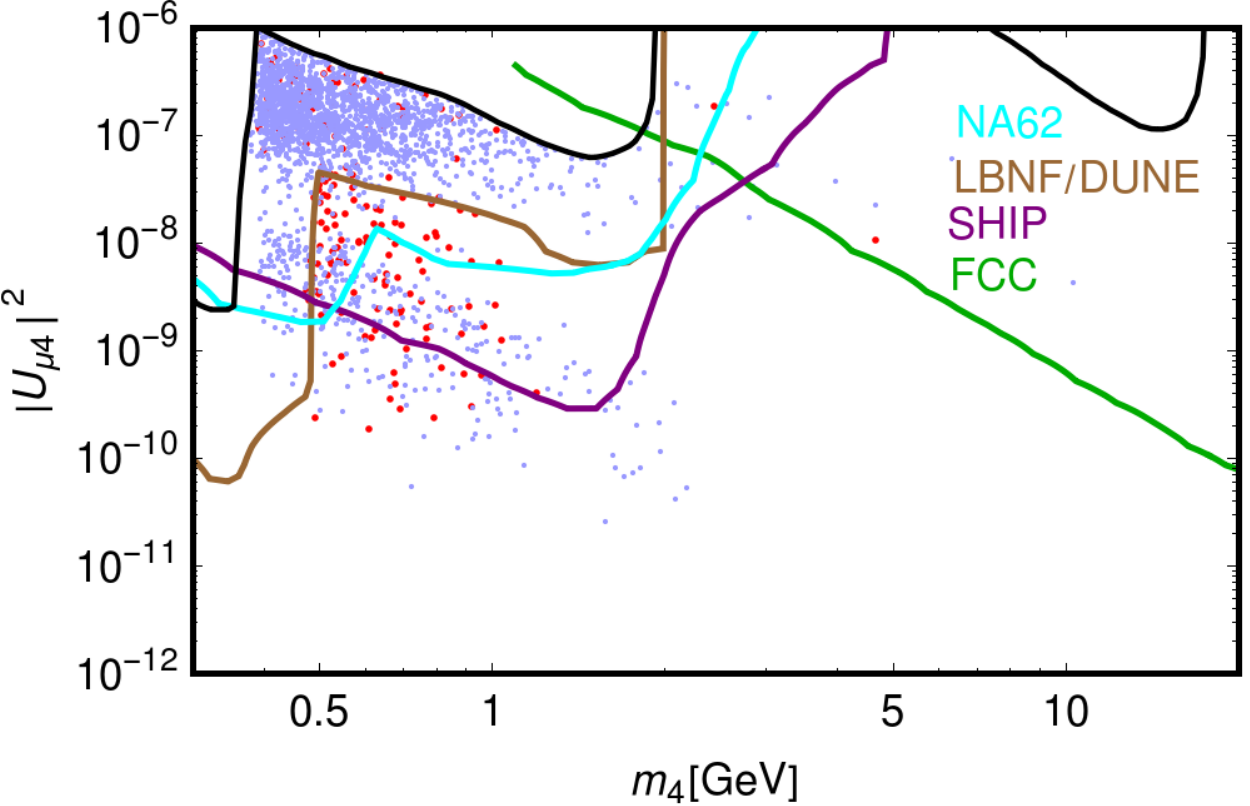} 
}
\caption{Mixing between active and singlet neutrino sector in the ISS(2,2) for viable leptogenesis solutions. The black line denotes the existing bounds discussed in Section~\ref{Sec:Constraints}, the coloured lines refer to the sensitivity curves of the planned future experiments NA62, LBNF/DUNE, FCC-ee and SHiP. Solutions corresponding to the normal (inverted) hierarchy are shown in blue (red). }
\label{fig_ISS22_mixing}
\end{figure}

\begin{figure}
 \centering
  \subfigure{%
    \includegraphics[width=8cm,height=5.2cm]{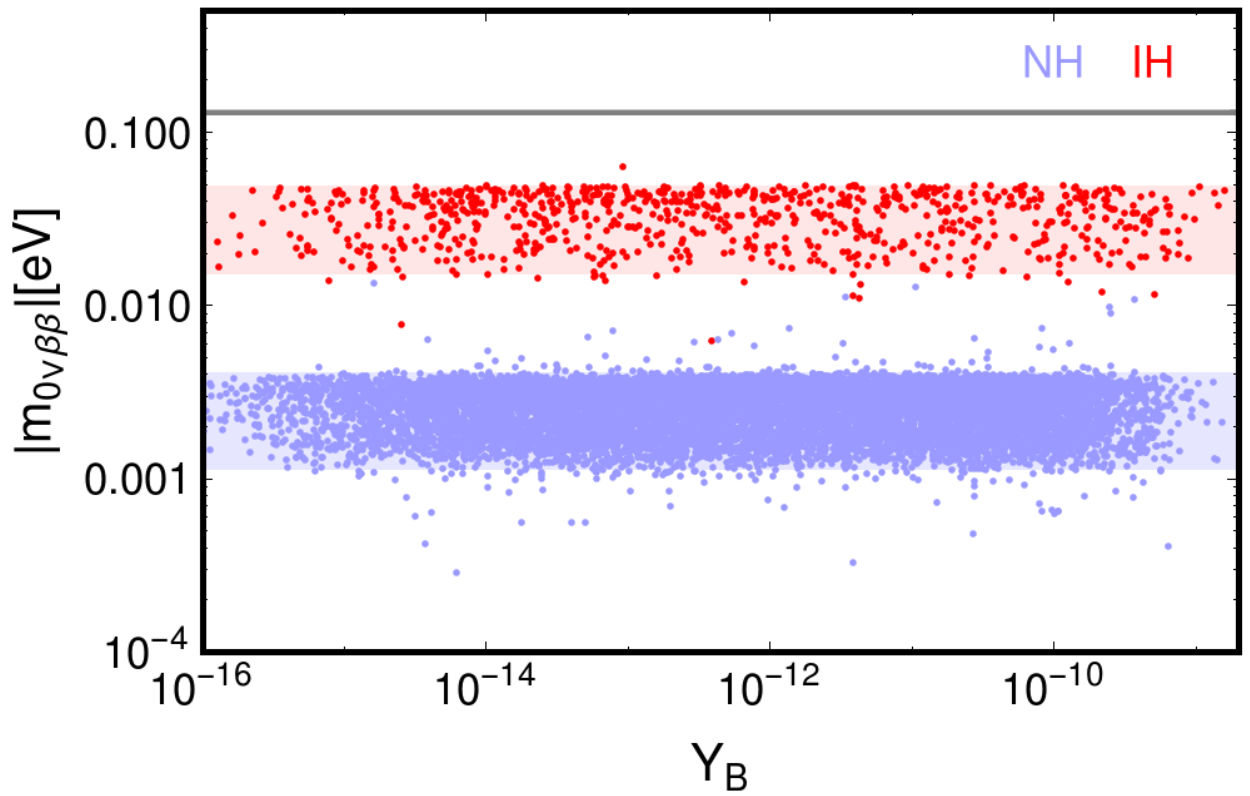}
 }\hfill
 \subfigure{%
   \includegraphics[width=8cm,height=5.2cm]{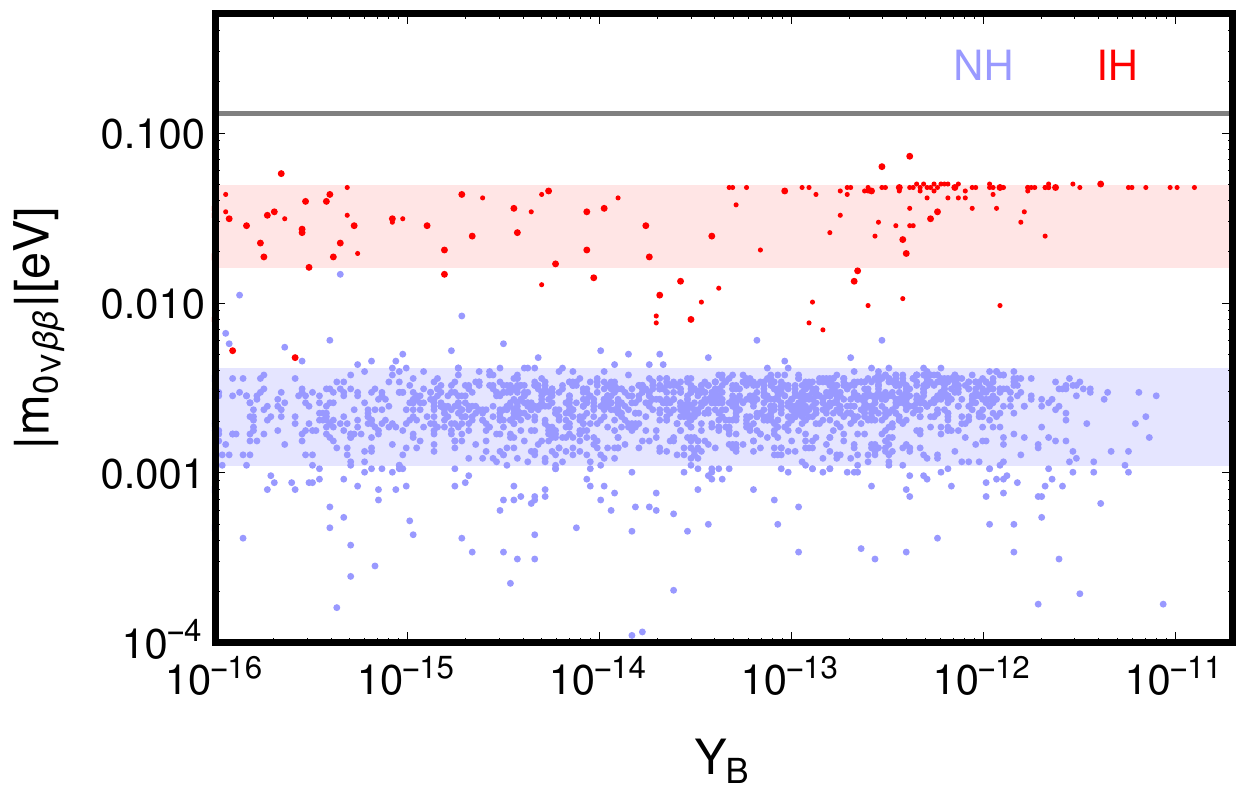} 
}
 \caption{Effective mass parameter of neutrinoless double beta decay in the ISS(2,2) (left panel) and ISS(2,3) (right panel) in terms of the generated baryon asymmetry. Blue (red) points denote solutions corresponding to the normal (inverted) hierarchy. The shaded bands denote the corresponding SM contributions, the horizontal line the current experimental upper bound~\cite{Albert:2014awa,KamLAND-Zen:2016pfg}. In the right panel, the condition of a cosmologically viable DM abundance has been imposed.}
 \label{fig_ISS_mnubb}
\end{figure}

 The effective mass in the amplitude of neutrinoless double beta decay, see Eq.~\eqref{eq:0vudbdecay}, is shown in Fig.~\ref{fig_ISS_mnubb}. Contrary to the LSS-ISS model, the ISS framework in principle allows for the possibility of sizeable contributions, detectable in upcoming experiments~\cite{Abada:2014vea}. However we do not observe this enhancement here for several reasons: firstly, for the contribution of the pseudo-Dirac pairs, an analogous cancellation to the one already discussed in the framework of the LSS-ISS model is at play. In addition the Inverse Seesaw strongly prefers a normal ordering for the light neutrinos, which, together with a massless state, results in the minimal possible contribution of active neutrinos to the effective mass $m_{0 \nu \beta \beta}$. In the ISS(2,3) the contribution of the isolated light sterile state could be sizeable, however cosmological constraints strongly limit the allowed values for its mixing with the active sector (see Sec.~\ref{sec:BAU-DM_ISS23}) in the keV mass range, resulting again in a suppressed contribution to $m_{0 \nu \beta \beta}$.

\subsubsection{Leptogenesis and dark matter in the ISS(2,3)}\label{sec:BAU-DM_ISS23}

Having established that the Inverse Seesaw mechanism can account for a neutrino spectrum suitable for leptogenesis, while simultaneously agreeing with all low-energy neutrino data, we now turn  to the question if the Inverse Seesaw mechanism can (simultaneously) account for dark matter in the form of sterile neutrinos. To this end, we consider the minimal ISS realisation which can account for the low-energy neutrino data and also provides a dark matter candidate, the ISS(2,3), see Sec.~\ref{Sec:MISS}. Here the ISS(2,2) mass spectrum is extended by an additional, mostly sterile state at an intermediate mass scale which can constitute (a fraction of) dark matter~\cite{Abada:2014vea}. The mass of this state is directly linked to the mass splitting within the lighter pseudo-Dirac pair, which is one of the key parameters determining the generated baryon asymmetry. For what concerns the analysis of the viable parameter space for leptogenesis, the ISS(2,3) closely resembles the ISS(2,2) model of the previous sections. Here we hence focus on the role played by the additional intermediate scale sterile state.

Any stable new physics neutrino state with a non-vanishing mixing to the active neutrinos will be produced through active - sterile neutrino conversions according to the so-called Dodelson - Widrow (DW) mechanism~\cite{Dodelson:1993je}. The resulting abundance is proportional to the active-sterile mixing and can be expressed as~\cite{Asaka:2006nq,Abazajian:2001nj}:
\begin{align}
\Omega_\text{DM} h^2 & = 1.1 \cdot 10^{7} \, \sum_\alpha C_\alpha(m_\text{DM}) \,  |{\cal U}_{\alpha 4}|^2 \left( \frac{m_\text{DM}}{\text{keV}} \right)^2 \,, \quad \alpha = e, \mu, \tau \nonumber \\
& \simeq 0.3\left(\frac{\sin^2 2\theta_\text{DM}}{10^{-8}}\right) {\left(\frac{m_{\rm DM}}{10\,\mbox{keV}}\right)}^2\ \,.
\label{eq:DW}
\end{align}
where the coefficients $C_\alpha$ can be determined numerically and are found to be of order~0.5~\cite{Asaka:2006nq}. Here ${\cal U}$ is the unitary mixing matrix introduced below Eq.~\eqref{eq:effYukawa} and ${\cal U}_{\alpha 4}$ parametrises the mixing between the DM candidate and the active sector, $\sin^2 2 \theta_\text{DM}= 4 \sum_{\alpha = e,\mu,\tau} |{\cal U}_{\alpha 4}|^2 $.}

The range of viable DM masses is restricted to $0.1\,\mbox{keV} \lesssim m_{\rm DM} \lesssim 50\,\mbox{keV}$. Smaller masses are forbidden by the Tremaine-Gunn bound~\cite{Tremaine:1979we} (derived by comparing the observed size of dwarf galaxies with a Fermi sphere of DM fermions, see also~\cite{Domcke:2014kla,Randall:2016bqw,DiPaolo:2017geq})  while above 50 keV, the DM candidate is no longer cosmologically stable. Taking into account additional observational constraints on the active sterile mixing, the DW mechanism can account for about $30\%$ of the total dark matter density today~\cite{Abada:2014vea}. In particular, to avoid overproduction of dark matter, the active-sterile mixing angle is required be very small,  $\sin^2 2\theta_\text{DM} < 10^{-(7 \div 10)}$.

\begin{figure}
\centering
\subfigure{%
\begin{tikzpicture}
\node at (0,0) {\includegraphics[width=7.8 cm]{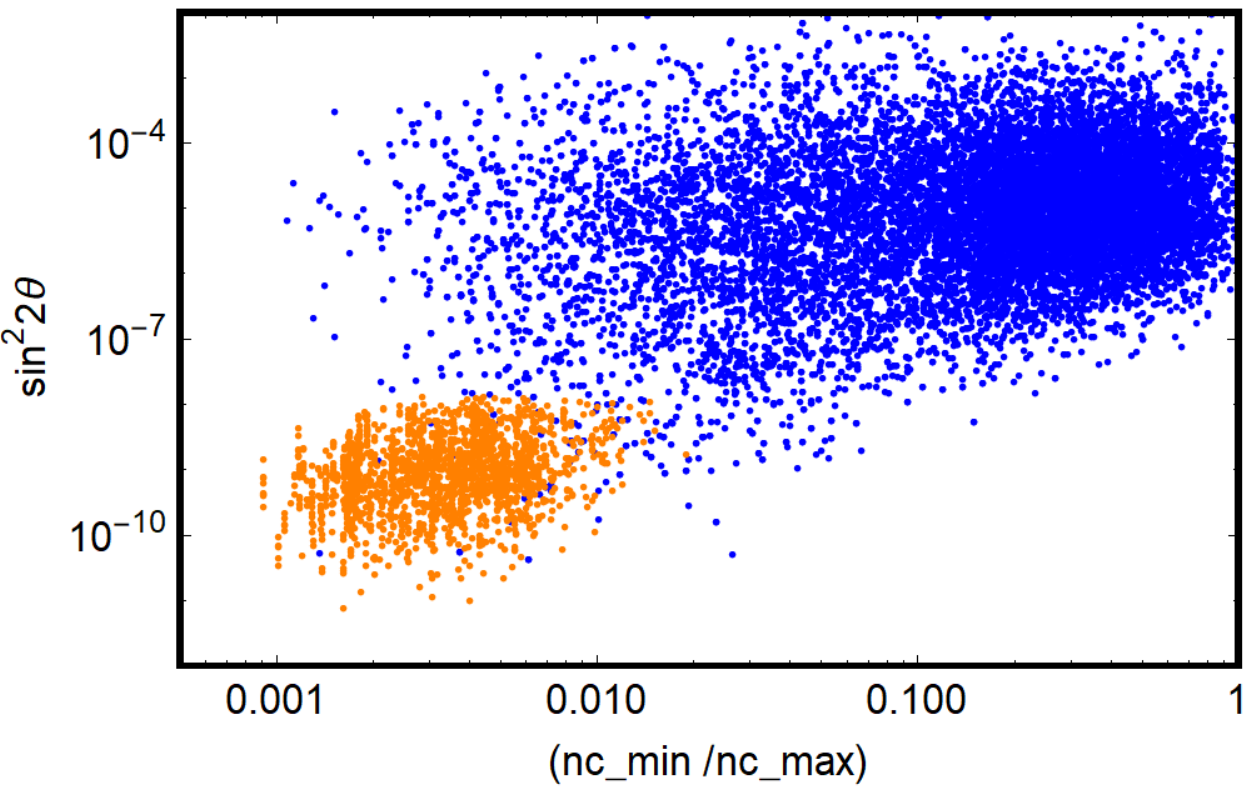}};
\node[rotate = 90, fill = white] at (-3.7,0.3) {\footnotesize $\sin^2(2 \theta_{DM})$};
\node[fill = white] at (0.5,-2.4) {\footnotesize $\text{min}_i(n_{\alpha i})/\text{max}_i(n_{\alpha i})$};
\end{tikzpicture}
}\hfill
\subfigure{%
\begin{tikzpicture}
\node at (0,0) {\includegraphics[width=7.8 cm]{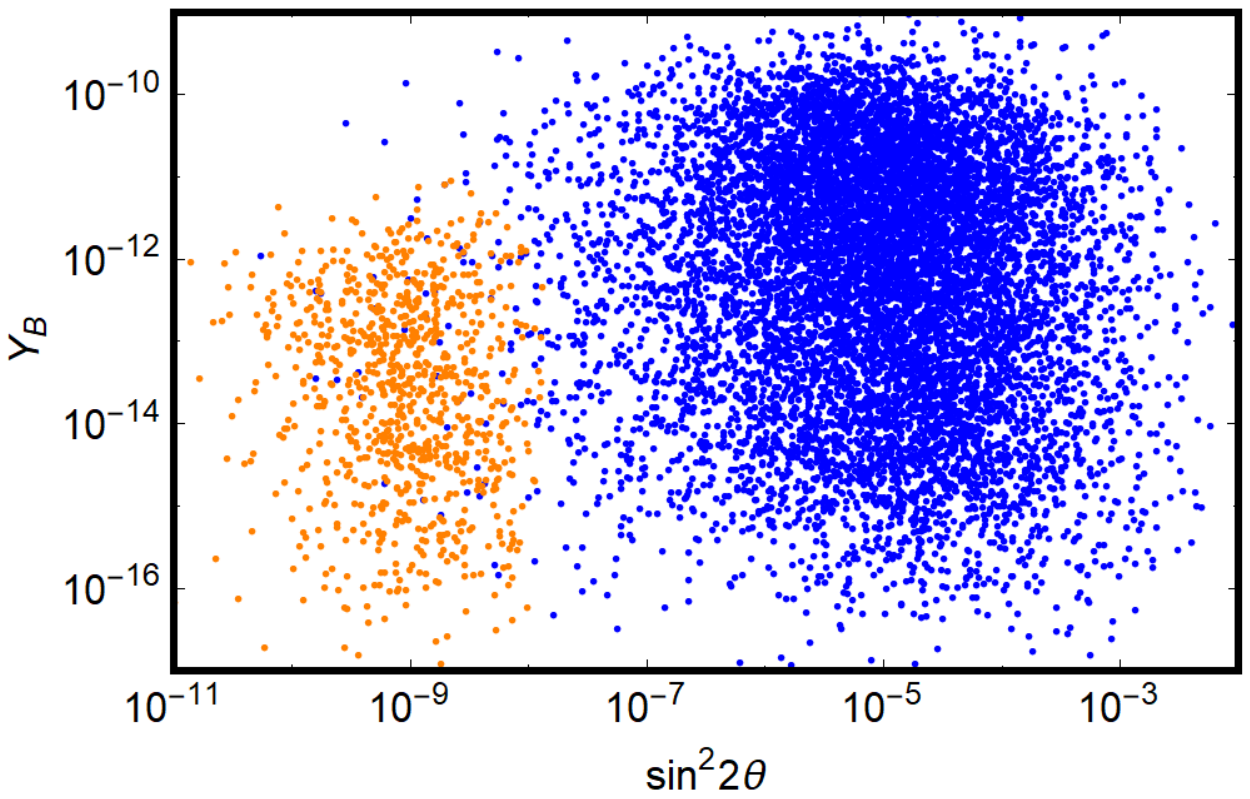}};
\node[fill = white] at (0.5,-2.4) {\footnotesize $\sin^2(2 \theta_{DM})$};
\node[rotate = 90, fill = white] at (-3.7,0.3) {\footnotesize $Y_B$};
\end{tikzpicture}
}
\caption{Mixing angle $\theta_\text{DM}$ between the DM candidate and the active neutrinos. The blue points comply with all the constraints mentioned in Sec.~\ref{Sec:Constraints} and provide a DM candidate in the suitable mass range, the yellow points in addition satisfy the DM related constraints of Ref.~\cite{Abada:2014zra}.
\textbf{Left panel}: Mixing angle versus hierarchy of the submatrix $n$ in Eq.~\eqref{general-iss}. \textbf{Right panel}: Generated baryon asymmetry versus mixing angle.}
\label{fig:sin_theta_DM}
\end{figure}
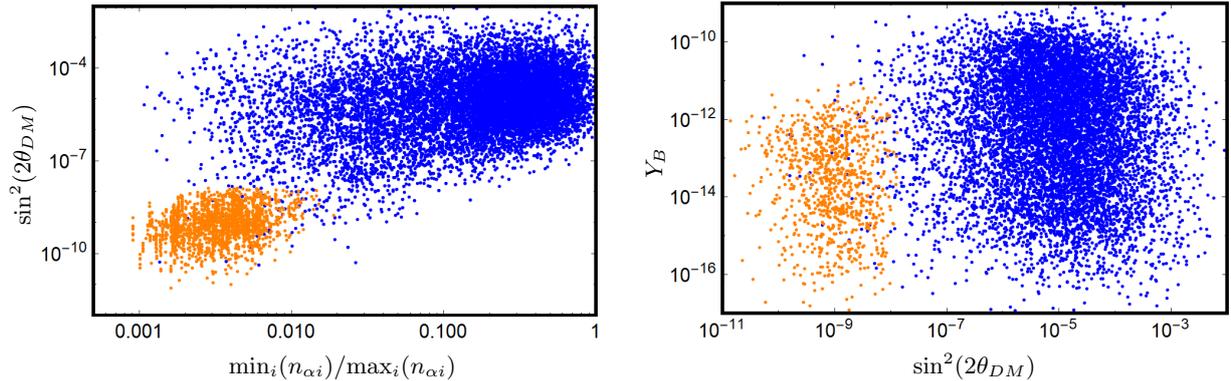

The generic value of the active-sterile mixing is given by $\sin^2 2\theta_\text{DM} = {\cal O}(Y^2 v^2/\Lambda^2)$, leading to an overproduction of DM in a wide range of the parameter space. 
This mixing angle is however suppressed if the entries of the submatrix $n$ in Eq.~\eqref{general-iss} feature a significant hierarchy, see appendix \ref{app:DM_mixing}. This is illustrated in the left panel of Fig.~\ref{fig:sin_theta_DM}. Here the yellow points avoid the overproduction of dark matter, typically requiring a hierarchy within the $n$ submatrix entries of about two orders of magnitude. While not a generic feature of the ISS, this part of the parameter space can be motivated by anthropological arguments to avoid the overclosure of the Universe.

The right panel of Fig.~\ref{fig:sin_theta_DM} depicts the distribution of the generated baryon asymmetry in terms of the mixing angle $\theta_\text{DM}$. We note that the small mixing angles required for reproducing the correct abundance of DM tend to generate a too small baryon asymmetry. This may be traced back to the tension between the preferred ranges for the mass splitting $\Delta m/m_\text{PD}$ and the Yukawa couplings (see Fig.~\ref{fig_ISS22_asymmetry}) and the relation~\eqref{eq:nogoISS}. For small mixing angles with a strongly hierarchical structure of the submatrix $n$ in Eq.~\eqref{general-iss}, the eight-neutrino ISS(2,3) model effectively reduces to a toy model with only one RH and two sterile neutrinos, in particular there can be no cancellations in the matrix equations related to the sterile sector. In this case, Eq.~\eqref{eq:nogoISS} becomes an exact relation, implying that it is difficult to simultaneously obtain a suitable mass splitting, Yukawa coupling and heavy neutrino mass scale $m_\text{PD}$. Both this analytical argument, as well as the numerical scan resulting in Fig.~\ref{fig:sin_theta_DM}, suggest that while there may be a tuned region in parameter space which can generate both the correct DM abundance and baryon asymmetry, 
\textit{generically} the ISS(2,3) cannot account for the baryon asymmetry of the Universe and its DM content simultaneously. We emphasize that the ISS(2,3) generically overproduces DM when successful leptogenesis is imposed, hence without any additions to its cosmological history, the ISS(2,3) with a DM candidate in the keV range  cannot be considered a successful setup for leptogenesis through neutrino oscillations.

Related work on the simultaneous explanation of the baryon asymmetry and DM of the Universe has been performed in the context of the $\nu$MSM~\cite{Asaka:2005an,Asaka:2005pn}, see e.g.~\cite{Canetti:2012kh} for a recent analysis. After producing the observed baryon asymmetry through neutrino oscillations, a second phase of leptogenesis is triggered at temperatures well below the EW phase transition. This generated lepton asymmetry is not transferred into the baryon sector,  but instead strongly enhances the production of DM sterile neutrinos in the keV range. This production mechanism is dubbed resonant production or Shi-Fuller mechanism~\cite{Shi:1998km,Shaposhnikov:2008pf,Laine:2008pg}. It allows for an efficient DM production for small enough mixing angles with the active neutrinos to comply with experimental limits. The Shi-Fuller mechanism requires a very efficient late time production of a lepton asymmetry and hence an extreme degeneracy for heavy RH neutrinos, corresponding in our notation to $\Delta m/m_\text{PD} \sim 10^{-14}$~\cite{Canetti:2012kh}. This mechanism is however not be at work in our framework, since these extremely small mass splittings
cannot be generated within the ISS.
Note that, given the systematically too large mixing-angles in the parameter region favoured by leptogenesis, a Shi-Fuller production, if active, would further worsen the already severe issue of DM overproduction. For analogous reasons the freeze-in production mechanism suggested in~\cite{Asaka:2005cn,Petraki:2007gq,Hall:2009bx,Abada:2014zra}, sourced by the decay of heavy sterile states, is not a viable option in our setup.

A possible solution to the DM problem could be a late time entropy injection~\cite{Asaka:2006ek}\footnote{Notice that DM is produced through the DW mechanism at temperatures of the order of 100 MeV. Entropy injection should occur at lower temperatures and, consequently, much later than leptogenesis.} diluting the DM abundance. This solution is however somewhat contrived since entropy injection would have analogous effect also on the baryon asymmetry. As shown in the right panel of Fig.~\ref{fig:sin_theta_DM}, the baryon asymmetry of the upper most points exceed the observed value by about two orders of magnitude, however only few blue points can be brought into the cosmologically viable region by reducing $\sin^2(2 \theta_\text{DM})$ by two orders of magnitude. Alternatively, one could consider the case in which the DM is driven to thermal equilibrium, for example by additional gauge interactions~\cite{Bezrukov:2009th,Nemevsek:2012cd}. Thermal keV Dark Matter would also be overabundant; however the amount of entropy injection needed to set the current abundance is more moderate and still potentially compatible with the correct amount of baryon asymmetry (notice however that the extra interactions thermalising the DM could also affect the leptogenesis process). A further option might be to suppress the DM - active neutrino oscillations in the early Universe by introducing a temperature-dependent neutrino mass term~\cite{Bezrukov:2017ike}. A full analysis of these possibilities is beyond the scope of this paper.

\section{Conclusion \label{sec:conclusions}}

A central piece of this work is a new linearised formulation of the set of Boltzmann equations describing the generation of the baryon asymmetry of the Universe from CP-violating oscillations of nearly mass degenerate neutrino pairs with a GeV mass scale. The small mass splitting at the origin of the leptogenesis mechanism naturally emerges in extensions of the SM involving extra sterile/right-handed fermions, based on a small violation of lepton number. 
The refined system of Boltzmann equation allows to study leptogenesis beyond the weak washout regime, extending and completing the results presented in~\cite{Abada:2015rta}. 

\noindent
Our study was conducted  in the framework of i) a minimal extension of the SM by two SM singlet fermions, the LSS-ISS, providing a natural explanation of their strong mass degeneracy based on two LNV parameters; ii) the Inverse Seesaw in its most minimal realisation, the ISS (2,2), which features two pseudo-Dirac neutrino pairs beyond the SM states; iii) the ISS (2,3), which leads to a similar spectrum with an additional  sterile state with mass around the keV, a possible DM candidate.

\noindent
We present the parametrisation and derivation of the new linearised kinetic equations based on Fermi-Dirac statistical distributions, including the impact of soft scatterings of gauge bosons in the thermal plasma and the presence of 
 small leptonic chemical potentials. We also take into account the re-distribution of the asymmetry in the active sector through spectator processes. 
This new treatment enables a strong simplification of the system of differential equations, empowering a fast numerical solution, allowing in particular a full coverage of the parameter space in both the strong and weak washout regimes. 

\noindent
In the case of the LSS-ISS model, we find that the parameter space relevant for viable leptogenesis in the strong washout regime shows a preference for a relative mass splitting between the heavy neutrinos of about $\Delta m/m_\text{PD} \sim 10^{-4} - 10^{-3}$ and for Yukawa couplings $ \lesssim 10^{-5}$. Contrary to the case of the weak washout regime which was the focus of~\cite{Abada:2015rta}, the viable model points in the strong washout regime lie within the expected sensitivity of planned future facilities like SHiP, FCC-ee and LBNF/DUNE. These experiments hence have the potential to discriminate between weak and strong washout regimes within this model. Our findings are in agreement with the recent bayesian parameter study of Ref.~\cite{Hernandez:2016kel} and the bounds on the active-sterile mixing derived e.g.\ in \cite{Drewes:2016jae} and demonstrate that the regime of small LNV studied here constitutes a significant part of the phenomenologically interesting parameter space.

\noindent
In the case of the ISS, we focus on the possibility that the lighter pseudo-Dirac pair generates the lepton asymmetry, whereas the mass scale of the second pseudo-Dirac pair is taken to be much heavier, so that it effectively decouples during 
leptogenesis. In this setup we find similar results as in the LSS-ISS model, 
however the range of viable masses for the neutrino pair responsible of the generation of the lepton asymmetry is sensitively reduced. This is due to a tighter relations between the masses of the new neutrinos and their Yukawa couplings in the ISS framework. Larger masses correspond to larger Yukawa couplings, implying too strong washout effects.

\noindent
In the final case of the ISS(2,3) model, viable leptogenesis is achieved in analogous regions of the parameter space as in the ISS(2,2). In addition, this model features the intriguing possibility of addressing at the same time the DM puzzle. This possibility appears however disfavoured in this minimal realisation of the ISS since the DM candidate is generically overproduced, implying an overclosure of the Universe unless the standard cosmological history is altered. Given the high dimensionality of the parameter space, we can however not exclude the existence of fine-tuned parameter combinations which might nevertheless achieve this task.

\vspace{1cm}
\subsubsection*{Acknowledgements}
We thank M.~Drewes, J.~Lopez-Pavon, B.~Mares and D.~Teresi for helpful discussions. {We are also thankful to P.~Hernandez and J.~Lopez-Pavon for their valuable comments on the manuscript.}
A.A. acknowledges partial support from the European Union Horizon 2020
research and innovation programme under the Marie Sk{\l}odowska-Curie: RISE
InvisiblesPlus (grant agreement n$^{\circ}$ 690575)  and 
the ITN Elusives (grant agreement n$^{\circ}$ 674896). V.D.\ acknowledges the financial support of the UnivEarthS Labex program at Sorbonne Paris Cit\'e (ANR-10-LABX-0023 and ANR-11-IDEX-0005-02), the Paris Centre for Cosmological Physics and the L'Or\'eal - Unesco program `For Women in Science'. 
M.L. acknowledges support by the Fonds de la Recherche Scientifique - FNRS under Grant n$^{\circ}$ IISN 4.4512.10.

\newpage
\appendix

\section{Some useful formulas for simplifying the kinetic equations \label{app:leptogenesis_details}}

This Appendix collects technical details and useful formulas supporting the derivations in Section~\ref{Sec:mechanism}.

\subsection{Decay rates \label{app:A1}}

The generic expression for the production process $a (E_1) + b(E_2) \rightarrow c(E_3) + N (p_N)$ involving one vertex associated with the Yukawa coupling $F$ is given by:
\begin{equation}
\label{eq:Gammadun}
\Gamma^d_N=\frac{1}{2 k_N} F^{\dagger}\left( \int \left(\prod_{f=1}^3 \frac{d^3 p_f}{{\left(2 \pi\right)}^3}\right) (2 \pi)^4 \delta(p_1+p_2-p_3-p_N) |M|^2 f_a(E_1) f_b (E_2) (1 \pm f_c(E_3))\right) F\ ,
\end{equation}
where $f_{f=a,b,c}$ are Fermi-Dirac ($f_F$) or Bose-Einstein ($f_B$) distributions. The destruction rate is given by:
\begin{equation}
\label{eq:Gammapun}
\Gamma^p_N=\frac{1}{2 k_N} F^{\dagger}\left( \int \left(\prod_{f}^3 \frac{d^3 p_f}{{\left(2 \pi\right)}^3}\right) (2 \pi)^4 \delta(p_1+p_2-p_3-p_N) |M|^2 (1 \pm f_a(E_1)) (1 \pm f_b (E_2))  f_c(E_3)\right) F\ .
\end{equation}
By using the following properties
\begin{eqnarray}
 f_B(E_1) f_B(E_2) &=& f_B(E_1+E_2) (1+f_B(E_1)+f_B(E_2)) \,,\nonumber\\
 f_B(-E_1) &=& -(1+f_B(E_1)) \,,\nonumber\\
 f_F(E_1) &=& -f_B(E_1+i \pi T) \,,
\end{eqnarray}
the products of Fermi-Dirac and Bose-Einstein distributions in Eqs.~(\ref{eq:Gammadun}-\ref{eq:Gammapun}) can be rewritten:
\begin{eqnarray}
\label{eq:trick}
 f_a(E_1) f_b (E_2) (1 \pm f_c(E_3))&=&f_F(x_N, \pm \mu) \hat f \tilde{f}\ , \noindent\\
 (1 \pm f_a(E_1)) (1 \pm f_b (E_2)) f_c(E_3)&=&(1-f_F(x_N, \pm \mu)) \hat f \tilde{f}\ .
\end{eqnarray}
 Notice that $\hat f$ and $\tilde f$ are also functions of the chemical potential. The functions $\Sigma$ and $\Psi$ defined in the main text represent the coefficients of the expansion of Eq.~(\ref{eq:trick}) with respect to the chemical potential $\mu$. By using the expression of the amplitudes given in~\cite{Anisimov:2010gy} and expanding Eq.~(\ref{eq:trick}), we obtain:
\begin{align}
& \Sigma(x_N)= \left(6 h_t^2 F_s^{q\,(0)} (x_N)+(3 g^2+g^{'\,2}) \left(F_s^{V\,(0)}(x_N)+F_{t_1}^{V\,(0)}(x_N)+F_{t_2}^{V\,(0)}(x_N)\right)\right)\ , \nonumber\\
& \Psi(x_N)=(3 g^2+g^{'\,2}) \left[(\delta F_{t_1,a}^V-\delta F_{t_1,b}^V)+(\delta F_{t_2,a}^V-\delta F_{t_2,b}^V)-(\delta F_{s,a}^V-\delta F_{s,b}^V)\right]-6 h_t^2 \delta F^q_s (x_N)\ ,
\end{align}

where:
\begin{eqnarray}
 F_s^{q\,(0)}(x_N)&=&\int_{x_N}^{\infty} d \chi^+ \int_0^{x_N} d \chi^- \left(f_B^0(\chi^0)+f^0_F(\chi^0-x_N)\right)\nonumber\\
&& \left(\chi+2 \left(\log(1-\exp(-\chi^+))-\log(1+\exp(\chi^-))\right)\right)\ ,
\end{eqnarray}
\begin{eqnarray}
 F_s^{V\,(0)}(x_N) &=& \int_{x_N}^\infty d \chi^+ \int_0^{x_N} d \chi^- \left(f_B^0(\chi^0-x_N)+f_F^0(\chi^0)\right)\nonumber\\
&& \left \{\frac{\chi}{2}-\frac{1}{\chi}\left[ (x_N-\chi^+) \left(\log(1-\exp(-\chi^+))-\log(1+\exp(-\chi^-))\right)\right. \right.\nonumber\\
&& \left. \left. +(x_N-\chi^-) \left(\log(1-\exp(-\chi^-))-\log(1+\exp(-\chi^+))\right)\right]\right.\nonumber\\
&& \left. -\frac{1}{\chi^2} (\chi^0-2 x_N) \left[Li_2 (\exp(-\chi^+))-Li_2 (\exp(-\chi^-))- \right.\right. \nonumber \\
&&\left.\left. Li_2 (-\exp(-\chi^+))+Li_2 (-\exp(-\chi^+))\right] \textcolor{white}{\frac{1}{1}} \hspace*{-3mm} \right \}\ ,\\
 F_{t_1}^{V\,(0)}(x_N)&=&\int_0^{x_N} d \chi^+ \int_{-\infty}^0 d\chi^- \left(1+f_B^0(x_N-\chi^0)-f_F^0(\chi^0)\right)\nonumber\\
&& \left \{\frac{1}{\chi}(x_N-\chi^-) \left[\log(1+\exp(-\chi^+))-\log(1-\exp(-\chi^-))\right]\right. \nonumber\\
&& \left. \frac{1}{\chi^2}(2 x_N-\chi^0) \left[- Li_2(\exp(-\chi^+))-Li_2(\exp(-\chi^-))\right]\right \}\ ,\\
 F_{t_2}^{V\,(0)}(x_N)&=&\int_0^{x_N} d \chi^+ \int_{-\infty}^0 d\chi^- \left(1+f_B^0(\chi^0)-f_F^0(x_N-\chi^0)\right)\nonumber\\
&& \left \{\frac{1}{\chi}(x_N-\chi^+) \left[-\log(1+\exp(-\chi^+))+\log(1-\exp(\chi^-))\right]\right. \nonumber\\
&& \left. \frac{1}{\chi^2}(2 x_N-\chi^0) \left[ Li_2(\exp(-\chi^+))-Li_2(-\exp(-\chi^-))\right]\right \}\ ,
\end{eqnarray}
where $\chi^0=\chi^+ +\chi^-$ and $\chi=\chi^+ - \chi^-$. Furthermore,
\begin{eqnarray}
\delta F^V_{s,a}&=&\int_{x_N}^{\infty} d \chi^+ \int_0^{x_N} d\chi^- f_F^{'}(\chi^0)\left \{\frac{\chi}{2}-\frac{1}{\chi}\left[(x_N-\chi^+)\left(\log(1-\exp(-\chi^+))-\log(1+\exp(-\chi^-))\right)\right. \right. \nonumber\\
&& \left. \left.+(x_N-\chi^-) \left(\log(1-\exp(-\chi^-))-\log(1+\exp(-\chi^+))\right)\right]\right.\nonumber\\
&& \left. -\frac{1}{\chi^2} (\chi^0-2 x_N) \left[Li_2 (\exp(-\chi^+))-Li_2 (\exp(-\chi^-))\right. \right. \nonumber \\
&& \left.\left.-Li_2 (-\exp(-\chi^+))+Li_2 (-\exp(-\chi^+))\right] \textcolor{white}{\frac{1}{1}} \hspace*{-3mm}  \right \}\ ,\\
\delta F^V_{s,b}&=&\int_{x_N}^{\infty} d \chi^+ \int_0^{x_N} d\chi^- \left(f_F^0(\chi^0)+f_B^0(\chi^0-x_N)\right)\nonumber\\
&& \left \{ -\frac{1}{\chi} f_B^0(\chi^+) f_B^0(\chi^-) \left[(x_N-\chi^+)\exp(\chi^-)+(x_N-\chi^-)\exp(\chi^+)+\exp(\chi)(\chi^0-2 x_N)\right]\right. \nonumber\\
&& \left. -\frac{1}{\chi^2} (\chi^0-2 x_N) \log \left[\frac{-1+\exp(\chi^-)}{-1+\exp(\chi^+)}\right]\right \}\ , \\
 \delta F^V_{t_1,a}&=&\int_0^{x_N} d \chi^+ \int_{-\infty}^0 d\chi^- f^{'}_F(\chi^0)\nonumber\\
&& \left \{\frac{1}{\chi}(x_N-\chi^-) \left[\log(1+\exp(-\chi^+))-\log(1-\exp(-\chi^-))\right]\right. \nonumber\\
&& \left. \frac{1}{\chi^2}(2 x_N-\chi^0) \left[- Li_2(\exp(-\chi^+))-Li_2(\exp(-\chi^-))\right]\right \}\ ,
\end{eqnarray}

\begin{eqnarray}
\delta F^V_{t_1,b}&=&\int_0^{x_N} d \chi^+ \int_{-\infty}^0 d\chi^- \left(1+f_B^0(x_N-\chi^0)-f_F^0(\chi^0)\right)\nonumber\\
&& \left \{\frac{1}{\chi} \left[f^0_F(\chi^-)(\chi^- - x_N)-\frac{\chi^+}{\chi}(\chi^0-2 x_N)\right]\right.\nonumber\\
&& \left. \frac{1}{\chi^2} (\chi^0-2 x_N) \log(1+\exp(\chi^+))\right \}\ ,\\
 \delta F^V_{t_2,a}&=&\int_0^{x_N} d \chi^+ \int_{-\infty}^0 d\chi^- f^{'}_B(\chi^0)\nonumber\\
&& \left \{\frac{1}{\chi}(x_N-\chi^+) \left[-\log(1+\exp(-\chi^+))+\log(1-\exp(\chi^-))\right]\right. \nonumber\\
&& \left. \frac{1}{\chi^2}(2 x_N-\chi^0) \left[ Li_2(\exp(-\chi^+))-Li_2(-\exp(-\chi^-))\right]\right \}\ ,\\
 \delta F^V_{t_2,b}&=& \int_0^{x_N} d \chi^+ \int_{-\infty}^0 d\chi^- \left(1+f_B^0(\chi^0)-f_F^0(x_N-\chi^0)\right)\nonumber\\
&& \left \{\frac{1}{\chi}f_F^0(\chi^-)\exp(-\chi^-)(\chi^- -x_N)+\frac{1}{\chi^2}(\chi^0-2 x_N)\log(1+\exp(-\chi^-))\right \}\ .
\end{eqnarray}

To rephrase the differential equations in terms of thermally averaged decay rates and to compare with earlier works based on the Maxwell-Boltzmann distribution $f_\text{B}$, some useful relations are
\begin{align}
\int \frac{d^3 k}{(2 \pi)^3} f^0_F(k/T)& =  \frac{3 T^3 \zeta(3)}{4 \pi^2} \,, \quad
&&\int \frac{d^3 k}{(2 \pi)^3} f_F'(k/T) = - \frac{T^3}{12} \,, \quad \int \frac{d^3 k}{(2 \pi)^3} \frac{f^0_F(k/T)}{k} =  \frac{T^2}{24} \,, \nonumber \\
\int \frac{d^3 k}{(2 \pi)^3} f^0_\text{B}(k/T)& =  \frac{ T^3 }{ \pi^2} \,,\quad  \quad &&\int \frac{d^3 k}{(2 \pi)^3} f_\text{B}'(k/T) = - \frac{T^3}{\pi^2}  \,.
\end{align}

\subsection{Weak washout limit \label{app:weakwo}}

In this subsection we will briefly present the derivation of the expression in Eq.~(\ref{eq:baryo_analytical}) for the baryon asymmetry in the weak washout regime. The procedure substantially coincides with the one already discussed in~\cite{Abada:2015rta}. Differently to this reference we will adopt the Fermi-Dirac distributions for neutrinos and active leptons and include in the interaction rates the processes relying on gauge interactions.

The weak washout limit solution is obtained through a perturbative expansion of the system of Eqs.~(\ref{eq:Rnstart}-\ref{eq:mustart}). As a first step, the equation for the neutrino density is solved at the lowest order, i.e. neglecting the chemical potential and approximating $R_N-I \approx I$:
\begin{equation}
\frac{d R_N}{dt}=-i \left[\langle H \rangle, R_N\right]+\langle \Gamma^0 \rangle F^\dagger F\ .
\end{equation}
Eliminating the oscillation term through the transformation $R_N=E(t) \tilde{R}_N E^{\dagger}(t)$, the equation is straightforwardly solved for:
\begin{equation}
\tilde{R}_N=\int_0^t dt_1 \langle \Gamma^0 \rangle (t_1) E(t_1) F^{\dagger}F E(t_1)\ .
\end{equation}  

\noindent
This solution is substituted in the leading order equation for the chemical potentials which read, taking again the lowest order contributions:
\begin{equation}
 \frac{d \mu_{\Delta_\alpha}}{dt}= -\frac{9 \zeta (3)}{2 N_D\, \pi^2} \langle \Gamma^0 \rangle \left(F R_N F^{\dagger}-F^{*} R_{\bar N} F^T\right)\ .
\end{equation}
This can be directly integrated:
\begin{eqnarray}
 \mu_{\Delta_\alpha}&=&-\frac{9 \zeta(3)}{2 N_D \pi^2}\int_0^t dt_1 \langle \Gamma^0 \rangle (t_1) \int_0^{t_1} d t_2 \langle \Gamma^0 \rangle (t_2) \left[F E(t_1) E(t_2)^\dagger F^\dagger F E(t_2) E(t_1)^\dagger F^\dagger \right.\nonumber\\ 
&& \left.- F^{*} E(t_1) E(t_2)^\dagger F^T F^{*} E(t_2) E(t_1)^\dagger F^T\right]_{\alpha \alpha}\nonumber\\
& =& -\frac{9 \zeta(3)}{2 N_D \pi^2} \delta_\alpha \int_0^t dt_1 \langle \Gamma^0 \rangle \int_0^{t_1} \langle \Gamma^0 \rangle (t_2) \sin\left( \int_{t_2}^{t_1} dt_3 E_2 (t_3)-E_3 (t_3)\right)\ ,
\end{eqnarray}
where
\begin{equation}
\int_{t_2}^{t_3} dt_3 E_2 (t_3)-E_3 (t_3)=z(T_1)-z(T_2)\ ,
\end{equation}
with
\begin{equation}
z(T)=\frac{T_L^3}{T^3},\,\,\,\,\,T_L={\left(\frac{1}{12}\frac{\pi^2}{9 \zeta(3)}M_0 \Delta M^2\right)}^{1/3}\ .
\end{equation}
Using this last result, we find
\begin{equation}
\mu_{\Delta \alpha}=-\frac{9}{2} \frac{\zeta(3)}{N_D \pi^2}\delta_\alpha {\left(\frac{M_0}{T_L}\right)}^2 {\left(\frac{\pi}{1152\zeta(3)}\right)}^2 \tilde{J}_{32}\left(\frac{T_L}{T}\right)\ ,
\end{equation}
where
\begin{eqnarray}
\tilde{J}_{32}(x)&=&\int_0^x dx_1 c_0(x_1) \int_0^{x_1} dx_2 c_0 (x_2) \sin\left(x_1^3-x_2^3\right)\ , \nonumber\\
 c_0(x)&=&c_Q^{(0)} h_t^2 +c_{\rm LPM}^{(0)}+\left(3 g^2 \left(\pi \frac{T_L}{x}\right)+g^{'\,2}\left(\pi \frac{T_L}{x}\right)\right) \times  \nonumber \\
 && \qquad \left(c_V^{(0)}+\log\left(\frac{1}{3 g^2 \left(\pi \frac{T_L}{x}\right)+g^{'\,2}\left(\pi \frac{T_L}{x}\right)}\right)\right)\ .\nonumber \\
\end{eqnarray}

\noindent
The last step is the solution for the asymmetry $\Delta R$ in the sterile neutrinos:
\begin{equation}
\frac{d\Delta R_{II}}{dt}=2 \langle \Gamma^{(1)}\rangle \left(F^{\dagger} A_{\alpha \beta} \mu_{\Delta \beta} F\right)_{II}\ ,
\end{equation}
which again can be directly integrated,
\begin{equation}
\left(\Delta R\right)_{II}=-9 \frac{\zeta(3)}{N_D \pi^2} {\left(\frac{\pi}{2304 \zeta(3)}\right)}^3 {\left(\frac{M_0}{T_L}\right)}^3 \left(F^{\dagger}A_{\alpha \beta}\delta_\beta F\right)_{II} \int_0^x dx_1 c_1(x_1)  \tilde{J}_{32}(x) \ ,
\end{equation} 
with
\begin{eqnarray}
c_1(x)&=& c_Q^{(1)} h_t^2 +c_{\rm LPM}^{(1)}+\left(3 g^2 \left(\pi \frac{T_L}{x}\right)+g^{'\,2}\left(\pi \frac{T_L}{x}\right)\right) \nonumber \\
&& \qquad \times \left(c_V^{(1)}+\log\left(\frac{1}{3 g^2 \left(\pi \frac{T_L}{x}\right)+g^{'\,2}\left(\pi \frac{T_L}{x}\right)}\right)\right)\ .
\end{eqnarray}
Assuming negligible variation with the temperature of the functions $c_0$ and $c_1$, and defining
\begin{equation}
\frac{{\left(c_0^2 \, c_1\right)}^{1/3}}{2304}=\frac{\sin \phi}{8}\ ,
\end{equation}

\noindent
the result simplifies to:
\begin{equation}
\left(\Delta R\right)_{II}(T)=-\frac{9 \pi^{7/6}}{512 \, \zeta(3)^{4/3} \,  \Gamma(5/6)} \frac{M_0}{T} \frac{M_0^{4/3}}{\Delta m^{4/3}} \sin^3\phi \left(F^{\dagger}A_{\alpha \beta}\delta_\beta F\right)_{II}\ ,
\end{equation}
so that the baryon abundance is given by:
\begin{equation}
Y_B=\frac{n_B}{s}=\frac{28}{79}Y_{N_0}\sum_I \left(\Delta R\right)_{II}(T_{\rm EW})\ ,
\end{equation}
with $Y_{N_0}=0.022$~\cite{Abada:2015rta}.

\subsection{Diagonalization of the equation for the sterile sector \label{app:A3}}

In this subsection, we derive the expression for the  unitary matrix $V_\alpha$, which describes a basis in which the equation~\eqref{eq:Rnstart0}  for the sterile sector is greatly simplified. We first perform a change of basis to absorb the oscillations induced by the vacuum Hamiltonian $H_N^0$:
\begin{equation}
R_N^{(0)} \mapsto \tilde R_N^{(0)} = E^\dagger(x) R_N^{(0)} E(x) \,,
\end{equation}
with
\begin{equation}
E(t) = \exp\left(- i \int_{t_i}^t \langle H_N^0 \rangle (t') dt'\right)\,,
\end{equation}
where the vacuum Hamiltonian of the sterile neutrinos  is given by $(H_N^0)_{ij} = \sqrt{k_N^2 + M_i^2} \delta_{ij}$.
This removes the vacuum commutator containing the vacuum Hamiltonian from Eq.~\eqref{eq:Rnstart}:
\begin{align}
\frac{d R_N}{dt} & = \frac{d}{dt} \left(E \tilde R_N E^\dagger\right) \\
&= - i \langle H_N^0 \rangle E \tilde R_N E^\dagger + E \left(\frac{d}{dt} \tilde R_N\right) E^\dagger + i E \tilde R_N E^\dagger \langle H_N^0 \rangle^\dagger \\
&=  E \left(\frac{d}{dt} \tilde R_N\right) E^\dagger  - i \left[ \langle H_N^0 \rangle, R_N \right] \,,
\end{align}
where we have exploited $[\langle H_N^0 \rangle, E] = 0$. In the ultra-relativistic limit, $H_N^0$ is given by
\begin{equation}
H_N^0 \rightarrow \frac{1}{2 k_N} \text{diag}(0 , \Delta M^2) \,,
\end{equation}
where we have omitted a contribution proportional to the unity matrix as this drops out in the commutator. After performing the thermal average,
\begin{equation}
\langle H_0 \rangle = \frac{x \pi^2}{36 \zeta(3) T_\text{EW} } \text{diag}(0 , \Delta M^2) \,,
\end{equation}
we find 
\begin{equation}
E(x) = \text{diag}\left(1, \exp(- i r^3 x^3)\right) \,, \quad r = T_L/T_\text{EW} \,, 
\end{equation}
with
\begin{equation}
T_L^3 = \frac{\pi^2}{108 \zeta(3)}M_0 \Delta M^2 \,.
\end{equation}
All remaining operators on the right-hand side of Eq.~\eqref{eq:Rnstart0} are now of the structure $E^\dagger(x) F^\dagger F E(x)$. We can thus perform a second change of basis by the unitary matrix $V(x)$ which diagonalises all these remaining operators. After removing the remaining ambiguity in the choice of $V(x)$ by requiring the second row to be real and positive, $V(x)$ can be calculated explicitly. It is of the form
\begin{equation}
V(x , \alpha) = \begin{pmatrix} e^{i(\alpha - x^3 r^3)} f_{11} & e^{i(\alpha - x^3 r^3)} f_{12} \\ f_{21} & f_{22} \end{pmatrix}\ ,
\end{equation}
where $f_{ij}$ are time-independent combinations of the absolute values of the matrix elements of $F^\dagger F$ and $\alpha$ denoting the phase of $(F^\dagger F)_{12}$. We see that in the total basis transformation by the matrix $E \cdot V$, the time (or equivalently temperature) dependence reduces to a global phase and hence cancels out in the unitary matrix transformation. We may thus replace $E(x) V(x,\alpha) \mapsto V_\alpha = V(x = 0, \alpha)$. 

Finally, exploiting 
\begin{equation}
V^\dagger \frac{d f}{dx} V = \frac{d}{dx} \left(V^\dagger f V\right) + \left[V^\dagger \frac{d V}{dx}, V^\dagger f V\right]\ ,
\end{equation}
which holds for any function $f(x)$ and unitary matrix $V(x)$, we arrive at Eq.~\eqref{eq:finalR0} quoted in the main text. As mentioned in the main text, this introduces the matrix $D$, which is defined by 
\begin{equation}
 V^\dagger \dot V = x^2 D \,.
\end{equation}

\section{The parameter space for DM in the ISS(2,3) \label{app:DM_mixing}}

In Section~\ref{sec:BAU-DM_ISS23}, we observed that a small mixing angle between the active sector and the DM candidate (required to avoid overproducing DM in the DW mechanism), can be achieved by allowing for a sizeable hierarchy within the submatrix $n$ of Eq.~(\ref{general-iss}). In this Appendix we explain this result analytically by considering a minimal toy model with one active flavour,
one right-handed neutrino and two sterile fermions:
\begin{equation}
{\mathcal{M}} = \begin{pmatrix}
0 & \frac{1}{2} Y v & 0 & 0 \\
\frac{1}{2} Y v & 0 & n_1 \Lambda & n_2 \Lambda\\
0 & n_1 \Lambda & \xi_1 \Lambda & 0 \\
0 & n_2 \Lambda & 0 & \xi_2 \Lambda
\end{pmatrix} \,.
\end{equation} 
For simplicity we will take all parameters to be real in the following.
To leading order in $Y$ and $\xi_{1,2}$, this mass matrix is diagonalised as
\begin{equation}
{\cal U}^T {\mathcal{M}} {\cal U} = \text{diag}(0, m_\text{DM} , m_\text{PD} - m_\text{DM},  m_\text{PD} + m_\text{DM})\ ,
\label{eq:egv}
\end{equation} 
with $m_{PD} = \sqrt{n_1^2 + n_2^2} \,\Lambda$ and $ m_{DM} = \frac{n_1^2 \xi_2 + n_2^2 \xi_1}{n_1^2 + n_2^2} \Lambda $. In this basis, the DM-active mixing is determined by the entry ${\cal U}_{12}$, i.e.\ by the first component of the (correctly normalised) eigenvector corresponding to the second eigenvalue in Eq.~\eqref{eq:egv}:
\begin{equation}
\sin^2(2 \theta_\text{DM}) = 4 {\cal U}_{12}^2 \simeq \frac{2 n_1^2 n_2^2 (\xi_1 - \xi_2)^2 }{(n_1^2 + n_2^2) (n_1^2 \xi^2 + n_2^2 \xi_1)^2 } \, \frac{v^2 Y^2}{\Lambda^2}\,.
\label{eq:mixing_analytical}
\end{equation}
If $n_{1,2}$ are order one parameters, this yields $\sin^2(2 \theta_\text{DM}) = {\cal O}(v^2 Y^2/\Lambda^2) = {\cal O}(10^{-10} - 10^{-4})$ for $Y = \mathcal{O}(10^{-7} - 10^{-4})$. If on the other hand $n_1 \gg n_2$ (or vice versa), the mixing angle (which depends on the product of both entries) is suppressed, whereas the mass eigenvalues (dependent on the sum of both entries) are governed by the larger entry.

\addcontentsline{toc}{part}{References}
\bibliographystyle{utphys}
\bibliography{biblio_AADL}{}

\end{document}